\documentclass{ws-ijmpb}

\begin{document}
\newcommand{\cdash}{${}^{\mbox{-}}$}

\markboth{G.~L.~Klimchitskaya, U.~Mohideen \& V.~M.~Mostepanenko}
{Control of the Casimir force using semiconductor test bodies}

%
\catchline{}{}{}{}{}
%

\title{CONTROL OF THE CASIMIR FORCE USING SEMICONDUCTOR TEST BODIES}

\author{G.~L.~Klimchitskaya}
\address{North-West Technical University, Millionnaya Street 5,
St.Petersburg, 191065, Russia}

\author{U.~Mohideen}
\address{Department of Physics and
Astronomy, University of California, Riverside, California 92521,
USA}

\author{V.~M.~Mostepanenko}
\address{Noncommercial Partnership ``Scientific Instruments'',
Tverskaya Street 11, Moscow, 103905, Russia}

\maketitle

\begin{history}
\received{Day Month Year}
\end{history}

\begin{abstract}
We describe experimental and related theoretical work on the
measurement of the Casimir force using semiconductor test bodies.
This field of research started in 2005 and several important and
interesting results have already been obtained. Specifically,
the Casimir force or its gradient were measured in the configuration
of an Au-coated sphere and different semiconductor surfaces.
It was found that the force magnitude depends significantly on the
replacement of the metal with a semiconductor and on the concentration
of charge carriers in the semiconductor material. Special attention
is paid to the experiment on the optical modulation of the
Casimir force. In this experiment the difference Casimir force
between an Au-coated sphere and Si plate in the presence and in
the absence of laser light was measured. Possible applications
of this experiment are discussed, specifically, for the
realization of the pulsating Casimir force in three-layer
systems. Theoretical problems arising from the comparison of the
experimental data for the difference Casimir force with the
Lifshitz theory are analyzed. We consider the possibility to
control the magnitude of the Casimir force in phase transitions
of semiconductor materials. Experiments on measuring the
Casimir force gradient between an Au-coated sphere and Si plate
covered with rectangular corrugations of different character
are also described. Here, we discuss the interplay between the
material properties and nontrivial geometry and the applicability
of the proximity force approximation. The review contains
comparison between different experiments and analysis of their
advantages and disadvantages.
\end{abstract}


\section{Introduction}

Historically the Casimir force was discovered\cite{1}
as a universal attraction between two parallel electrically
neutral ideal metal plates in vacuum separated with a gap of
width $a$. This force per unit area of the plates (i.e., the
pressure) is given by
\begin{equation}
P(a)=-\frac{\pi^2\hbar c}{240a^4},
\label{eqn1}
\end{equation}
\noindent
i.e., it depends only on the Planck constant $\hbar$,
velocity of light $c$ and does not depend on the electric
charge or any other interaction constants.
According to quantum field theory the energy of the
electromagnetic field in the vacuum state in free space is
infinitely large and all physical energies are measured from
the energy of vacuum. If two ideal metal planes are placed in free
space, the tangential component of the electric field and the
normal component of the magnetic induction must vanish on the
planes. As a result, not all zero-point oscillations, whose
energies taken together create the infinite vacuum energy,
are allowed. In spite of the fact that the energy of
allowed oscillations is still equal to infinity, after
subtraction of the vacuum energy in free space, one arrives
at the finite energy per unit area
\begin{equation}
E(a)=-\frac{\pi^2\hbar c}{720a^3}.
\label{eqn2}
\end{equation}
\noindent
It is obvious that the Casimir pressure (\ref{eqn1}) is obtainable
as the negative derivative of Eq.~(\ref{eqn2}) with respect to
separation.

The expressions (\ref{eqn1}) and (\ref{eqn2}) are of somewhat
academic character because they are derived using the idealization
of ideal metal planes. Nevertheless, in a few decades after
Casimir's discovery a lot of work has been done on
the generalization
of these results on, for instance, ideal metal rectangular
boxes, spheres, cylinder and more complicated configurations,
including fields of different spin.\cite{2}\cdash\cite{5}
In so doing, it was found that the Casimir force can be both
attractive and repulsive (as, for instance, for spheres and
rectangular boxes with some specific ratio of sides).
A breakthrough was achieved by Lifshitz\cite{6,7} who
generalized the Casimir force for the case of two
thick plates made of
real materials decribed by the frequency-dependent dielectric
permittivities $\varepsilon^{(1,2)}(\omega)$ and developed the unified
theory of the van der Waals and Casimir force.
According to the Lifshitz theory, the free energy
(per unit area) of the dispersion
interaction of two material semispaces separated with a gap of
width $a$ at temperature $T$ in thermal equilibrium is
given by
\begin{equation}
{\cal F}(a,T)=\frac{k_BT}{2\pi}\sum_{l=0}^{\infty}
{\vphantom{\sum}}^{\prime}
\int_{0}^{\infty}k_{\bot}dk_{\bot}\sum_{\alpha}
\ln\left[1-r_{\alpha}^{(1)}r_{\alpha}^{(2)}\,e^{-2aq_l}\right].
\label{eqn3}
\end{equation}
\noindent
Here, $k_B$ is the Boltzmann constant,
$k_{\bot}$ is the magnitude of the projection of the wave
vector on the boundary plane $(x,y)$,
$q_l^2=k_{\bot}^2+\xi_l^2/c^2$,
$\xi_l=2\pi k_BTl/\hbar$ with $l=0,\,1,\,2,\,\ldots$ are
the Matsubara frequencies, and the primed summation sign
means that the term with $l=0$ is multiplied by 1/2.
The reflection coefficients on the first and second plates
($n=1,\,2$) for the electromagnetic waves
with transverse magnetic ($\alpha={\rm TM}$) and
transverse electric ($\alpha={\rm TE}$) polarizations
calculated at the imaginary Matsubara frequencies are
\begin{eqnarray}
r_{\rm TM}^{(n)}&\equiv&
r_{\rm TM}^{(n)}(i\xi_l,k_{\bot})=
\frac{\varepsilon^{(n)}(i\xi_l)q_l-k_l^{(n)}}{\varepsilon^{(n)}(i\xi_l)q_l+k_l^{(n)}},
\nonumber \\
r_{\rm TE}^{(n)}&\equiv&
r_{\rm TE}^{(n)}(i\xi_l,k_{\bot})=
\frac{q_l-k_l^{(n)}}{q_l+k_l^{(n)}},
\label{eqn4}
\end{eqnarray}
\noindent
where
\begin{equation}
k_l^{(n)}\equiv k_l^{(n)}(i\xi_l,k_{\bot})=\left[k_{\bot}^2+
\varepsilon^{(n)}(i\xi_l)\frac{\xi_l^2}{c^2}\right]^{1/2}.
\label{eqn5}
\end{equation}
\noindent
The pressure of the dispersion interaction is expressed as
\begin{eqnarray}
P(a,T)&=&-\frac{\partial{\cal F}(a,T)}{\partial a}
\nonumber \\
&=&
-\frac{k_BT}{\pi}\sum_{l=0}^{\infty}
{\vphantom{\sum}}^{\prime}
\int_{0}^{\infty}q_lk_{\bot}dk_{\bot}\sum_{\alpha}
\left[\frac{e^{2aq_l}}{r_{\alpha}^{(1)}r_{\alpha}^{(2)}}-1\right]^{-1}.
\label{eqn6}
\end{eqnarray}

The Lifshitz formulas (\ref{eqn3}) and  (\ref{eqn4}) were
originally derived using an assumption that dielectric materials
are characterized by randomly fluctuating sources of
electromagnetic fields. The pressure between semispaces was
found as the $zz$-component of the Maxwell energy-momentum
tensor statistically averaged using the fluctuation-dissipation
theorem. Later Eq.~(\ref{eqn3}) was
rederived\cite{5,8}\cdash\cite{10} in Casimir's spirit as the
finite difference of two infinite free energies of the
fluctuating electromagnetic field in the presence of two
semispaces with the standard continuity boundary conditions on
their surfaces and in the free space.
It was shown\cite{5,7} that in the limiting case of short
separation distances between the semispaces (well below the
characteristic absorption wavelength of the semispace materials)
the free energy and the force of dispersion interaction coincide
with the commonly known nonrelativistic van der Waals energy and
force.

When materials of semispaces are characterized by the zero
temperature, i.e., $T=0\,$K, the Matsubara frequencies $\xi_l$ are
replaced with a continuous frequency $\xi$ and the sum over the
Matsubara frequencies is replaced with an integral:
\begin{equation}
{k_BT}\sum_{l=0}^{\infty}{\vphantom{\sum}}^{\prime}
\to\frac{\hbar}{2\pi}\int_{0}^{\infty}d\xi.
\label{eqn6a}
\end{equation}
\noindent
In so doing the quantities $q_l$ and $k_l^{(n)}$ are replaced for
continuous quantities $q$ and $k^{(n)}$, and the Lifshitz formula
for the free energy (\ref{eqn3}) turns into the energy at zero
temperature
\begin{equation}
E(a)=\frac{\hbar}{4\pi^2}\int_{0}^{\infty}d\xi
\int_{0}^{\infty}k_{\bot}dk_{\bot}\sum_{\alpha}
\ln\left[1-r_{\alpha}^{(1)}r_{\alpha}^{(2)}\,e^{-2aq}\right].
\label{eqn6b}
\end{equation}
\noindent
In a similar way, Eq.~(\ref{eqn6}) for the Casimir pressure turns
into
\begin{equation}
P(a)=-\frac{\hbar}{2\pi^2}\int_{0}^{\infty}d\xi
\int_{0}^{\infty}qk_{\bot}dk_{\bot}\sum_{\alpha}
\left[\frac{e^{2aq}}{r_{\alpha}^{(1)}r_{\alpha}^{(2)}}-1\right]^{-1}.
\label{eqn6c}
\end{equation}
\noindent
Here, the reflection coefficients preserve their form (\ref{eqn4})
with the replacements indicated above.
For metallic semispaces at large separations,
where relativistic retardation effects are dominant, the
force per unit area (\ref{eqn6c}) and energy (\ref{eqn6b})
of dispersion interaction
coincide\cite{4,5,11} with the Casimir results (\ref{eqn1})
and (\ref{eqn2}), respectively.
Note that Eqs.~(\ref{eqn6b}) and (\ref{eqn6c}) are sometimes
used for the interpretation of experiments performed at room
temperature $T=300\,$K. This approach is not self-consistent
because the dielectric permittivities at room temperature
are substituted into the Lifshitz formulas at zero temperature.
In addition it was shown\cite{19} that the quantity obtained from
(\ref{eqn6b}) in this way does not coincide with the energy,
as it is defined in thermodynamics.

Most of theoretical and experimental work in the Casimir effect
was done for metallic test bodies.\cite{3}\cdash\cite{5}
It was experimentally demonstrated\cite{12}\cdash\cite{15}
that at separations below a micrometer the Casimir force is
strongly influenced by conduction electrons (the effect of skin
depth) and by the surface roughness.
By and large metallic bodies have major advantages in comparison
with dielectric ones as they ensure low electrostatic charges
and low residual potential difference between the surfaces.
Actually, a test body coated with a metal film of a few tens of
nanometers thickness acts like it is made of a thick metal.
Experiments on measuring the gradient of the Casimir force
between an Au coated sphere and an Au coated
plate\cite{5,15}\cdash\cite{18a} (using the proximity force
approximation, this quantity can be recalculated as the
Casimir pressure between two parallel plates) led to important
conclusions on the nature of interaction between thermal
electromagnetic fluctuations and metals. Specifically, it was
demonstrated that the experimental data exclude theoretical
predictions of the Lifshitz theory at a 99.9\% confidence
level if relaxation properties of conduction electrons are
taken into account. On the theoretical side, it was shown\cite{19,20}
that inclusion of relaxation properties of conduction electrons
into the Lifshitz theory results in a violation of the Nernst
heat theorem. {}From this it was concluded\cite{21} that
there may be some deep differences in the response of a
physical system to real and fluctuating electromagnetic fields.

Although metals are most convenient for high precision experiments
on the Casimir force, measurements of dispersion forces using
semiconductor test bodies open new opportunities for both
fundamental physics and numerous applications. It is common
knowledge that semiconductors are the most important materials
used in nanotechnology, and their conductivity properties range
from metallic to dielectric.
An aim of considerable scientific and technological promise
is to control the magnitude of the Casimir force varying it
from large to small, and, if possible, even changing attraction
for repulsion and vice versa. The use of semiconductor test
bodies suggests several ways on how to achieve this aim.
The reflectivity of a semiconductor surface can be changed over
a wide frequency range by changing the density of charge
carriers. The latter can be done in a number of ways, for
instance, through variation of the temperature, using different
kinds of doping, or via illumination of the semiconductor
surface with laser light. This makes it possible to examine the
influence of semiconductor material properties on the Casimir
force and use the obtained results to modulate the magnitude
and separation dependence of the force. We note that in
comparison to dielectrics, semiconductors with a reasonably high
density of charge carriers have the same advantage as metals,
i.e., they do not have problems such as
 accumulation of charges and screening effects.

An attempt to measure dispersion forces on a semiconductor
surface and to modify them by light was undertaken long
ago\cite{22} in the configuration of a glass lens and a Si plate
and between  a glass lens coated with amorphous Si and a Si plate.
However, no force change was found on illumination at separations
below 350\,nm, where it should be most pronounced. This might be
explained by the fact that the use of glass (and also high
resistivity Si) leads to uncontrolled electric forces.
The influence of laser light on the Casimir force between two
Si plates and between an Au sphere and a Si plate was
investigated theoretically.\cite{22a,48}

The modern stage in the investigation of dispersion forces using
semiconductor surfaces started in 2005 when the Casimir
attraction between a B-doped Si plate and a metal coated sphere
was measured\cite{23} using an atomic force microscope (AFM).
At this stage, most of experiments were performed in the configuration
of an Au-coated sphere (or a spherical lens) above a semiconductor
plate drawing on experience accumulated in previous measurements
of the Casimir force between metallic test bodies. The present review
collects and discusses all main results on the Casimir force
obtained using semiconductor surfaces.
Section 2 is devoted to the Casimir force between a spherical
surface and a plate made of different semiconductors.
We begin with the experiment which revealed\cite{23,24} that the
measured Casimir force for a plate made of $p$-type Si is
markedly different from the calculation results for
high-resistivity dielectric Si. In this experiment it was
demonstrated that the density of charge carriers in a semiconductor
material influences the Casimir force between metallic and
semiconductor test bodies. Then we discuss the experiment where
the Casimir force between an Au-coated sphere and two $n$-type
Si plates with different densities of charge carriers was
successively measured.\cite{25} Through this, the difference in the
Casimir forces for different Si plates was calculated, and, thus,
 more clear evidence for the effect of the charge carrier
density on the Casimir force was provided.
We next consider the experiment on measuring the Casimir force
between a metal coated sphere and an indium tin oxide (ITO)
plate.\cite{26,29a} In this experiment, the reduction of the Casimir
force by up to a factor of two, when compared with the case of two test
bodies made of good metal, was reported.
In the end of the section we describe the experiment\cite{27} on
measuring the Casimir force between a Ge spherical lens
with the curvature radius 15.1\,cm and Ge plate. The measurement
data of this experiment were found consistent\cite{27} with five
different theoretical approaches to the thermal Casimir force.
In this connection we indicate a possible source of noncontrolled
systematic errors inherent to all measurements of the Casimir
force using lenses of centimeter-size curvature radii.

In Section 3 the experiment on the optical modulation\cite{28,28a} of
the Casimir force is considered. Here, the difference in the
Casimir forces between an Au-coated sphere and a Si plate in the
presence and in the absence of laser light on the plate is
measured. The frequency and power of the laser pulses are selected in
such a way that the charge carrier density in Si during the bright
phase is almost 5 orders of magnitude higher than in the dark
phase. In so doing, Si in the bright phase was a metal-type
semiconductor (conductivity of such a material goes to a nonzero
limit with vanishing temperature). In the dark phase Si was a
semiconductor of dielectric-type with vanishing conductivity
in the limit of zero temperature. The comparison between the
measured difference Casimir force and the Lifshitz theory has
led to an unexpected conclusion. It was demonstrated that the
Lifshitz theory with inclusion of dc conductivity of
Si in the dark phase is excluded by the data at a 95\% confidence
level. At the same time the measurement data were found
consistent with the Lifshitz theory if the dc conductivity of
Si in the dark phase is omitted. Keeping in mind that dc
conductivity is connected with the relaxation properties of free
charge carriers, one can conclude that for dielectric-type
semiconductors the optical modulation experiment plays the same
role as the experiments\cite{16}\cdash\cite{18a} mentioned above
for metals. In both cases the experimental data are inconsistent
with the Lifshitz theory taking into account relaxation properties
of free charge carriers. {}From the theoretical side it was
proved\cite{29}\cdash\cite{31} that the inclusion of the  dc
conductivity in the model of the dielectric response for dielectrics
and dielectric-type semiconductors results in contradiction
between the Lifshitz theory and the Nernst heat theorem.
This is also somewhat analogous with the respective situation for
metals. In Sec.~3 we also discuss recent
attempts\cite{32}\cdash\cite{34} to modify the Lifshitz theory
and show that they are excluded\cite{35}\cdash\cite{38}
by the results of the optical modulation experiment at a 70\%
confidence level. Thus, the optical modulation experiment,
which demonstrated the possibility to periodically change the
Casimir force with light, is important not only for applications,
but for fundamental physics as well.

Section 4 is devoted to the investigation of the changes of the Casimir
force in phase transitions. Experiments on this subject involve
an Au-coated sphere and a plate made of a phase change material.
In the previous experiment\cite{39} an amorphous
sample of AgInSbTe was used as the plate. Under annealing
 the plate material
was switched from the amorphous to crystalline
phase. In the proposed experiment\cite{40} the VO${}_2$ film
deposited on a sapphire plate undergoes dielectric-to-metal
phase transition with an increase in temperature.

In Sec.~5 we discuss measurements of the gradient of the Casimir
force between an Au-coated sphere and a Si surface covered
with rectangular corrugations.\cite{41,42}
These measurements allow to observe effects that include the role
of geometry and the optical properties of a semiconductor
material.

The possibility to obtain a pulsating Casimir force in a three layer
system, where at least one layer is made of a semiconductor, is
considered in Sec.~6. A pulsation, i.e., periodic change of
attraction with repulsion and vice versa, can be achieved\cite{43}
through the illumination of a semiconductor
layer with laser pulses.

In Sec.~7 the reader will find our conclusions and discussion.

The presentation below is based on original publications.
In several cases, however, we make a more exact analysis of the
experimental data and their comparison with theory. These cases
are indicated in the text. Many figures illustrating
the comparison between experiment and theory are made
specially for this review. We write equations
in the Gaussian system of units. Some values of the experimentally
measured quantities  are given, however,
in the International System of units.

\section{Casimir force between a spherical surface and a plate made
of different semiconductor materials}

Here, we discuss the results of four experiments, where a
semiconductor plate interacts via the Casimir force with an
Au-coated sphere or a Ge spherical lens. In these experiments,
different semiconductors are used. In all cases it was confirmed
that the replacement of the metallic plate with a semiconductor one
markedly diminishes the magnitude of the Casimir force.
However, as is shown below, different experiments, depending
on their precision, contain different information with respect
to theoretical models consistent with the experimental data.

\subsection{{\rm B}-doped Si plate}

\begin{figure}[t]
\vspace*{0.3cm}
\hspace*{-6.1cm}
\psfig{file=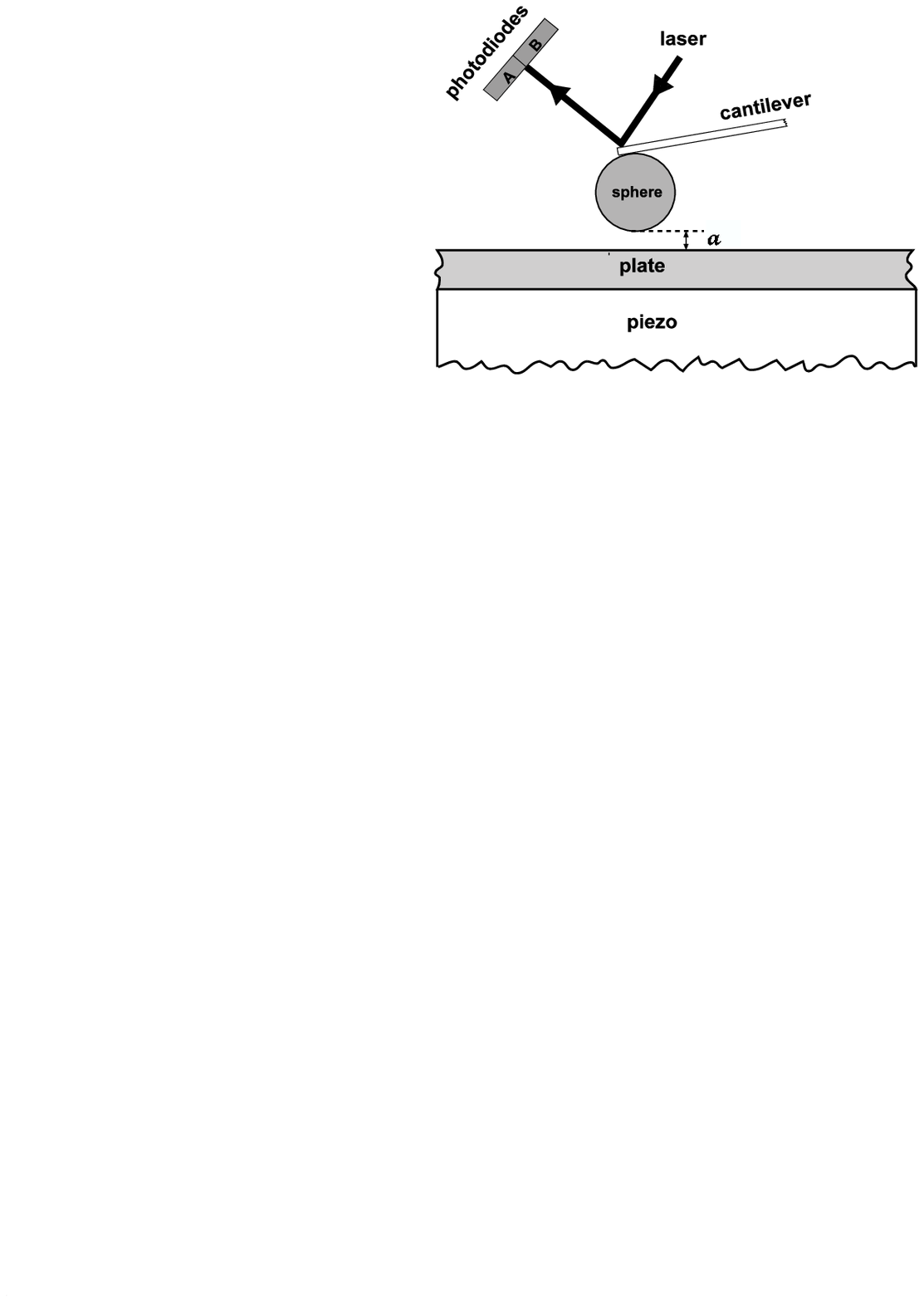,width=6.5in}
\vspace*{-20.cm}
\caption{Schematic diagram of the experimental setup using an
atomic force microscope.}
\label{aba:fig1}
\end{figure}

In this experimen,\cite{23,24} the Casimir force was measured
between an Au-coated sphere of diameter
$2R=202.6\pm 0.3\,\mu$m and a single-crystal Si plate
doped with B. The $p$-type Si plate had an area
$5\times 10\,\mbox{mm}^2$ and a thickness of $350\,\mu$m.
The resistivity of the plate at room temperature
$T=300\,$K was $\rho=0.0035\,\Omega\,$cm which corresponds to
a charge carrier density
$n\approx (2.9-3.2)\times 10^{19}\,\mbox{cm}^{-3}$.
This is a metal-type semiconductor whose conductivity does not
go to zero when the temperature vanishes. For typical metals,
however, resistivities are two or three orders of magnitude
lower. The same experimental setup\cite{12}\cdash\cite{15}
as for two metallic test bodies (with much higher vacuum
of $2\times 10^{-7}\,$Torr) was used to measure the Casimir force.
The schematic of the setup is shown in Fig.~\ref{aba:fig1},
where force between the sphere and the plate causes the cantilever
of an AFM to flex. This flexing is detected by the deflection
of the laser beam leading to a difference signal between the
photodiodes A and B. The difference signal is calibrated by means
of an electrostatic force. Here we only discuss the obtained
experimental results and their comparison with theory and do not
consider details of the setup and calibration procedures which are
fully described in the original publications\cite{23,24} and
in Refs.~\refcite{5,15}.

\begin{figure}[t]
\vspace*{-8.2cm}
\hspace*{-1.9cm}
\psfig{file=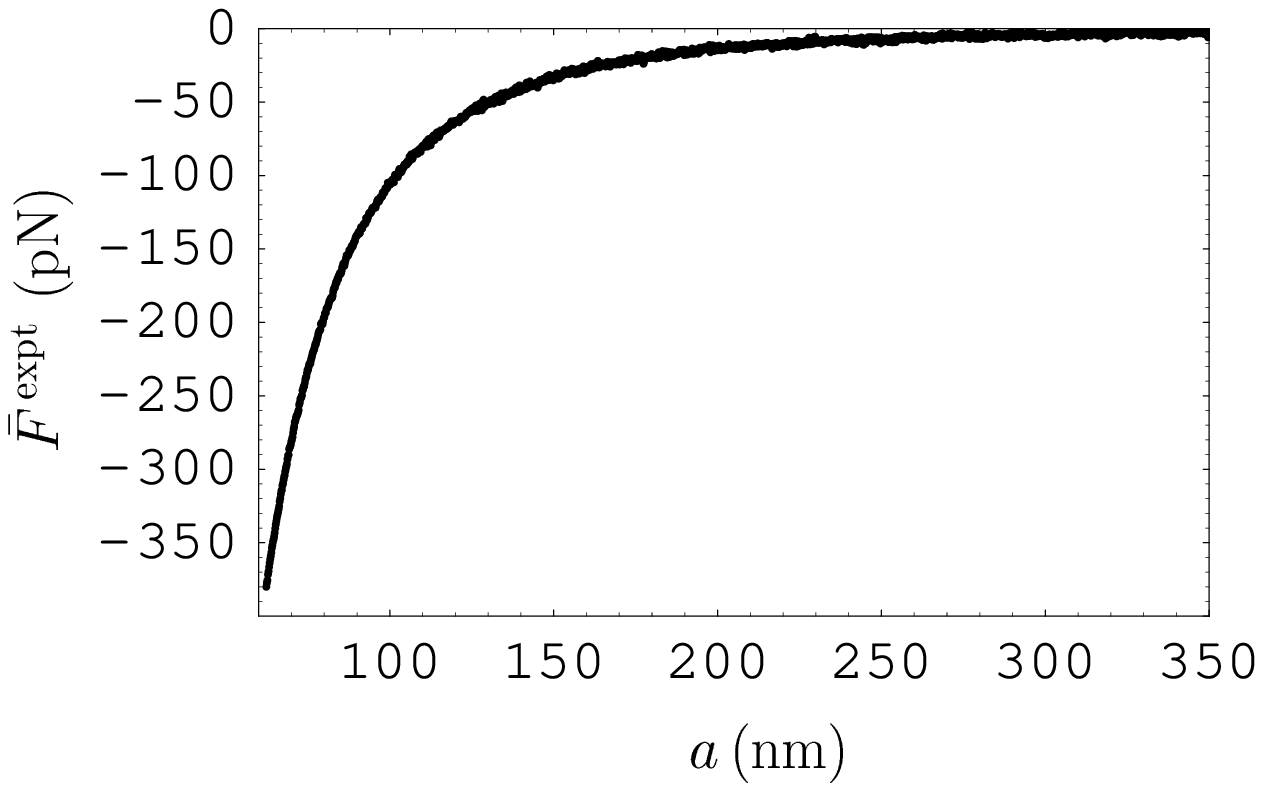,width=6.5in}
\vspace*{-9cm}
\caption{The mean measured Casimir force between an Au-coated sphere
of $101.3\,\mu$m radius and Si plate of $0.0035\,\Omega\,$cm resistivity
is shown as the function of separation.}
\label{aba:fig2}
\end{figure}

The Casimir force between the sphere and the plate was measured
as a function of separation. In so doing, the sphere was kept
grounded while a compensating voltage was applied to the plate
to cancel the residual potential difference
$V_0=-0.114\pm 0.002\,$V (the latter was found to be independent
of separation). The separation was varied continuously from
about $6\,\mu$m to small distances by applying continuous
triangular voltages at 0.02\,Hz to the piezoelectric actuator.
However, the measurement data becomes meaningful only below
350\,nm (at $a=350\,$nm the total experimental error is
approximately equal to the force magnitude). The force data were
collected at equidistant points $a\geq 62.33\,$nm separated by
0.17\,nm. The mesurement was repeated  65 times. The mean
force $\bar{F}^{\,\rm expt}$ is shown in Fig.~\ref{aba:fig2} as
a function of separation. The total experimental error of the
force measurements, $\Delta^{\!\rm tot}{F}^{\,\rm expt}$,
determined at a 95\% confidence level, was equal to 3.33\,pN.
It is mostly determined by the random error. The error in the
measurement of absolute separation was $\Delta a=0.8\,$nm.

Now we compare the experimental results with theory.
In the original publications,\cite{23,24}
as well as in Refs.~\refcite{5,15},
the theoretical Casimir forces for the needs of this experiment
were computed at $T=0$, whereas the measurements were performed
at $T=300\,$K.
This was justified by the smallness of thermal effects at short
separations.
Keeping in mind increased attention to the problem of thermal
Casimir force during the last few years, here we recalculate
theoretical results using the Lifshitz formula (\ref{eqn3}) at
$T\neq 0$. According to the PFA,\cite{5,15,44} the Casimir force
between a sphere and a plate is given by
\begin{equation}
F(a,T)=2\pi R{\cal F}(a,T),
\label{eqn7}
\end{equation}
\noindent
where the free energy ${\cal F}(a,T)$ is given in Eq.~(\ref{eqn3}).
Errors arising from the use of an approximate Eq.~(\ref{eqn7})
are less\cite{45} than $a/R\approx 0.1$\% at separations of
about 100\,nm.

For the dielectric permittivity of Au,
$\varepsilon^{(1)}(i\xi_l)$ in Eq.~(\ref{eqn3}), we use the
generalized plasma-like model
\begin{equation}
\varepsilon^{(1)}(i\xi_l)=\frac{{\omega_p^{(1)}}^2}{\xi_l^2}+
\varepsilon_{\rm ce}^{(1)}(i\xi_l),
\label{eqn8}
\end{equation}
\noindent
where $\varepsilon_{\rm ce}^{(1)}(i\xi_l)$ is the dielectric permittivity
determined by the core electrons, and $\omega_p^{(1)}=9.0\,$eV is
the plasma frequency.
At frequencies below $\xi=15\,$eV the analytical expression for
$\varepsilon_{\rm ce}^{(1)}(i\xi_l)$ in terms of the six-oscillator
model is available.\cite{5,15,18a} This allows precise computations
of the Casimir force (\ref{eqn7}) at $a\geq 100\,$nm.
At shorter separations, a more precise representation for
$\varepsilon_{\rm ce}^{(1)}(i\xi_l)$ is required. To obtain this,
one should first consider ${\rm Im}\varepsilon_{\rm ce}^{(1)}(\omega)$
which is the difference between $2n_1(\omega)n_2(\omega)$
[where $n_1(\omega)$ and $n_2(\omega)$ are the tabulated data for
real and imaginary parts of the complex index of refraction of Au
at different frequencies\cite{46}] and the respective contribution
from conduction electrons. Note that ${\rm Im}\varepsilon_{\rm ce}^{(1)}(\omega)$
vanishes at low frequencies. Then $\varepsilon_{\rm ce}^{(1)}(i\xi_l)$
is found from the Kramers-Kronig relation
\begin{equation}
\varepsilon_{\rm ce}^{(1)}(i\xi_l)=1+\frac{2}{\pi}\int_0^{\infty}
\frac{\omega\,{\rm Im}\varepsilon_{\rm ce}^{(1)}(\omega)}{\omega^2+\xi_l^2}\,
d\omega.
\label{eqn9}
\end{equation}

A frequently used alternative method to find the dielectric
permittivity of Au along the imaginary frequency axis is based
on the extrapolation of the optical data by means of the Drude
dielectric function
\begin{equation}
\varepsilon_{D}^{(1)}(\omega)=1-
\frac{{\omega_p^{(1)}}^2}{\omega(\omega+i\gamma^{(1)})},
\label{eqn10}
\end{equation}
\noindent
where $\gamma^{(1)}=0.035\,$eV is the relaxation parameter.
In the framework of this method\cite{47} the tabulated data for
$2n_1(\omega)n_2(\omega)$ are extrapolated for low frequencies
by using the imaginary part of $\varepsilon_{D}^{(1)}(\omega)$
defined in Eq.~(\ref{eqn10}).
Then the dielectric permittivity of Au at the imaginary Matsubara
frequencies, $\tilde{\varepsilon}^{(1)}(i\xi_l)$, is obtained
by means of the standard Kramers-Kronig relation.

\begin{figure}[t]
\vspace*{-7.7cm}
\hspace*{-1.9cm}
\psfig{file=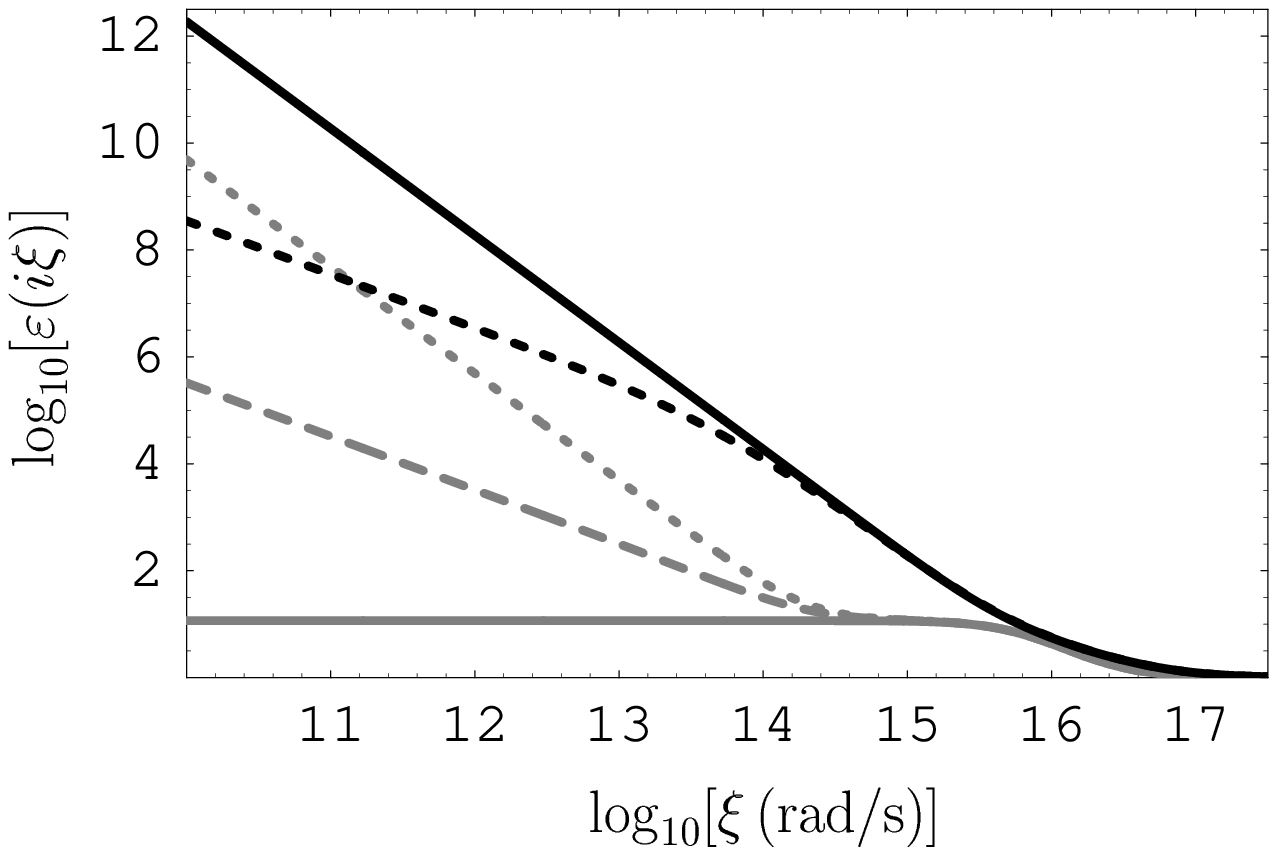,width=6.5in}
\vspace*{-9.cm}
\caption{The logarithms of the dielectric permittivity as a function
of the imaginary frequency for Au described by the generalized
plasma-like model (the solid black line), for Au described by the
tabulated optical data extrapolated by the Drude model (the dashed
black line), for high-resistivity Si (the solid grey line),
for Si described by the generalized
plasma-like model (\ref{eqn11}) (the dotted grey line), and
for Si described by the Drude-like model (\ref{eqn12})
(the dashed grey line).
}
\label{aba:fig3}
\end{figure}
The dielectric permittivities ${\varepsilon}^{(1)}(i\xi_l)$
and $\tilde{\varepsilon}^{(1)}(i\xi_l)$ for Au are shown in
Fig.~\ref{aba:fig3} by the solid and dashed black lines,
respectively. The dashed line takes into account the relaxation
properties of conduction electrons, whereas the solid line
disregards relaxation of free charge carriers.
As was noted in Sec.~1, experiments with metallic test
bodies\cite{5,15}\cdash\cite{18a} exclude the permittivity
$\tilde{\varepsilon}^{(1)}(i\xi_l)$.
Keeping in mind that the latter describes correctly the
dielectric response of Au for a real electromagnetic field, this
might be connected with some deep unexplored differences
between real and fluctuating fields.
In the experiment under consideration in this section the
separation region of main interest is from 62 to 100\,nm
(using the force magnitudes in Fig.~\ref{aba:fig2},
one can conclude that the relative total
experimental error varies from only 0.87\% at the shortest
separation to 5.3\% at $a=120\,$nm and reaches 64\%
at $a=300\,$nm). In this separation region the use of the
dielectric permittivities ${\varepsilon}^{(1)}(i\xi_l)$
and $\tilde{\varepsilon}^{(1)}(i\xi_l)$ leads to only
negligible differences in the Casimir force specified
below.

The dielectric permittivity of dielectric Si with
resistivity $\rho_0=1000\,\Omega\,$cm is obtained using
the tabulated optical data\cite{46} for the complex index
of refraction and the Kramers-Kronig relation.
It is notated as ${\varepsilon}_{\rm ce}^{(2)}(i\xi_l)$
and shown by the solid grey line in Fig.~\ref{aba:fig3}.
Note that the analytical approximations to
${\varepsilon}_{\rm ce}^{(2)}(i\xi_l)$
suggested\cite{48,49} lead to errors of less than 1\%
in the magnitude of the Casimir force.
The resistivity of a B-doped Si plate used in the measurements
was much lower, and the doping  concentration indicated above
corresponds to a plasma frequency
$\omega_p^{(2)}\approx 7\times 10^{14}\,$rad/s and relaxation
parameter $\gamma^{(2)}\approx 1.5\times 10^{14}\,$rad/s.
Thus, disregarding the relaxation processes of charge carriers,
the dielectric permittivity of the Si plate
${\varepsilon}^{(2)}(i\xi)$ in Eq.~(\ref{eqn3}) can be
represented in the form
\begin{equation}
\varepsilon^{(2)}(i\xi)=\frac{{\omega_p^{(2)}}^2}{\xi^2}+
\varepsilon_{\rm ce}^{(2)}(i\xi).
\label{eqn11}
\end{equation}
\noindent
This is the generalized plasma-like model for metallic-type Si
with resistivity $\rho$. Alternatively, preserving the role of
relaxation properties of free charge carriers, instead of
Eq.~(\ref{eqn11}), one obtains
\begin{equation}
\tilde{\varepsilon}^{(2)}(i\xi)=\frac{{\omega_p^{(2)}}^2}{\xi(\xi+\gamma^{(2)})}+
\varepsilon_{\rm ce}^{(2)}(i\xi).
\label{eqn12}
\end{equation}
\noindent
The dielectric permittivities of Si $\varepsilon^{(2)}(i\xi)$ and
$\tilde{\varepsilon}^{(2)}(i\xi)$ are shown in Fig.~\ref{aba:fig3}
by the dotted and dashed grey lines, respectively.

Finally, for the comparison between experiment and theory one should
take the surface roughness into account which is not included in
Eqs.~(\ref{eqn3}) and (\ref{eqn7}). For this purpose the topographies
of both Au-coated sphere and Si plate were investigated using
an AFM. The fractions of the Au-coating on the sphere with
different heights were determined. The same was done for a Si
plate. Then the Casimir force was computed using
Eqs.~(\ref{eqn3}) and (\ref{eqn7}) for all fractions of the
surfaces and the results were geometrically averaged\cite{5,15,23,24}
to obtain the theoretical Casimir force between rough surfaces
$F^{\rm theor}(a)$. Here, $a$ is the separation distance between
the zero levels of the roughness profiles on both surfaces.
The contributions to the force from surface roughness were very
minor because the rms roughness was only 3.446 and 0.111\,nm
on the sphere and on the plate, respectively.

\begin{figure}[b]
\vspace*{-3.5cm}
\hspace*{-2.6cm}
\psfig{file=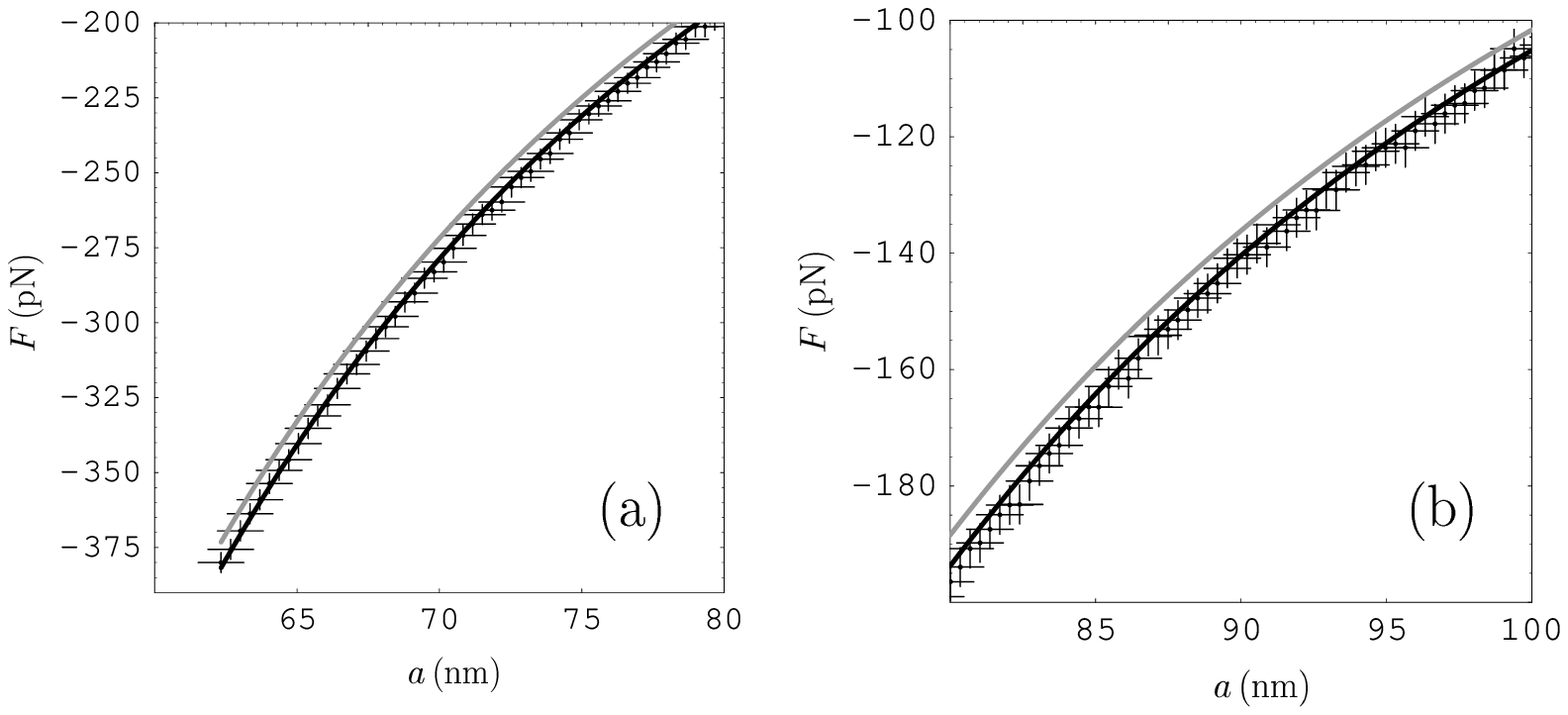,width=6.5in}
\vspace*{-14.5cm}
\caption{Comparison between the experimental data for the Casimir force
in the configuration of an Au-coated sphere and Si plate shown as crosses and
two theoretical approaches over the separation regions (a) from 62 to
80\, nm and (b) from 80 to 100\, nm (only each third cross is shown).
The arms of the crosses are determined at a 95\% confidence level.
For both the black and grey theoretical lines Au is described by
means of the generalized plasma-like model. Si is described either
by the generalized plasma-like model (the black line) or as a
high-resistivity material (the grey line).}
\label{aba:fig4}
\end{figure}
\begin{figure}[t]
\vspace*{-3.5cm}
\hspace*{-2.6cm}
\psfig{file=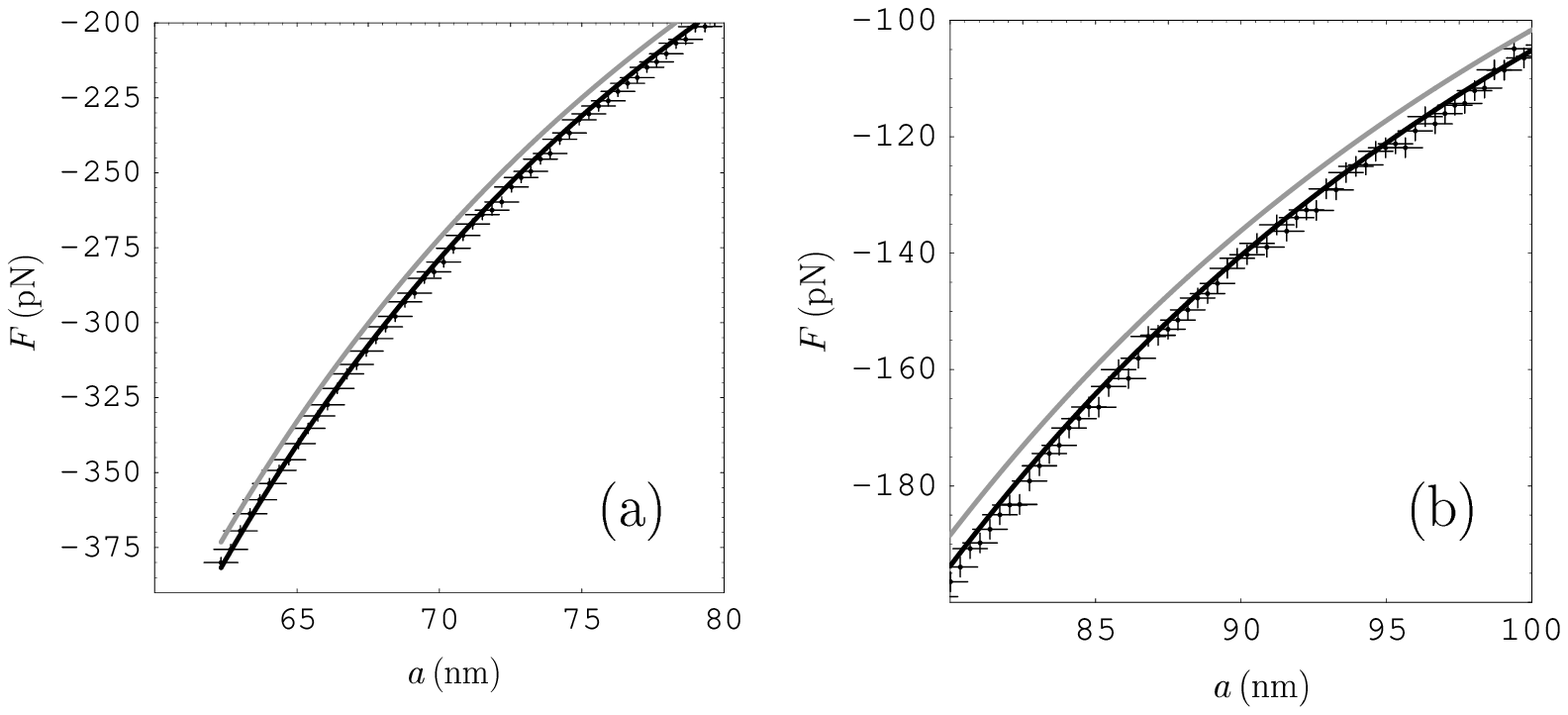,width=6.5in}
\vspace*{-14.5cm}
\caption{The same as in Fig.~\ref{aba:fig4}, but the
arms of the crosses are determined at a 70\% confidence level.}
\label{aba:fig5}
\end{figure}

In Fig.~\ref{aba:fig4} we present comparison between the experimental
data and different theoretical approaches over the separation
regions (a) from 62 to 80\,nm and (b) from 80 to 100\,nm.
The experimental data for the mean Casimir forces are shown as
crosses.
The arms of the crosses are equal to twice the respective absolute
errors determined at a 95\% confidence level.
For both black and grey lines representing the computational
results Au sphere is described by the generalized plasma-like
model in Eq.~(\ref{eqn8}) (the solid black line in Fig.~\ref{aba:fig3}).
The Si plate is described either by the generalized plasma-like
model of Eq.~(\ref{eqn11}) (the black line) or by the dielectric
permittivity $\varepsilon_{\rm ce}^{(2)}(i\xi)$ of a high resistivity
Si (the grey line).
As can be seen in Fig.~\ref{aba:fig4}(a), many experimental
crosses at least touch the grey line at $a<75\,$nm.
This means that at separations below 75\,nm both theoretical
approaches are consistent with the data within a 95\% confidence
interval. {}From Figs.~\ref{aba:fig4}(a) and \ref{aba:fig4}(b),
however, one can conclude that within the separation interval
from 75 to 100\,nm the experimental data are consistent with
the description of Si using Eq.~(\ref{eqn11}) (the black line),
but exclude the model of high-resistivity Si (the grey line).
In Fig.~\ref{aba:fig5}(a,b) a similar comparison between experiment
and theory is shown, but the experimental crosses are plotted
at a 70\% confidence level.
For this purpose the vertical arms of all crosses determined by
the random errors were divided by two as follows from the
normal or Student distribution. The horizontal arms determined by
the systematic errors were divided by 1.357 based on the
uniform distribution (see below for a more detailed discussion).
{}From this figure it can be
concluded that the model of high-resistivity dielectric Si
is excluded by the data over a wide interval from 70 to
100\,nm.

Another method for the comparison between experiment and theory
considers differences between the theoretical and mean experimental
Casimir forces, $F^{\rm theor}(a_i)-\bar{F}^{\rm expt}(a_i)$,
and the confidence intervals for these differences taking into
account both the experimental and theoretical errors.
\begin{figure}[t]
\vspace*{-5.cm}
\hspace*{-2.6cm}
\psfig{file=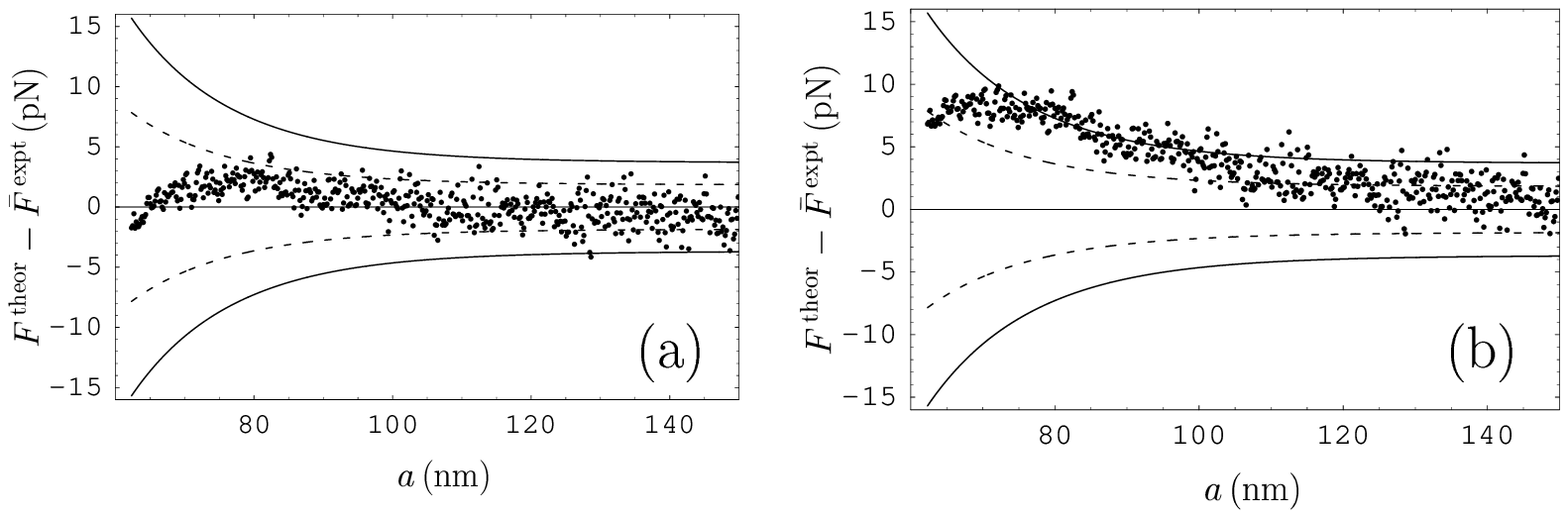,width=6.5in}
\vspace*{-14.5cm}
\caption{The differences between theoretical and mean experimental
Casimir forces in the configuration of an Au-coated sphere and Si plate
are indicated as dots. The borders of the confidence intervals
determined at a 95\% and 70\% confidence levels are shown as the
solid and dashed lines, respectively. The theoretical Casimir
forces are computed using the generalized plasma-like model for Au
and describing Si by (a) the generalized plasma-like model and
(b) the model of high-resistivity material.}
\label{aba:fig6}
\end{figure}
In Fig.~\ref{aba:fig6}(a,b) the force differences are indicated as
dots and the borders of the confidence intervals determined at a 95\%
and 70\% confidence levels are shown as the solid and dashed lines,
respectively.
Theoretical Casimir forces are computed using the description
 of Si (a) by the generalized plasma-like model (\ref{eqn11}) and
 (b) the model of high-resistivity Si. Here, the half-widths
 of the confidence intervals ad a 95\% and 70\% confidence levels
 are related as for a normal distribution, i.e.,
 $\Xi_{0.95}/\Xi_{0.70}=2$ because the distribution of the force
 differences was found to be close to the normal one (this is in
 fact a conservative assumption
 as for the Student distribution with different
 numbers of degrees of freedom $\Xi_{0.95}/\Xi_{0.70}>2$ holds).
 As can be seen in Fig.~\ref{aba:fig6}(a), the most of dots over
 the entire separation interval from 62 to 150\,nm lie inside
 both 70\%  and 95\% confidence intervals. This means that the
description of Si using the generalized plasma-like model (\ref{eqn11})
is consistent with the data. {}From Fig.~\ref{aba:fig6}(b) it is
seen that within the interval from 75 to 115\,nm more than 5\% of dots
are outside the 95\% confidence intervals. Thus, within this
interval the model of dielectric Si is excluded by the data at
a 95\% confidence level. {}From Fig.~\ref{aba:fig6}(b) one can
conclude also that at the 70\% confidence level the model of
high-resistivity Si is excluded by the data within a wider range
of separations from 65 to 145\,nm because here more than 30\%
of dots are outside the 70\% confidence intervals.

Now we demonstrate that at separations considered the generalized
plasma-like model used above for the description of dielectric
properties of Au and Si leads to almost the same results as the
tabulated optical data extrapolated to low frequencies by means
of the Drude model. As an example, at $a=62.33$, 104.83, and
147.33\,nm the magnitudes of theoretical Casimir forces between
a sphere and a plate, $|F^{\,\rm theor}|$, used in
Figs.~\ref{aba:fig4},\ \ref{aba:fig5} and \ref{aba:fig6}(a) are
the following: 382.760, 92.4342, and 36.0303\,pN.
If, alternatively, the tabulated optical data for both Au and
Si were extrapolated by the Drude model, we would get at
respective separations $|\tilde{F}^{\,\rm theor}|=382.393$,
91.8658, and 35.5928\,pN. This leads to only --0.09\%,
--0.61\% and --1.21\% relative differences at the same
respective separations. These differences are well below the
experimental errors in the experiment under consideration.

To conclude this section, the first measurement of the Casimir
force with semiconductor plate\cite{23,24} convincingly
demonstrated that by using semiconductors one can control
the force magnitude. If the plate in this experiment
were made of Au instead of Si with all other conditions and
parameters preserved, the magnitude of the Casimir force at
$a=62.33$, 104.83, and 147.33\,nm would be equal to
$|F_{\rm Au-Au}^{\,\rm theor}|=515.948$, 134.456 and
54.8289\,pN, respectively.
Comparing these values with the above magnitudes of the
Casimir force between an Au sphere and Si plate, we obtain
that through the replacement of Au with Si force is decreased
by 25.8\%, 31.3\% and 34.3\% at respective separations.
It was demonstrated also that charge carrier density in
metallic materials is an important characteristic for the
determination of the Casimir force. The experiment described
in the next section provides further confirmation to
these conclusions.

\subsection{Two {\rm P}-doped Si plates with different dopant
concentrations}

In the next experiment\cite{25} on measuring the Casimir force
between an Au-coated sphere and a semiconductor plate, two
P-doped Si plates  which possess
radically different charge carrier densities were compared.
The scheme of the
setup was already described in Sec.~2.1 (see Fig.~\ref{aba:fig1}).
In this case the diameter of the Au-coated polystyrene sphere
was equal to $2R=201.8\pm 0.6\,\mu$m. The two Si plates were
placed next to each other and had an area $4\times 7\,\mbox{mm}^2$
and a thickness of $500\,\mu$m. Two identically polished
single-crystal Si samples were chosen as plates. They were
$n$-type and doped with P. The resistivity of both plates was
measured using the four-probe techniques. The plates had a
$\rho_a\approx 0.43\,\Omega\,$cm (i.e., more than two orders
of magnitude higher than in the previous experiment with one Si
plate). This corresponds to a concentration of charge carriers
$n_a\approx 1.2\times 10^{16}\,\mbox{cm}^{-3}$.

One of the plates was  used in the
experiment as the first Si plate. The other plate was subjected
to thermal-diffusion doping in order to decrease the resistivity and
to increase the concentration of charge carriers.\cite{5,25}
As a result, the resistivity and the carrier density were
measured to be $\rho_b\approx 6.7\times 10^{-4}\,\Omega\,$cm  and
$n_b\approx 3.2\times 10^{20}\,\mbox{cm}^{-3}$.
This plate was used as the second Si plate in the experiment.
Its resistivity was more than 50 times less than the resistivity
of the plate in the experiment described in Sec.~2.1. In fact the
first plate was a semiconductor of dielectric-type and the
second of metallic-type (in the sense that conductivity of the
first goes to zero with vanishing temperature, whereas
conductivity of the second has a nonzero limit in the limit of
zero temperature).

\begin{figure}[t]
\vspace*{-6.cm}
\hspace*{-2.8cm}
\psfig{file=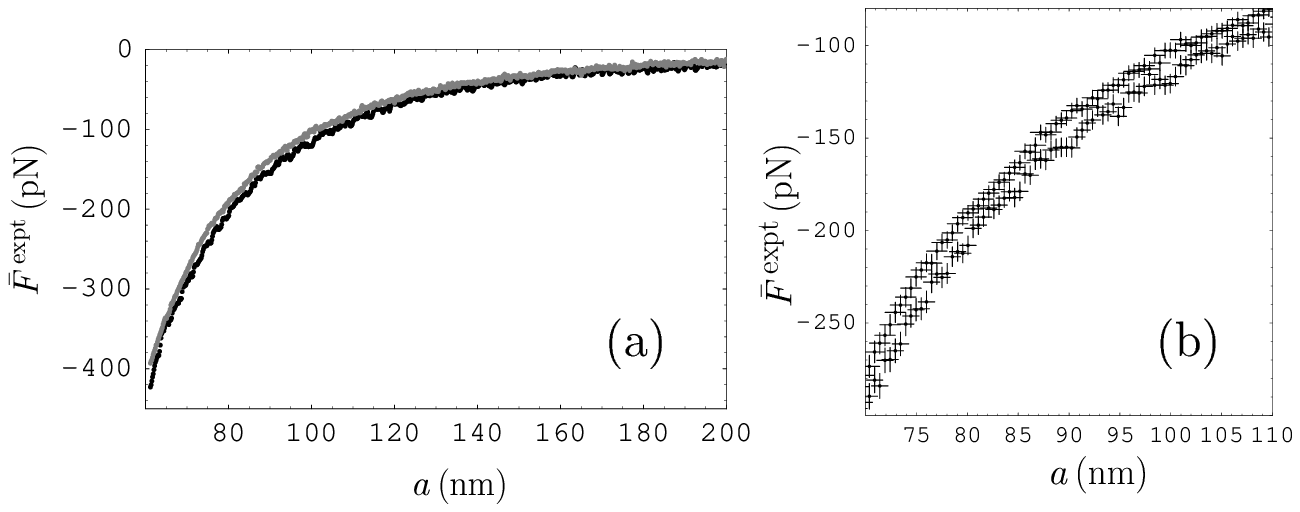,width=7.8in}
\vspace*{-17.2cm}
\caption{(a) Mean measured Casimir forces between an Au-coated sphere and
Si plates with high (the grey dots) and low (the black dots) resistivity
are shown as a function of separation.
(b) The same mean forces are shown as crosses with arms determined at a
95\% confidence level over a narrower range of separations (only every third
cross is plotted).}
\label{aba:fig7}
\end{figure}

The calibration procedures using the measurements of electric
forces were significantly improved as compared with previous
experiments on the Casimir force. Specifically, the parabolic
dependence of the signal on the applied voltage was used
to obtain the residual potential difference $V_0$ between the
grounded sphere and the plate.\cite{25,50}
For the first and second samples the residual potential
differences were found to be $V_{0a}=-0.341\pm 0.002\,$V
and $V_{0b}=-0.337\pm 0.002\,$V. Both were demonstrated
to be independent on separation within the limits of an
experimental error. The separation was varied continuously
in the same way as was described in Sec.~2.1 up to some
minimum value. The measurement of the Casimir force between
the sphere and the first plate was repeated 40 times over
the separation region from 61.19\,nm to 400\,nm.
The mean values of the obtained force $\bar{F}_a^{\,\rm expt}$
over the separation region below 200\,nm, where the results
are most meaningful, are shown in Fig.~\ref{aba:fig7}(a)
as the grey dots. Similar measurements of the force between
the sphere and the second plate was repeated 39 times
over the region from 60.51\,nm to 400\,nm.
The mean values $\bar{F}_b^{\,\rm expt}$, as a function of
separation, are shown in Fig.~\ref{aba:fig7}(a) as the black
 dots.

The total experimental error in both measurements depends on
separation and is determined by the random errors (the systematic
error determined at a 95\% confidence level is equal to only
1.2\,pN for both measurements and is separation-independent).
For example, for a measurement with the first plate (which is of
higher resistivity) the total experimental error determined at
a 95\% confidence level is equal to 8, 6, and 4\,pN at the
separations 61.19\,nm, 70\,nm, and $a\geq 80\,$nm, respectively.
For a measurement with the second plate (which is of
lower resistivity) the total experimental error determined at
the same confidence is equal to 11, 7, and 5\,pN at the
respective separations 61.51\,nm, 70\,nm, and $a\geq 80\,$nm.
For the first and second plates, separations were measured with
the absolute errors equal to 1 and 0.8\,nm, respectively.

To make sure that within some separation distance the
measurement results for the first and second plates do not
overlap, in Fig.~\ref{aba:fig7}(b) the data are plotted
as crosses. The arms of the crosses are determined by the
absolute errors, as explained in Sec.~2.1. {}From
Fig.~\ref{aba:fig7}(b) it can be seen that at most separations
within the interval from 70 to 110\,nm the measured Casimir
forces between the sphere and the first and second plates
belong to two individual lines separated with some gap.
Note that only each third experimental point is plotted in
Fig.~\ref{aba:fig7}(b) because in the presence of all points
the figure becomes unreadable. All crosses are plotted at
a 95\% confidence level.

Now we are in a position to compare the experimental data with
theory. This can be conveniently done using the second method
of comparison considered in Sec.~2.1 which is based on the
consideration of differences between theoretical and experimental
Casimir forces. The theoretical Casimir forces are computed using
Eqs.~(\ref{eqn3}) and (\ref{eqn7}) and taking surface roughness
into account by means of the geometrical averaging, as discussed
in Sec.~2.1. In so doing the dielectric permittivity of Au,
$\varepsilon^{(1)}(i\xi_l)$, is presented by Eq.~(\ref{eqn8}).
The dielectric permittivity of high-resistivity Si (the first plate)
along the imaginary Matsubara frequencies is given by
$\varepsilon_{\rm ce}^{(2)}(i\xi)$ obtained using the tabulated
optical data for the complex index of refraction (see Sec.~2.1).
The role of low concentration of charge carriers $n_a$ in this
experiment is negligibly small. It can be investigated in more
precise difference force measurement (see Sec.~3).
As to the dielectric permittivity of low-resistivity Si
(the second plate), one can describe it either by
$\varepsilon_{\rm ce}^{(2)}(i\xi)$ or by the dielectric permittivity
of the generalized plasma-like model $\varepsilon^{(2)}(i\xi)$
defined in Eq.~(\ref{eqn11}) [this experiment, as well as the
experiment discussed in Sec.~2.1, is not of sufficient precision
to discriminate between theoretical descriptions given by
 Eqs.~(\ref{eqn11}) and (\ref{eqn12})].
 The value of the plasma frequency
 $\omega_p^{(2)}=2.0\times 10^{15}\,$rad/s should be used in this
 case in accordance with the concentration of charge carriers.
Computations below are done at the laboratory temperature
$T=300\,$K specially in this review (in the original
publication\cite{25} computations were performed at $T=0$
keeping in mind the smallness of thermal effects at short
separations).

\begin{figure}[t]
\vspace*{-5.5cm}
\hspace*{-2.4cm}
\psfig{file=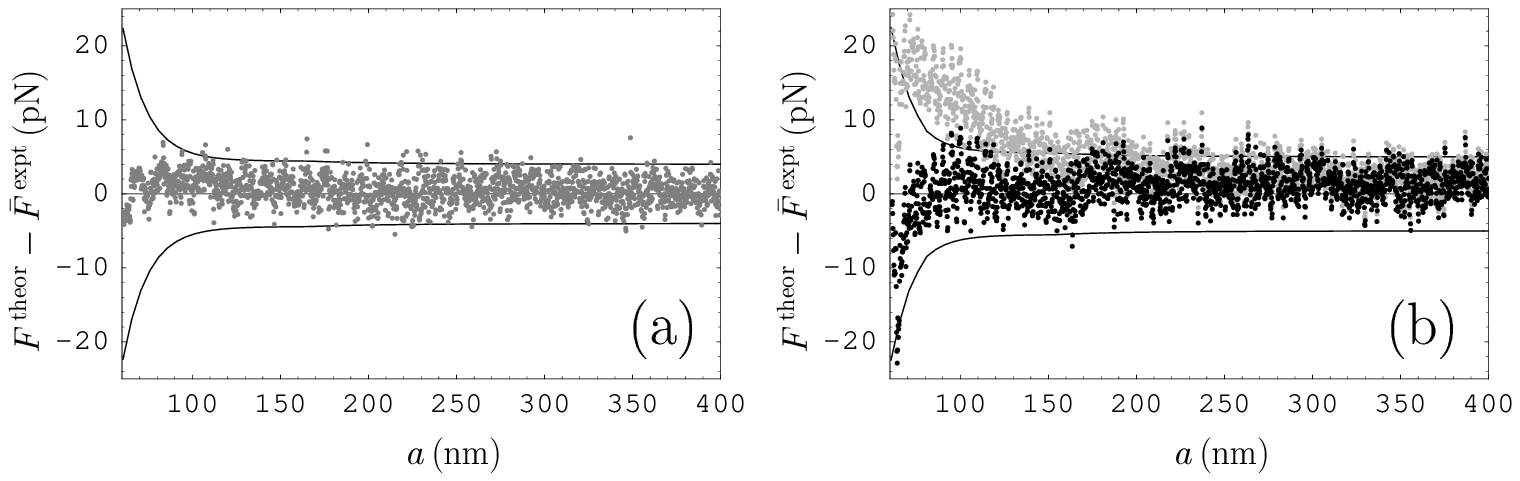,width=6.5in}
\vspace*{-14.5cm}
\caption{The differences between theoretical and mean measured Casimir
forces in the configuration of
an Au-coated sphere and two Si plates are
shown as dots for (a) high-resistivity and (b) low-resistivity plate.
Au is described by the generalized plasma-like model. Si of the plates
is described as a high-resistivity material (the grey dots) or
by the generalized plasma-like model (the black dots).}
\label{aba:fig8}
\end{figure}
In Fig.~\ref{aba:fig8}(a) the differences between computed and mean
measured Casimir forces,
$F_a^{\,\rm theor}-\bar{F}_a^{\,\rm expt}$, for the
high-resistivity plate versus separation are shown as grey dots.
The pairs of solid lines in Fig.~\ref{aba:fig8}(a) indicate the
borders of the confidence intervals, to which the differences
between theoretical and mean experimental Casimir forces should belong
with a 95\% probability. These confidence intervals take into
account both the experimental and theoretical errors combined using
the conventional statistical rules.\cite{5,15,24}
As can be seen in Fig.~\ref{aba:fig8}(a), more than 95\% of
the grey dots belong to the confidence intervals.
This means that theoretical description of high-resistivity Si
by the dielectric permittivity $\varepsilon_{\rm ce}^{(2)}(i\xi)$
with charge carriers disregarded is consistent with the
experimental data within a 95\% confidence interval.

Quite a different situation holds for a low-resistivity sample.
In Fig.~\ref{aba:fig8}(b) we show as grey dots the theoretical and
mean measured force differences,
$F_b^{\,\rm theor}-\bar{F}_b^{\,\rm expt}$,
when the theoretical Casimir forces are computed using the
dielectric permittivity $\varepsilon_{\rm ce}^{(2)}(i\xi)$.
In the same figure the black dots represent the force
differences, $F_b^{\,\rm theor}-\bar{F}_b^{\,\rm extp}$,
when the theoretical Casimir forces are computed using the
dielectric permittivity of the generalized plasma-like model
(\ref{eqn11}). As can be seen in Fig.~\ref{aba:fig8}(b),
this theoretical description of low-resistivity Si is
consistent with the data within a 95\% confidence interval.
At the same time, more than 5\% of grey dots lie outside the
95\% confidence interval within the separation region from 62
to 200\,nm. This means that theoretical description of the
second plate by means of the dielectric permittivity of
high-resistivity Si $\varepsilon_{\rm ce}^{(2)}(i\xi)$ is
excluded by the data at a 95\% confidence level.

\begin{figure}[t]
\vspace*{-1.5cm}
\hspace*{-1.2cm}
\psfig{file=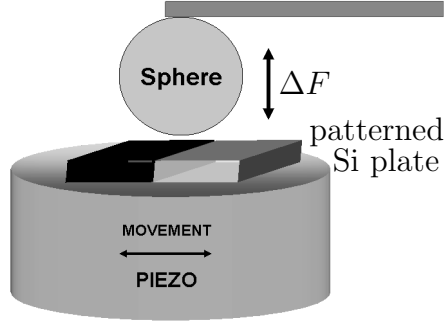,width=8.5in}
\vspace*{-24.1cm}
\caption{Schematic diagram of an experimental setup for
the measurement of the difference Casimir force between an
Au-coated sphere and a patterned Si plate.}
\label{aba:fig8a}
\end{figure}

As was noted above, successive measurements of the Casimir force
in the configuration of an Au-coated sphere and two Si plates
of different resistivities are not enough precise to
discriminate between the descriptions of low-resitivity Si by
means of Eqs.~(\ref{eqn11}) and (\ref{eqn12}) (the generalized
plasma-like and Drude-like models). There is a proposal\cite{40}
in the literature that this can be done by means of a
patterned Si plate with two sections of different dopant
concentrations. Such a plate is mounted on a piezoelectric
actuator below an Au-coated sphere attached to the cantilever
of an AFM (see Fig.~\ref{aba:fig8a}).
The actuator oscillates in the horizontal
direction, and the cantilever flexes in response to the
Casimir force above different regions of the plate.
Thus, the sphere in such an experiment is subjected to
the difference Casimir force which is an immediately measured
quantity. In this case individual forces between the
sphere and each of the two sections of the plate are not
measured.

The patterned plate is composed of single-crystal Si specially
fabricated to have adjacent sections with two different charge
carrier densities. In so doing both $p$- and $n$-type dopants
can be used (B and P, respectively). A sharp transition boundary
between these sections with a width less than 200\,nm can be
achieved. Identically prepared but unpatterned samples can be
used to measure the properties which are needed for the
theoretical computations (with Hall probes for measuring the
charge carrier density, and a four-probe technique for
measuring the conductivity). During the measurements of the
difference Casimir force, the distance between the sphere and
the patterned plate is kept fixed and the plate oscillates in
the horizontal direction, such that the sphere crosses the
boundary between the two sections in the perpendicular direction
during this oscillation. Note that a similar approach has been
exploited\cite{53a} for constraining new forces using the
oscillations of an Au-coated sphere above two dissimilar metals.
The Casimir force on the sphere changes as the sphere crosses
the boundary. This change corresponds to the difference force
$F_{\rm diff}$, equal to the difference between the Casimir
forces due to the different charge carrier densities
$n_a$ and $n_b$ in different sections.
Preliminary estimations show\cite{40} that the increased sensitivity
of this experimental scheme should be sufficient to
discriminate between the plasma-like and Drude-like models of
dielectric response.

The above results again demonstrate that for the semiconductor
of metallic type the magnitude of the Casimir force depends
on the density of charge carriers which should be included in
the model of the dielectric response in the Lifshitz theory.
It is important also that the use of different semiconductor
materials provides a way of controlling the magnitude of the
Casimir force. One example was already provided in the end of
Sec.~2.1. Here, for an Au sphere interacting with
the first plate
(high-resistivity Si), the magnitudes of the Casimir force
are equal to 377.696, 89.3329, and 34.2372\,pN at separations
$a=62.38$, 104.88 and 147.38\,nm, respectively.
This should be compared with the magnitudes of the
Casimir force, $|F_{\rm Au-Au}^{\,\rm theor}|$, between the
same Au sphere and Au plate 520.831, 134.635, and 54.7224\,pN
at the same respective separations. {}From this one can conclude
that the replacement of an  Au plate with a high-resistivity Si one
results in a decrease of the magnitude of the Casimir force
by 27.5\%, 33.6\%, and 37.4\% at the respective separations
indicated above. Smaller changes occur when an Au plate
is replaced with the second Si plate (low-resistivity Si).
Here the force magnitudes are decreased by 22.6\%, 26.4\%, and
28.0\% at $a=62.38$, 104.88 and 147.38\,nm, respectively.

\subsection{ITO plate}

One more demonstration that the magnitude of the Casimir force
depends on the dielectric properties of metallic-type
semiconductors has been performed\cite{26,29a} in the configuration
of an Au-coated sphere and an ITO plate (In${}_2$O${}_3$:Sn).
Advantages of using ITO in the Casimir physics were
proposed in Ref.~\refcite{19}.
The experimental setup used an AFM schematically shown in
Fig.~\ref{aba:fig1} with a sphere of $R=100\,\mu$m radius.
Main differences from experiments described above are that the
measurements were performed in an ambient environment in the
dynamic mode. This means that the AFM cantilever with the
sphere attached to it was considered as a harmonic oscillator
with a natural resonant frequency
$\omega_0=2\pi\times 1900\,$rad/s. Under the influence of
the Casimir force $F(a)$ the value of the resonant frequency
was changed to some $\omega_r$. In the linear regime we
get\cite{5}
\begin{equation}
\frac{\partial F}{\partial a}=k_0-k_{\rm eff},
\label{eqn13}
\end{equation}
\noindent
where $k_0=m\omega_0^2$, $k_{\rm eff}=m\omega_r^2$, and $m$ is the
mass of the oscillator.

 Earlier the frequency shift $\omega_0-\omega_r$ from the dynamic scheme
was  measured directly using the frequency modulation
technique in experiments on the determination
of the Casimir pressure between two metallic
surfaces by means of a micromechanical torsional
oscillator.\cite{16}\cdash\cite{18a,51,51a}
Note that the PFA allows recalculation of the gradient of the
Casimir force between a sphere and a plate into the Casimir
pressure between the two parallel plates according
to\cite{5,15}
\begin{equation}
P(a)=-\frac{1}{2\pi R}\,\frac{\partial F}{\partial a}.
\label{eqn14}
\end{equation}
\noindent
In the experiment\cite{26,29a} the frequency modulation technique
was not used. Instead, the oscillation amplitude at a fixed
frequency was monitored. Then the resonant frequency shift due
to the Casimir force led to a change in amplitude which can be
related to the force gradient (similar dynamic scheme was used
in the measurements\cite{52} of the gradient of the Casimir
force between metallic surfaces by means of an AFM, but the
phase of the oscillator was monitored instead of the
amplitude). However, it is necessary to be cautious with both
amplitude and phase detection methods. The problem is that
oscillator amplitude and phase can change not only because of
frequency shift, but also because of change in dissipation
(oscillator energy loss). Dissipation is of particular concern
if the experiments are done in air as the air layer thickness
between plate and sphere keeps changing when the plate is
moved closer to the sphere.

Another feature of this experiment is that it
consisted of the two measurements. In the first measurement the
force between an Au coated sphere and an Au-coated sapphire plate
was measured. In the second measurement this plate was replaced
with a glass substrate coated with an ITO film of thickness
of 190\,nm, resistivity
$\rho\approx 1.6\times 10^{-4}\,\Omega\,$cm and charge carrier
density $n\approx 1.2\times 10^{21}\,\mbox{cm}^{-3}$.
Thus, the measure of reduction in the force magnitudes
due to the replacement of Au with the semiconductor can be
found with no comparison to the theoretical results.

The relative random error of the total force gradient
measurements (electric plus Casimir) at a separation 95\,nm
was estimated\cite{29a} as about 3.5\% of force gradient.
Keeping in mind that the random errors in both total and
electric forces are distributed normally, the relative random
error in the measurements of the Casimir force cannot be
smaller than 3.5\%. In addition, the measured Casimir force
gradients were affected\cite{29a} by a 3\% systematic error.
The absolute random error of 0.5\,nm in the determination
of initial separation (chosen at about $8.5\,\mu$m)
was reported.\cite{29a} During the 580 measurement runs,
the mechanical drift of initial separation for 52\,nm was observed.
The absolute systematic error in the initial separation
was of about\cite{29a} 1.4\,nm.
 {}From the comparison of the measured
gradient of the Casimir force in Au-Au and Au-ITO configurations
it was concluded that the replacement of an Au plate with an
ITO plate leads to roughly 40\%--50\% smaller Casimir pressure
between two parallel plates at separations from 80 to 120\,nm.
Note that in an ambient environment used in these measurements
the residual potential difference $V_0$ between the sphere
and the plate varied with time significantly in the
Au-ITO configuration (from 72 to 50\,mV at the separation
of 100\,nm).

The comparison of the experimental data with theory was performed
using the Lifshitz formula (\ref{eqn6b}) at zero temperature
and Eq.~(\ref{eqn7}) with energy instead of the free energy, although
measurements were performed at $T=300$K. In this comparison
the contribution of
the surface roughness to the gradient of the Casimir force
was disregarded. In fact an Au-coating on the polystyrene
sphere was characterized\cite{26,29a} by stochastic roughness with the
variance $\delta=3.8\,$nm. For the Au and ITO coatings on the
plates  $\delta=0.8$ and 4\,nm, respectively.
Note that surface roughness increases the magnitude of the
Casimir force. If corrections due to surface roughness were
included in computations of the Casimir force, this would
lead to about 2\% correction in the case of Au-Au surfaces
and 4\% correction for Au-ITO surfaces at the separation
$a=80\,$nm.
The dielectric permittivity of Au along the imaginary
frequency axis was described by the tabulated optical
data\cite{46} for $2n_1(\omega)n_2(\omega)$ (see Sec.~2.1)
extrapolated to low frequencies by the imaginary part of the
Drude function (\ref{eqn10}). For the dielectric permittivity
of ITO Eq.~(\ref{eqn12}) was used where the representation
for $\varepsilon_{\rm ce}^{(2)}(i\xi)$ was taken from
Ref.~\refcite{53}.
As a result, different powers of separations for data and
theory were obtained.\cite{29a} This can be explained by
the use of the Lifshitz theory at zero temperature for the
comparison with the room-temperature measurement data
(for Au test bodies this approach was experimentally
excluded\cite{81} at a 70\% confidence level).
Further refinement of the dielectric properties of ITO
is also needed to bring the experimental data in
agreement with theory.

To conclude,  this measurement is another
confirmation  that using different semiconductor
materials one can control the magnitude of the Casimir force.

\subsection{Ge spherical lens above a Ge plate}

All the experiments using semiconductor test bodies, which were
discussed in Secs.~2.1--2.3, have been performed with the help
of an AFM and spheres of about $100\,\mu$m radius.
In fact the cantilever of an AFM with attached sphere can be
considered as a micromechanical device having a very high force
sensitivity. This made the use of an AFM in measurements of the
Casimir force, suggested for the first time for two metallic
test bodies in Ref.~\refcite{12}, very productive.
An alternative approach to measuring the Casimir force by using
a torsion pendulum and a spherical lens of more that 10\,cm
curvature radius was suggested for two metallic test bodies in
Ref.~\refcite{54}. This approach has an advantage in that it deals
with large lenses of centimeter-size curvature radii and, thus,
much larger forces allowing measurements at separations above
$1\,\mu$m. Below it is shown, however, that the use of large
lenses leads to serious problems regarding the reproducibility of
measurement results and the comparison between experiment and
theory.

Now we consider the measurement of the Casimir force between a
crystalline intrinsic Ge plate and a crystalline intrinsic Ge
lens of curvature radius $R=15.10\pm 0.05\,$cm performed\cite{27} by
means of a torsion pendulum. The Ge lens was mounted on a
piezoelectric $xyz$ motion stage and the Ge plate on one arm
of the torsion pendulum with a body of 15\,cm length in
vacuum of $5\times 10^{-7}\,$Torr.
A pendulum body was suspended by a tungsten wire of length
2.5\,cm and diameter $25\,\mu$m.
The other arm of the
pendulum played the role of the center electrode situated in
between two fixed compensator plates forming two parallel
plate capacitors $C_1$ and $C_2$. The Casimir force between
a Ge lens and Ge plate resulted in a torque which rotated
the pendulum body. This resulted in changes in the capacitances
$C_1$ and $C_2$, which were detected with a phase sensitive
circuit. Then, compensating voltages were applied to the
capacitances through a feedback circuit to counteract the
change in the angle of the torsion pendulum and to keep the
system in equilibrium. These compensating voltages were
a measure of the Casimir force.

The calibration of the setup was performed by means of the
measurements of electric forces.
It was  found, however, that the residual potential difference
$V_0$ depends on separation where it was measured. According to
Ref.~\refcite{27}, this might be explained by the polishing
stresses or the curvature of the lens surface changing the crystal
plane orientation at the surface. In addition, it was suggested
that there was the contribution to electric force due to random
short-scale patches. As a result, even after the application of
the compensating potential, there was a residual electrostatic
force of a complicated nature. An expression for this force
containing three fitting parameters was obtained at large
separations, where the Casimir force was assumed to be negligibly small.
The values of the fitting parameters were determined from the
fit to the experimental data of electric force measurements at
large separations. Then, the obtained expression for the electric
force was extrapolated to lower separation distances and subtracted
from the experimental data for the total measured force at all
separations in order to get the measured data for the Casimir
force alone. This procedure seems to be not enough justified
because, even if the found phenomenological expression for the
electric force is applicable at large separations, it might
incorrectly describe the residual electric force at short
separations.

The obtained experimental Casimir forces were compared with five
theoretical approaches. Within the first approach intrinsic Ge
was considered as a high-resistivity dielectric described by
the dielectric permittivity of core electrons
$\varepsilon_{\rm ce}^{(1)}(i\xi)=\varepsilon_{\rm ce}^{(2)}(i\xi)
\equiv\varepsilon_{\rm Ge}(i\xi)$.
In the second approach, the free charge carriers (electrons and
holes) were taken into account by means of the Drude model
using Eq.~(\ref{eqn12}), but with account of charge carriers
of two types:
\begin{equation}
\tilde\varepsilon^{(1)}(i\xi)=\tilde\varepsilon^{(2)}(i\xi)=
\frac{\omega_{p,n}^2}{\xi(\xi+\gamma_n)}+
\frac{\omega_{p,p}^2}{\xi(\xi+\gamma_p)}+
\varepsilon_{\rm Ge}(i\xi).
\label{eqn15}
\end{equation}
\noindent
Here, for intrinsic Ge we have\cite{55}
$\omega_{p,n}\approx 7.8\times 10^{11}\,$rad/s,
$\omega_{p,p}\approx 5.9\times 10^{11}\,$rad/s,
and $\gamma_n\approx\gamma_p\approx 2.6\times 10^{11}\,$rad/s.
The respective charge carrier densities at $T=300\,$K are
$n_n=n_p\approx 2.3\times 10^{13}\,\mbox{cm}^{-3}$.
In the framework of the third approach free charge carriers were
taken into account by means of the plasma model similar to
Eq.~(\ref{eqn11}):
\begin{equation}
\varepsilon^{(1)}(i\xi)=\varepsilon^{(2)}(i\xi)=
\frac{\omega_{p,n}^2}{\xi^2}+
\frac{\omega_{p,p}^2}{\xi^2}+
\varepsilon_{\rm Ge}(i\xi).
\label{eqn16}
\end{equation}
\noindent
Finally, as the fourth and fifth approaches, the
modifications\cite{32,33} of the Lifshitz theory have been used
in the comparison with the experimental data (these
modifications were mentioned in Sec.~1 and are discussed in more
detail in Sec.~3.4). In all cases computations of the theoretical
Casimir force were performed using the PFA in Eq.~(\ref{eqn7}).

The experimental data for the Casimir force were compared with
theoretical results computed using the above five approaches and
were found consistent with all of them within the limits of
experimental errors. According to Ref.~\refcite{27},
``The error bars take into account all statistical uncertainties
(2\%--3\%) as well as fitting uncertainties from the electrostatic
force analysis (10\%).'' The confidence level is not reported.
However, these estimates of Ref.~\refcite{27} do not correspond
and match to those reported in rest of the paper.
For example, using the given expression for the residual
electrostatic force,\cite{27} one obtains at separations of
1, 2, and $3\,\mu$m that it is determined with the minimum absolute
errors equal to 67, 33, and 22\,pN, respectively.
As the Casimir force is obtained from the subtraction of the
residual electrostatic force from  the
total measured force, its error
cannot be less than these at the corresponding distances.
Thus, the correct
relative errors of the Casimir force at
1, 2, and $3\,\mu$m are no less than 40\%, 124\%, and 211\%,
respectively.

There is a discussion in the literature concerning the use of
spherical\cite{56}\cdash\cite{63a} and cylindrical\cite{60}
metallic test bodies of centimeter-size radii of curvature
for the measurements of the Casimir force. Problems emerged
when an anomalous force-distance relation for the electric
force between an Au-coated spherical lens of $R=3.09\,$cm
curvature radius and an Au-coated plate was
observed,\cite{56} distinct from that predicted by classical
electrodynamics. As discussed above, anomalous electric
forces of unclear origin emerged also between a Ge lens and
Ge plate. It was shown\cite{58} that the anomalous
behavior of the electrostatic force can be explained due to
deviations from a perfect spherical shape
of the mechanically polished and ground surface
for lenses with
centimeter-size curvature radii (later this possibility
was recognized\cite{60} as a crucial point to be taken into
account in future experiments for a cylinder of
centimeter size radius near a plate as well).
Different kinds of surface imperfections (bubbles, pits
and scretches) allowed by the optical surface specification
data\cite{64a}\cdash\cite{61} can lead to significant deviations of the
force-distance relation from the form predicted by
classical electrodynamics under an assumption of perfect
spherical surface.

\begin{figure}[t]
\vspace*{-4.3cm}
\hspace*{-3.5cm}
\psfig{file=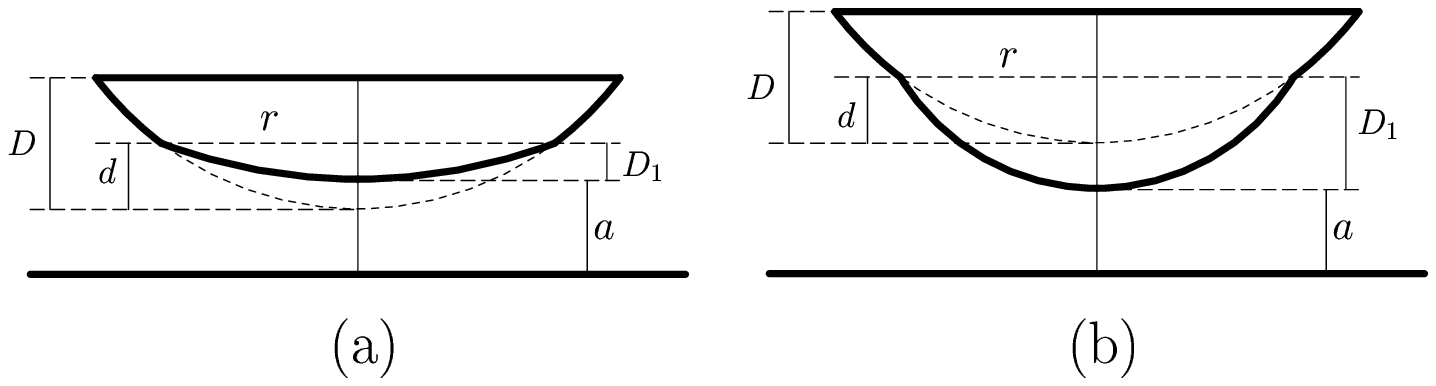,width=7.1in}
\vspace*{-18.2cm}
\caption{The configuration of a spherical lens with radius of curvature $R$
possessing surface imperfection around the point of closest
approach to a plate. (a) The bubble radius of curvature is $R_1>R$.
(b) The bubble radius of curvature is $R_1<R$.
The relative sizes of the lens and imperfection are not shown
to scale.}
\label{aba:fig9}
\end{figure}

According to recent results,\cite{59,63a} bubbles and pits, that
are unavoidably present on spherical surfaces of
centimeter-size curvature radii, make the PFA inapplicable
in the form of Eq.~(\ref{eqn7}) used to calculate the Casimir
force in Ref.~\refcite{27}. As two of the simplest examples we
consider a Ge lens of thickness $D$ with the curvature radius
$R=15.1\,$cm having a bubble either of the radius of
curvature $R_1=22\,$cm and thickness $D_1=0.09\,\mu$m or
$R_1=10\,$cm and $D_1=0.2\,\mu$m near the point of the closest
approach to a Ge plate [see Fig.~\ref{aba:fig9}(a) and (b),
respectively].
The radii of the bubbles are determined from
$r^2=2R_1D_1-D_1^2$. This leads to bubble diameters
$2r=0.4\,$mm for the bubbles shown in Fig.~\ref{aba:fig9}(a,b).
The obtained value should be compared with constraints imposed
by the scratch/dig optical surface specifications of the
used\cite{27} Ge lens of ISP optics\cite{65a}, GE-PX-25-50
with surface quality 60/40. The latter means that 0.4\,mm is just
the maximum diameter of bubbles allowed. It is easily seen
also that the flattening of the lens surface in
Fig.~\ref{aba:fig9}(a) or the swelling up in Fig.~\ref{aba:fig9}(b)
are much less than the absolute error of $R$ equal to\cite{27}
$\Delta R=0.05\,$cm. Really, the quantity $d$ defined in
Fig.~\ref{aba:fig9}(a,b) is equal to
$d\approx r^2/(2R)\approx 0.13\,\mu$m.
This results in the flattening of the lens surface in
Fig.~\ref{aba:fig9}(a) given by $d-D_1\approx 0.04\,\mu$m or
to the swelling up in Fig.~\ref{aba:fig9}(b) given by
$D_1-d\approx 0.07\,\mu$m.

For a spherical lens with bubbles shown in Fig.~\ref{aba:fig9}(a,b)
and a plate, the Casimir force is presented\cite{59,63a}
by the following generalization of the PFA:
\begin{equation}
F(a,T)=2\pi(R-R_1){\cal F}(a+D_1,T)+2\pi R_1{\cal F}(a,T).
\label{eqn17}
\end{equation}
\noindent
Computations using the dielectric permittivity\cite{62}
of intrinsic Ge, $\varepsilon_{\rm ce}(i\xi)$, show that for the above
parameters of the bubble in Fig.~\ref{aba:fig9}(a)
Eq.~(\ref{eqn17}) leads to the larger magnitudes of the Casimir force
by 15\% and 10\% than Eq.~(\ref{eqn7}) at separations $a=0.6$ and
$1\,\mu$m, respectively. For the bubble of Fig.~\ref{aba:fig9}(b)
the use of Eq.~(\ref{eqn17}) instead of Eq.~(\ref{eqn7}) results
in smaller magnitudes of the Casimir forces by 19\% and 14\%
at the same respective separations. Keeping in mind that
experimentally it is hard to determine the point of the closest
approach between the two surfaces with sufficient precision
and to investigate the character of surface imperfection at
this point, the possibility of using lenses of
centimeter-size radius of curvature in measurements of the Casimir
force becomes problematic.
Specifically, it was shown\cite{63a} that the Casimir force
between a perfectly spherical lens and a plate described by the
Drude model can be made approximately equal to the force between
a sphere with some surface imperfection and a plate described
by the plasma model, and vice versa. This makes uncertain the
results of such experiments as purported observation\cite{66a}
of the thermal Casimir force using a spherical lens of
15.6\,cm radius of curvature.

The problem of surface imperfections is not relevant to
polystyrene spheres of about $100\,\mu$m radii made by the
solidification from the liquid phase. The surface quality of
such spheres after metallic coating was investigated using a
scanning electron microscope\cite{5,14,15} and did not
reveal any bubbles or scratches. Thus, in experiments
discussed in Secs.~2.1--2.3 the PFA
in its simplest form presented in Eq.~(\ref{eqn7}) is applicable.
It is applicable also in the proposed experiment\cite{66b}
using a cylinder-plate configuration with cylinders of about
$100\,\mu$m radii.

\section{Optically modulated Casimir force}

As discussed in Sec.~1, there are two experiments on the Casimir
effect dealing with two macroscopic bodies which are inconsistent
with the Lifshitz theory under some conditions. One of these
experiments was performed three times\cite{5,15}\cdash\cite{18a}
with metallic test bodies. It is outside the framework of our
review. The other experiment\cite{28,28a} is a measurement of
the optically modulated difference Casimir force between an
Au-coated sphere and a semiconductor (Si) plate illuminated with
laser pulses. It is considered in more detail in this section.
The experiment on the optically modulated Casimir force is
important in two aspects. On the one hand, it provides new
fundamental insights by demonstrating that the Lifshitz theory
and its modifications are incompartible with the inclusion of
dc conductivity of dielectric materials in the model of
the dielectric response. On the other hand,
technologically the optical modulation experiment
opens opportunities to periodically change the magnitude of
the Casimir force and to realize the pulsating regime without use of
mechanical springs. This is discussed in Sec.~6.

\subsection{Experimental scheme and measurements results}

\begin{figure}[t]
\vspace*{-2.cm}
\hspace*{-2.3cm}
\psfig{file=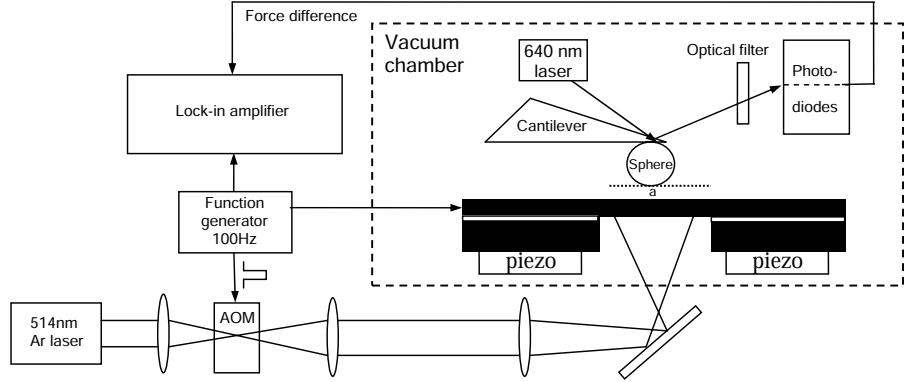,width=6.5in}
\vspace*{-14.1cm}
\caption{Schematic of the experimental setup for studying optical
modulation of the Casimir force. The main components are shown.}
\label{aba:fig10}
\end{figure}

The experiment\cite{28,28a} on the optical modulation of the Casimir
force is a direct measurement of the change in the force between
an Au-coated sphere of $197.8\pm 0.3\,\mu$m diameter and a Si
plate in the presence and in the absence of a laser light on it.
Being a difference measurement (the Casimir forces with and
without light on the plate are not measured separately but only
their difference), this experiment is characterized by a very high
force sensitivity of a few tens of pN. The experimental scheme is
shown in Fig.~\ref{aba:fig10}. Measurements were made by means
of an AFM in an oil-free vacuum chamber with a pressure of
around $2\times 10^{-7}\,$Torr. The Si plate was mounted on the
top of a piezoelectric actuator, which was used to change the
separation $a$ between the sphere and the plate from contact
to $6\,\mu$m. The Si used was $p$-type with charge carrier
density equal to $n\approx 5\times 10^{14}\,\mbox{cm}^{-3}$.
The excitation of additional carriers in the Si plate was
done with 5\,ms wide light pulses obtained from Ar laser at
514\,nm wavelength, modulated at a frequency of 100\,Hz.
The laser pulses were focused on the bottom surface of
the Si plate (see Fig.~\ref{aba:fig10}).

The cantilever of the AFM flexed when the Casimir force between
the sphere and the plate changed depending on the presence or
absence of laser light on the plate. Similar to
Fig.~\ref{aba:fig1}, the cantilever deflection was monitored
with a 640\,nm beam from a second laser shown in
Fig.~\ref{aba:fig10}.
An optical fiber was used to prevent interference of the beams
from both lasers. The change in the Casimir force due to the
light incident on the plate led to a difference signal between
the two photodiodes (see Fig.~\ref{aba:fig10}). This
signal was measured with a lock-in amplifier.

The most important part of the described setup is the Si plate
colored black in Fig.~\ref{aba:fig10}. It should be sufficiently
thin and of appropriate resistivity to ensure that the density of
charge carriers increases by several orders of magnitude under the
influence of laser pulses (note that the charge carrier density
indicated above corresponds\cite{46} to relatively high
resistivity $\rho=10\,\Omega\,$cm).
At the same time Si plate should be thick enough to make as low
as possible
the photon pressure of the transmitted light on the sphere.
The Si plate shown in Fig.~\ref{aba:fig10} was fabricated in
several steps.
\begin{figure}[t]
\vspace*{-1.3cm}
\hspace*{-4.5cm}
\psfig{file=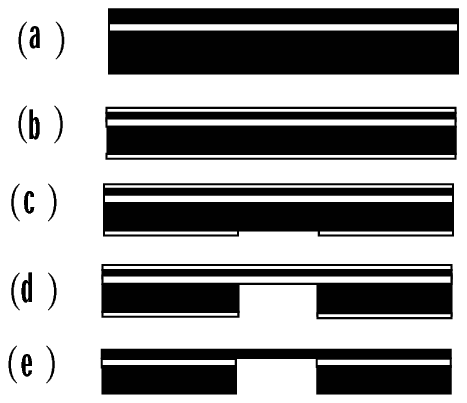,width=9in}
\vspace*{-26.3cm}
\caption{Fabrication process of the Si membrane.
(a) The  Si substrate (colored black) with a buried
SiO${}_2$ layer (white). (b) The substrate was mechanically
polished and oxidized, and
(c) a window in the bottom SiO${}_2$ layer was etched with
HF. (d) Next, TMAH was used to etch the Si. (e)
Finally, the SiO${}_2$ layer was etched away in HF solution to
form a clean Si surface.}
\label{aba:fig11}
\end{figure}
As an initial sample, the commercial wafer of Si grown on an
insulator (SiO${}_2$) was used. The wafer consisted of a Si
substrate of thickness $600\,\mu$m and a Si top layer of
thickness $5\,\mu$m with a buried intermediate SiO${}_2$
of thickness 400\,nm [see Fig.~\ref{aba:fig11}(a) where Si
is colored black and SiO${}_2$ white]. The thickness of the
Si substrate was reduced to about $200\,\mu$m and then,
after RCA cleaning of the surface, the wafer was oxidized
at high temperature in a dry O${}_2$ atmosphere.
As a result, a thermal oxide layer with a thickness of
about $1\,\mu$m was formed on the bottom and top sides
of the wafer [see Fig.~\ref{aba:fig11}(b)]. Then a hole with a
diameter of 0.85\,mm was etched with HF in the center of the
bottom oxide layer [Fig.~\ref{aba:fig11}(c)].
Next, TMAN was used at 363\,K to etch the Si substrate
(note that TMAN etching rate for Si is 1000 times greater than
for SiO${}_2$). As a result, a hole was formed as shown in
Fig.~\ref{aba:fig11}(d). Finally, all the thermal oxidation
layers and the buried oxidation layer in the hole were
etched away in HF solution to form a clean Si plate
(membrane) over
a hole [see Fig.~\ref{aba:fig11}(e)].
The thickness of this plate was measured to be $d=4.0\pm 0.3\,\mu$m.

All calibrations were done by applying different potentials to
the Si plate while keeping the sphere grounded. The calibrations
were made in a similar way to other experiments with
semiconductor surfaces. As  a result the signal calibration
constant and the residual potential differences
$V_0=-0.225\pm 0.002\,$V and $V_0^l=-0.303\pm 0.002\,$V
in the absence and in the presence of the laser light on
the Si plate were determined. The residual potential
differences were shown to be independent of separation
(all details of calibration procedures can be found in
Refs.~\refcite{5,28a} and \refcite{50}).

The measurement of the difference Casimir force in the presence
and in the absence of a light pulse on a Si plate as a function of
separation was performed as follows. During the bright phases of
the pulse train, a voltage $V^l$ was applied to the plate, and
during the dark phases, a voltage $V$. Then the difference of
the total force, $F_{\!\rm diff}^{\,\rm tot}(a)$ (Casimir and
electric), was measured.
This was done using a lock-in amplifier with an integration time
of 100\,ms, which corresponds to a bandwidth of 0.78\,Hz.
The difference between the Casimir
forces in the presence and in the absence of light,
\begin{equation}
F_{\!\rm diff}^{\,\rm expt}(a)=F_l(a)-F(a),
\label{eqn20a}
\end{equation}
\noindent
was found by subtracting the
contribution of the electric forces:
\begin{equation}
F_{\!\rm diff}^{\,\rm expt}(a)=F_{\!\rm diff}^{\,\rm tot}(a)-
X\left(\frac{a}{R}\right)\,\left[(V^l-V_0^l)^2-(V-V_0)^2\right],
\label{eqn18}
\end{equation}
\noindent
where $X(z)$ is the function whose explicit form is known
from the exact solution of the electrostatic problem in the
sphere-plate geometry.\cite{5,15,24,28a,63}
The difference Casimir forces as a function of separation were
found for different absorbed laser powers:
$P^{\,\rm eff}=9.3$, 8.5, and 4.7\,mW.
At each absorbed power measurements were repeated with different
pairs of applied voltages $(V^l,V)$ between the sphere and the
plate in the bright and dark phases, respectively, and the
mean difference Casimir force, $\bar{F}_{\!\rm diff}^{\,\rm expt}$,
was obtained. Below we consider the experimental data for the
maximum, $P^{\,\rm eff}=9.3\,$mW, and minimum, 4.7\,mW, absorbed
powers averaged over 31 and 33 repititions with different pairs
of applied voltages. With these absorbed laser powers, charge
carrier densities in Si in the bright phase were equal to
$n_l=(2.1\pm 0.4)\times 10^{19}\,\mbox{cm}^{-3}$ and
$(1.4\pm 0.3)\times 10^{19}\,\mbox{cm}^{-3}$, respectively,
where all errors were determined at a 95\% confidence
level.\cite{5,15,28,28a}
The lifetime of the charge carriers excited in the Si plate
by the pulses from the Ar laser was measured using a
noninvasive optical pump-probe technique.\cite{64,65}
This time represents both surface and bulk recombination
and is consistent with that expected for Si.
The measured values of the carrier lifetime were used in
theoretical computations of the change in the Casimir force
for different incident laser powers.

\begin{figure}[t]
\vspace*{-4.7cm}
\hspace*{-2.4cm}
\psfig{file=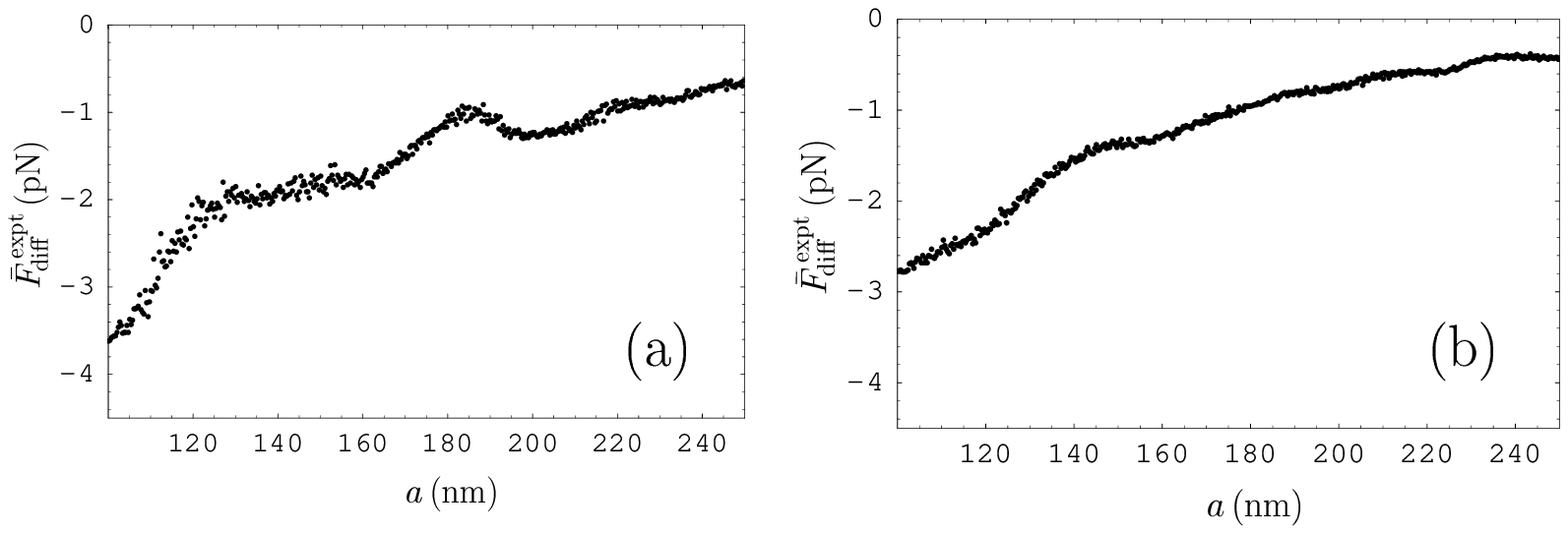,width=6.4in}
\vspace*{-14.2cm}
\caption{Mean measured differences of the Casimir force between
an Au-coated sphere and Si plate in the presence and in the
absence of light for the absorbed power of (a) 9.3\,mW and
(b) 4.7\,mW versus separation are shown by dots.}
\label{aba:fig12}
\end{figure}

In Fig.~\ref{aba:fig12}(a,b) the experimental data for
$\bar{F}_{\!\rm diff}^{\,\rm expt}(a)$ as a function of
separation are shown by dots for absorbed powers
$P^{\!\rm eff}=9.3$ and 4.7\,mW, respectively.
The corresponding incident powers were 15.0 and 7.6\,mW.
As expected, the magnitude of the Casimir force difference
has the largest values at the shortest separations and
decreases with the increase of separation.
Some oscillations of the data with the distance
in Fig.~\ref{aba:fig12}(a,b) come from the interference of
one beam of light of 640\,nm laser scattered by the plate
with another beam scattered from the sphere.
Below we present the results of error analysis and the
comparison of the experimental data with theory.

\subsection{Analysis of errors and uncertainties}

\begin{figure}[t]
\vspace*{-4.7cm}
\hspace*{-2.4cm}
\psfig{file=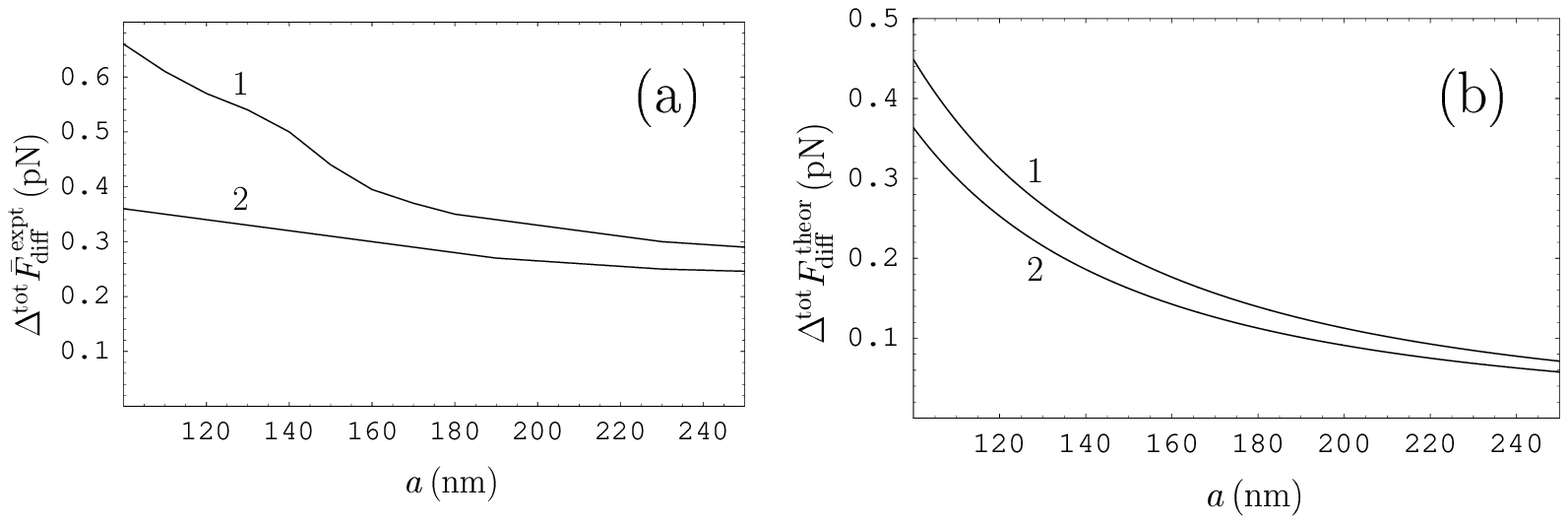,width=6.4in}
\vspace*{-14.2cm}
\caption{(a) Total experimental errors
and (b) total theoretical errors, versus separation. The cases of different
absorbed powers 9.3 and 4.7\,mW are labeled 1 and 2, respectively.}
\label{aba:fig13}
\end{figure}

Here, we discuss both the experimental and theoretical errors
arising in measurements and computations of the difference
Casimir forces. The main role in the experimental error in this
experiment is played by the random error. The absolute random
error in this experiment as a function of separation was
found\cite{28,28a} by the standard procedure using Student's
$t$-distribution with the number of degrees of freedom 30 and
32 for the measurements with two different absorbed powers.
In so doing the 95\% confidence level was chosen. The obtained
random errors decrease from 0.65 to 0.29\,pN and from 0.32 to
0.24\,pN for the absorbed powers 9.3 and 4.7\,mW, respectively,
when the separation increases from 100 to 250\,nm.
The main systematic
errors were from the instrumental noise,
$\Delta_1^{\!\rm syst}{F}_{\!\rm diff}^{\,\rm expt}\approx 0.08\,$pN,
from the resolution error in data acquisation,
$\Delta_2^{\!\rm syst}{F}_{\!\rm diff}^{\,\rm expt}\approx 0.02\,$pN,
and from calibration,
$\Delta_3^{\!\rm syst}{F}_{\!\rm diff}^{\,\rm expt}\,$.
The latter was equal to 0.6\% of the measured difference Casimir
force, whereas the two former were independent on separation.
The three systematic errors were combined at a 95\% confidence
level by using the standard rule valid for quantities distributed
uniformly (this is the most conservative approach\cite{5,15,66}).
As a result, the total systematic error varied from 0.092 to
0.095\,pN for measurements with different absorbed powers.
The total experimental error determined by the random error was
found at a 95\% confidence level. For this purpose the statistical
rule\cite{5,15,66} adapted for the case when the
quantities to be combined  are described by the normal (or Student)
and uniform distributions was used.
In Fig.~\ref{aba:fig13}(a) the total experimental error as a
function of separation is shown by the lines 1 and 2 for the
absorbed powers 9.3 and 4.7\,mW, respectively.
As can be seen in Fig.~\ref{aba:fig13}(a), the relative total
experimental error varies from 10\% to 20\% at a separation
$a=100\,$nm and from 25\% to 33\% at a separation
$a=180\,$nm for different absorbed powers. The absolute error
in the measurements of separations was $\Delta a=1\,$nm.

Now we discuss the theoretical errors. The main source of the
theoretical uncertainty in this experiment is the error in the
concentration of charge carriers $n_l$ when the light is on.
{}From Sec.~3.1, this error is of about 20\%.
This leads to respective errors in the values of the plasma
frequency (see Sec.~3.3) and results in the relative error in
computed difference Casimir forces,
$\delta_1F_{\!\rm diff}^{\,\rm theor}$, approximately equal
to 12\%. This error does not depend on separation.
Another theoretical error is due to the uncertainty of the
experimental separations $a_i$ at which the values of the
theoretical force $F_{\!\rm diff}^{\,\rm theor}$ should be
computed. For each of the forces in the absence,
$F^{\,\rm theor}(a_i)$, and in the presence,
$F_l^{\,\rm theor}(a_i)$, of light this relative error is
equal to $3\Delta a_i/a_i$ and takes its maximum value 3\%
of the Casimir force at $a=100\,$nm. This leads to only
$\delta_2F_{\!\rm diff}^{\,\rm theor}=2$\%  error
in the difference Casimir force at $a=100\,$nm and to
smaller errors at larger separations. Note that the
other theoretical errors, such as due to sample-to-sample
variations of the optical data, patch potentials, and the use
of the PFA are negligibly small in comparison with the two
errors listed above.
The two errors $\delta_1F_{\!\rm diff}^{\,\rm theor}$ and
$\delta_2F_{\!\rm diff}^{\,\rm theor}$ were combined at
a 95\% confidence level using the statistical rule for the
combination of the quantities distributed
uniformly.\cite{5,15,24,28a,66}
The resulting absolute total theoretical errors are shown
as functions of separation in Fig.~\ref{aba:fig13}(b) by the
lines 1 and 2 for the maximum and minimum absorbed powers,
respectively.
When calculating the absolute errors shown in
Fig.~\ref{aba:fig13}(b) we used the Drude-like behavior of
the dielectric permittivity at low frequencies because the
results obtained are applied below to demonstrate an
inconsistency between the Lifshitz theory and the Drude
model approach. The use of the plasma-like behavior results in
slightly larger absolute errors.
The absolute total theoretical errors
shown in Fig.~\ref{aba:fig13}(b) can be employed for the
determination of the confidence interval of the quantity
$F_{\!\rm diff}^{\,\rm theor}(a_i)-F_{\!\rm diff}^{\,\rm expt}(a_i)$
similar to those shown in Figs.~\ref{aba:fig6} and \ref{aba:fig8}.
It should be kept in mind, however, that the theoretical error
$\delta_2F_{\!\rm diff}^{\,\rm theor}$ is not applicable to
the widths of the theoretical bands computed over the entire
measurement range, but only at each experimental
point (see below Figs.~\ref{aba:fig15} in Sec.~3.3 and
\ref{aba:fig17} in Sec.~3.4).

In the original publications\cite{28,28a} one more major theoretical
error was considered connected with the pressure of light
transmitted through the plate and incident on the bottom
of the sphere. This effect is present only during the bright
phase of the pulse train and at a separation $a=100\,$nm led
to 2.3\% and 1.5\% errors in the difference Casimir force for
the two absorbed powers. At $a=200\,$nm the relative total
theoretical error due to light pressure was estimated as
8.9\% and 5\% for different absorbed powers.\cite{5,28a}
The repulsive force on the sphere due to the light pressure
can be, however, computed and subtracted from the measured
difference Casimir force at all experimental separations.
In this way the {\it corrected} values of the difference
Casimir force are restored which would be measured in the
case when all 100\% of the light from a 514\,nm laser were
absorbed in the Si plate. The magnitudes of corrected
values are a bit larger than the original ones.
Such a procedure in metrology is called a
{\it correction}.\cite{66} Below we correct the
experimental data for the difference Casimir force due to the
presence of the light pressure, instead of including its
role in the theoretical error.

If $I$ is the intensity of light incident on an area element
$dS$ at an angle $\theta$, the magnitude of respective
force due to the light pressure acting perpendicular to
the plate takes the form
\begin{equation}
|d\mbox{\boldmath$F$}_{\rm lp}|=2\,\frac{I}{c}\,\cos\theta\,dS.
\label{eqn19}
\end{equation}
\noindent
Let the $z$ axis be perpendicular to the plate and passes
through the center of the sphere. The intensity of laser light
incident on the spherical ring of radius $\rho=R\sin\theta$
and width $Rd\theta$ on the soarce side of the plate is
given by
\begin{equation}
I(\rho)=\frac{2\alpha_0P^{\,\rm eff}}{\pi w^2}\,
e^{-2\rho^2/w^2},
\label{eqn20}
\end{equation}
\noindent
where $\alpha_0$ is the fraction of the absorbed power
transmitted through the plate and $w=0.23\pm 0.01\,$mm
is the Gaussian width of the focused beam on the
plate.\cite{28}
The value of $\alpha_0$ can be calculated as
\begin{equation}
\alpha_0=r_{\rm tr}e^{-d/l_{\rm opt}}\approx 0.00641.
\label{eqn21}
\end{equation}
\noindent
It is obtained using the transmission coefficient
$r_{\rm tr}\approx 0.35$.

The area of a spherical ring of radius $\rho$ on a sphere
surface set at an angle $\theta$ to the $z$ axis is equal to
\begin{equation}
dS=2\pi\rho Rd\theta=2\pi R^2\sin\theta d\theta.
\label{eqn22}
\end{equation}
\noindent
Then the force, $F_{\rm lp}$, acting on a sphere owing to the
light incident on the source side of the plate (i.e., in the
$z$ direction) is the following:
\begin{equation}
F_{\rm lp}=\int_{0}^{\pi/2}dF_{{\rm lp},z}(\theta)=
\int_{0}^{\pi/2}\cos\theta |d\mbox{\boldmath$F$}_{\rm lp}(\theta)|,
\label{eqn23}
\end{equation}
\noindent
where $|d\mbox{\boldmath$F$}_{\rm lp}(\theta)|$ is given by Eqs.~(\ref{eqn19})--(\ref{eqn22}).
Introducing the new variable $t=\cos\theta$, Eq.~(\ref{eqn23})
can be rearranged to the form
\begin{equation}
F_{\rm lp}=\frac{8\alpha_0R^2P^{\,\rm eff}}{cw^2}\,e^{-2R^2/w^2}
\int_{0}^{1}t^2e^{2R^2t^2/w^2}dt.
\label{eqn24}
\end{equation}
\noindent
The last expression can be rewritten in terms of the imaginary
error function ${\rm Erfi}(z)$:
\begin{equation}
F_{\rm lp}=\frac{2\alpha_0 P^{\,\rm eff}}{c}\left[1-
f\left(\frac{\sqrt{2}R}{w}\right)\right],
\label{eqn25}
\end{equation}
\noindent
where
\begin{equation}
f(x)=e^{-x^2}\frac{\sqrt{\pi}\,{\rm Erfi}(x)}{2x}.
\label{eqn26}
\end{equation}

Now we determine the error in calculation of the force
$F_{\rm lp}$ using Eqs.~(\ref{eqn25}) and (\ref{eqn26}).
The error in the quantity $f(x)$ in Eq.~(\ref{eqn26})
is determined by the systematic errors in the measurements
of sphere radius $R$ and the width of the beam $w$
($\delta R=0.15$\% and $\delta w=4$\%, respectively).
Using the conservative assumption that these quantities
take any value around their mean values in the limit of errors
with equal probability (any other assumption leads to smaller
error) we find the error of $x=\sqrt{2}R/w$ determined at
a 95\% confidence level, $\delta x\approx 0.04$, from the same
combination rule\cite{5,15,24,28a,66} as was used above.
Then from Eq.~(\ref{eqn26}) one obtains
$1-f(x)=0.214\pm 0.015$.
The error of the factor $2\alpha_0P^{\,\rm eff}/c$ in
Eq.~(\ref{eqn25}) is mostly determined by the error of
$\alpha_0$ defined in Eq.~(\ref{eqn21}). The latter, in its
turn, is determined by $\delta d\approx 7.5$\%.
{}From Eq.~(\ref{eqn21}) this leads to
$\alpha_0=r_{\rm tr}(0.019\pm 0.005)$, i.e., to
$\delta\alpha_0=25$\%. Combining the two errors with the help of
the same combination rule, one obtains
$\delta F_{\rm lp}=28$\%.
This leads to an additional small
systematic error in the corrected measured difference
Casimir force. For example, at $a=100\,$nm the new
systematic error is equal to 0.64\% and 0.42\% of the
difference Casimir force for the two absorbed powers,
i.e., of the same order as other systematic errors
considered in Sec.~3.2. Although this results in a slightly
different combined systematic error, the total
experimental error shown in Fig.~\ref{aba:fig13}(a)
remains primarily determined by the random error and,
thus, is pretty much unchanged.

\begin{figure}[t]
\vspace*{-4.7cm}
\hspace*{-2.4cm}
\psfig{file=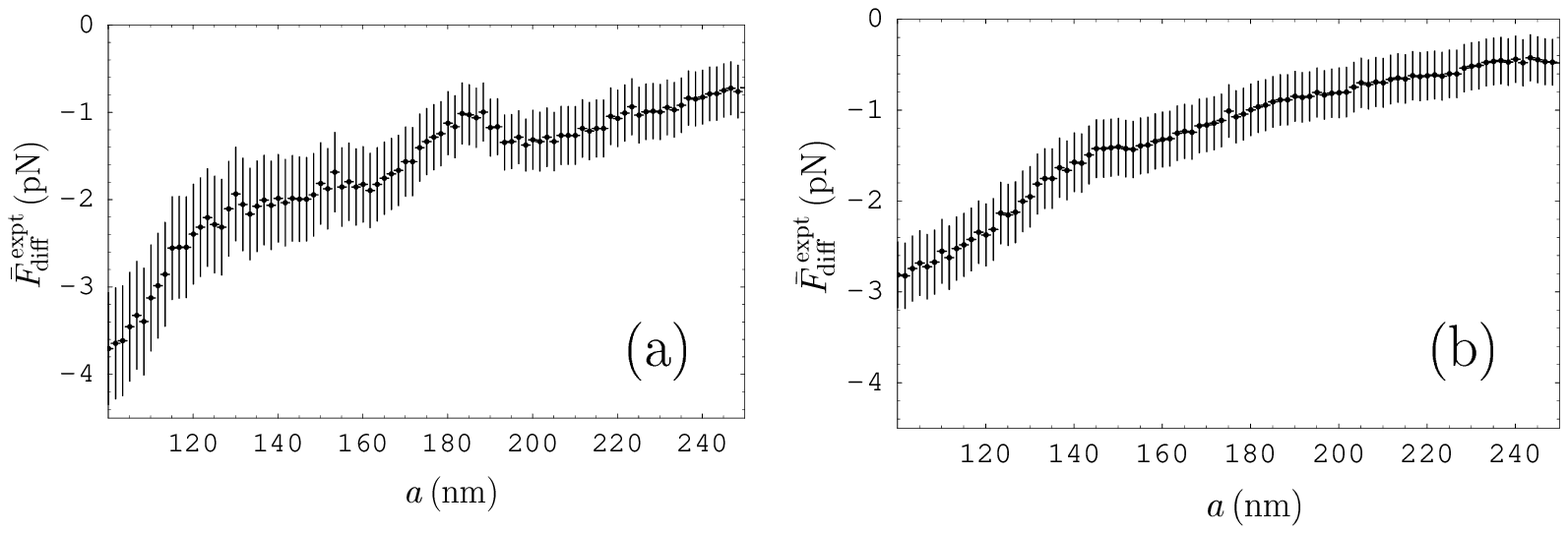,width=6.4in}
\vspace*{-14.2cm}
\caption{(a) The experimental differences in the Casimir force
in the presence and in the absence of light corrected for the
light pressure are shown as crosses versus separation.
The arms of the crosses are determined at a 95\% confidence
level.
The absorbed powers are (a) 9.3\,mW and (b) 4.7\,mW.}
\label{aba:fig14}
\end{figure}
In Fig.~\ref{aba:fig14}(a,b), the experimental data for the
difference Casimir force for
the absorbed powers of 9.2 and 4.7\,mW, respectively,
 with the subtracted
force (\ref{eqn25}) due to the light pressure are shown
as crosses.
Here, the vertical arms of the crosses are equal to
$2\Delta^{\!\rm tot}\bar{F}_{\!\rm diff}^{\,\rm expt}$
determined at a 95\% confidence level
[see Fig.~\ref{aba:fig13}(a)]. The horizontal arms of each
cross are equal tp $2\Delta a=2\,$nm. This should be
compared with Fig.~\ref{aba:fig12}(a,b) where the
uncorrected data for the difference Casimir force are
presented. To make Fig.~\ref{aba:fig14}(a,b) readable,
we have plotted only each fifth dot shown in
Fig.~\ref{aba:fig12}(a,b). In the following two sections,
the experimental data of Fig.~\ref{aba:fig14} are
compared with computations using different theoretical
approaches.

\subsection{Comparison with the Lifshitz theory}

Now we compare the experimental data for the difference in the
Casimir forces between an Au-coated sphere and Si plate in the
presence and in the absence of light on the plate with
theoretical predictions of the Lifshitz theory.
Computations of the Casimir force $F(a)=F(a,T)$ were performed
at $T=300\,$K using the Lifshitz formula (\ref{eqn3}) and
the PFA (\ref{eqn7}). For the dielectric permittivity of Au
either the generalized plasma-like model (\ref{eqn8}) or
the tabulated optical data\cite{46} extrapolated to low
frequencies by the imaginary part of the Drude model
(\ref{eqn10}) have been used. For Si in the absence of light
the dielectric permittivity along the imaginary frequency
axis, $\varepsilon_{\rm ce}^{(2)}(i\xi)$, is determined by the
contribution of the core electrons (see Sec.~2.1). For
high-resistivity, dielectric materials the role of dc
conductivity was usually neglected (see, for instance,
computations of the Casimir-Polder force between Rb atoms
and SiO${}_2$ plate\cite{67}). If, however, we would like
to take into consideration the dc conductivity  of the Si
plate in the absence of light, the dielectric permittivity
along the imaginary frequency axis takes the form\cite{46}
\begin{equation}
\varepsilon^{(2)}(i\xi)=\frac{4\pi\sigma_0^{(2)}(T)}{\xi}+
\varepsilon_{\rm ce}^{(2)}(i\xi),
\label{eqn27}
\end{equation}
\noindent
where $\sigma_0^{(2)}(T)$ is the static conductivity.
As was mentioned in Sec.~1, the substitution of the dielectric
permittivity (\ref{eqn27}) into the Lifshitz theory results in
the violation of the Nernst heat theorem.\cite{29}\cdash\cite{31}

Both the violation of the Nernst theorem and significantly larger
magnitudes of the Casimir force obtained when the dc conductivity
is taken into account are explained by different contributions
from the TM reflection coefficient (\ref{eqn4}) into the
zero-frequency term of the Lifshitz formula (\ref{eqn3}).
Thus, from Eqs.~(\ref{eqn3}) and (\ref{eqn7}), for a metal
sphere above a dielectric plate with dc conductivity neglected
[Si is described by $\varepsilon_{\rm ce}^{(2)}$ and Au either by
the generalized plasma-like model (\ref{eqn8}) or by the
tabulated optical data extrapolated by the Drude model]
the contribution of the zero-frequency term to the force is:
\begin{equation}
F_0(a,T)=-\frac{k_BTR}{8a^2}\,{\rm Li}_3\left[
\frac{\varepsilon_0^{(2)}-1}{\varepsilon_0^{(2)}+1}\right].
\label{eqn27a}
\end{equation}
\noindent
Here $\varepsilon_0^{(2)}\equiv\varepsilon_{\rm ce}^{(2)}(0)$
and ${\rm Li}_n(z)$ is the polylogarithm function.
Note that Eq.~(\ref{eqn27a}) represents the contribution
of the TM mode alone because
$r_{\rm TE}^{(2)}(0,k_{\bot})=0$.
If the dc conductivity of a dielectric plate is taken into
account, i.e., the dielectric permittivity
$\varepsilon^{(2)}(i\xi)$ in Eq.~(\ref{eqn27}) is used
instead of $\varepsilon_{\rm ce}^{(2)}(i\xi)$,
the contribution of the zero-frequency term to the force is
given by
\begin{equation}
\tilde{F}_0(a,T)=-\frac{k_BTR}{8a^2}\,\zeta(3),
\label{eqn27b}
\end{equation}
\noindent
where $\zeta(z)$ is the Riemann zeta function. In so doing
the contribution from the TE mode at zero frequency remains
equal to zero. It is easily seen that
$|\tilde{F}_0(a,T)|>|F_0(a,T)|$.
The measurements of the optically modulated Casimir force were
of sufficient precision to check the predictions of the
Lifshitz theory with dc conductivity included experimentally.

For Si in the presence of light the dielectric permittivity along
the imaginary frequency axis was described using either the
generalized Drude-like model
\begin{equation}
\tilde\varepsilon_l^{(2)}(i\xi)=\frac{\omega_{p,n}^2}{\xi(\xi+\gamma_n)}+
\frac{\omega_{p,p}^2}{\xi(\xi+\gamma_p)}+
\varepsilon_{\rm ce}^{(2)}(i\xi)
\label{eqn28}
\end{equation}
\noindent
or the generalized plasma-like model
\begin{equation}
\varepsilon_l^{(2)}(i\xi)=\frac{\omega_{p,n}^2}{\xi^2}+
\frac{\omega_{p,p}^2}{\xi^2}+
\varepsilon_{\rm ce}^{(2)}(i\xi).
\label{eqn29}
\end{equation}
\noindent
Here, the values of the plasma frequencies obtained from the
densities of charge carriers indicated above are the following:
$\omega_{p,n}=(5.1\pm 0.5)\times 10^{14}\,$rad/s,
$\omega_{p,p}=(5.7\pm 0.6)\times 10^{14}\,$rad/s
for the absorbed power $P^{\,\rm eff}=9.3\,$mW
and
$\omega_{p,n}=(4.1\pm 0.4)\times 10^{14}\,$rad/s,
$\omega_{p,p}=(4.6\pm 0.4)\times 10^{14}\,$rad/s
for the absorbed power $P^{\,\rm eff}=4.7\,$mW.
The values of the relaxation parameters were:\cite{68}
$\gamma_n\approx 1.8\times 10^{13}\,$rad/s and
$\gamma_p\approx 5.0\times 10^{12}\,$rad/s.
They do not depend on the absorbed power.

The calculated differences of the Casimir  force in the presence
and in the absence of light, (\ref{eqn20a}), were corrected
for the presence of surface roughness using the nonmultiplicative
approach of geometrical averaging, as discussed in Sec.~2.1.
The force differences with inclusion of surface roughness are
notated as $F_{\!\rm diff}^{\,\rm theor}(a)$.
In the experiment on optically modulated Casimir force the
contribution from the roughness correction was very small.
Thus, at $a=100\,$nm, it contributed only 1.2\% of the
calculated $F_{\!\rm diff}^{\,\rm theor}(a)$.
At $a=150\,$nm, the contribution from the surface roughness
decreased to only 0.5\% of the calculated force difference.

\begin{figure}[t]
\vspace*{-4.7cm}
\hspace*{-2.4cm}
\psfig{file=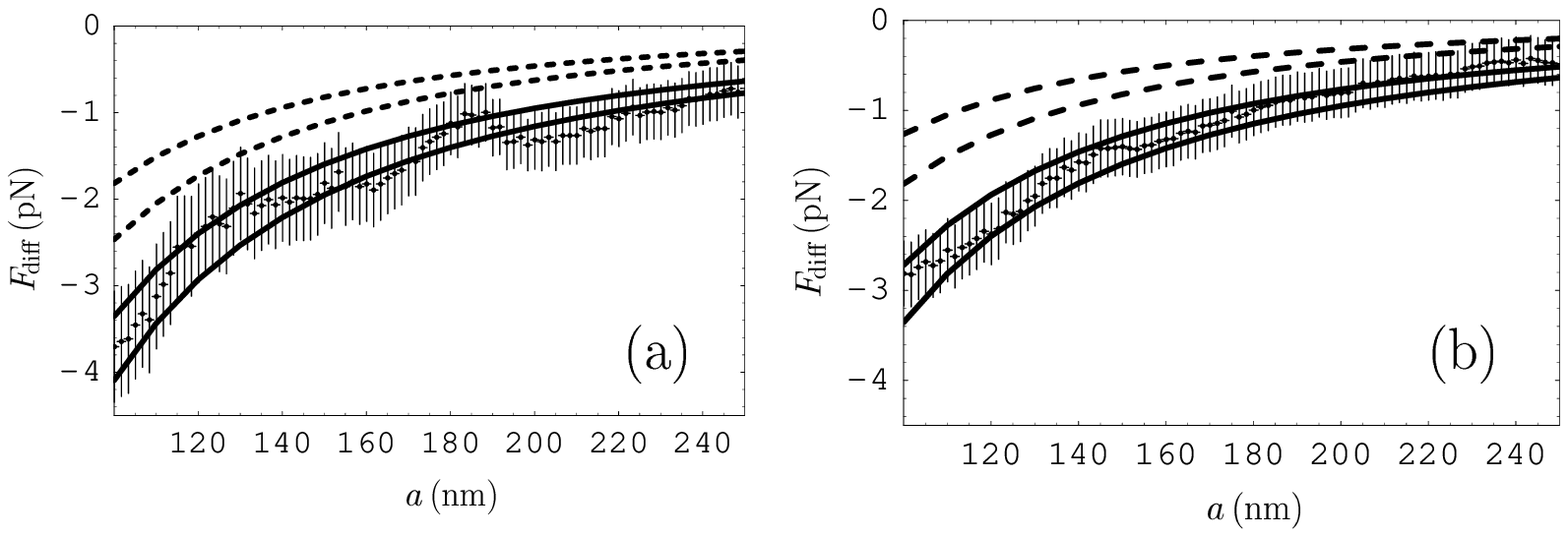,width=6.4in}
\vspace*{-14.2cm}
\caption{The experimental differences in the Casimir force
(crosses plotted at a 95\% confidence level) are reproduced
from Fig.~\ref{aba:fig14}.
The theoretical bands confined between the pairs of solid
lines and dashed lines are found at a 95\% confidence level
with the dc conductivity of a Si plate in the dark phase
disregarded and included, respectively.
The absorbed powers are (a) 9.3\,mW and (b) 4.7\,mW.}
\label{aba:fig15}
\end{figure}
The comparisons between the experimental data and the Lifshitz
theory for the absorbed powers of 9.3 and 4.7\,mW are presented
in Fig.~\ref{aba:fig15}(a,b), respectively. The experimental data
for the difference in the Casimir forces in the presence and
in the absence of light are shown as crosses. These crosses,
whose arms are determined at a 95\% confidence level, are the
same as in Fig.~\ref{aba:fig14} (each fifth experimental data
point is shown). The theoretical bands confined between the
two solid lines are computed using the generalized plasma-like
model for Au. For Si the dielectric permittivity of core
electrons $\varepsilon_{\rm ce}^{(2)}(i\xi)$ was used in the dark
phase and the generalized plasma-like model (\ref{eqn29}) in
the bright phase. Note that the use of the tabulated optical
data extrapolated to low frequencies by the Drude model
for Au and the generalized Drude-like model for Si in the
bright phase is illustrated in the next figure.
The widths of the bands are determined by the errors in the
plasma frequency of Si in the bright phase originating from
the errors in the concentration of charge carriers.
These bands are calculated not at the experimental separations,
but over the entire measurement range. Because of this, their
widths do not include uncertainties due to the experimental error
$\Delta a$ in the measurement of separations. As can be seen in
Fig.~\ref{aba:fig15}(a,b), both theoretical bands confined
between the pairs of two solid lines are consistent with the data.

The theoretical bands confined between the two dashed lines in
Fig.~\ref{aba:fig15}(a,b) are computed accounting for
dc conductivity of a Si plate in the absence of light by means
of Eq.~(\ref{eqn27}). In accordance with this Si plate in the
bright phase is described by the generalized Drude-like
model (\ref{eqn28}) and Au by the tabulated optical
data extrapolated by the Drude model to low frequencies.
The widths of the dashed bands are again determined by large
errors of the plasma frequencies of Si  in the bright phase.
As can be seen in Fig.~\ref{aba:fig15}(a,b), the Lifshitz
theory taking into account dc conductivity of high-resistivity
Si in the dark phase is experimentally excluded by the data over
the entire range of separations from 100 to 250\,nm
(for the absorbed power 9.3\,mW) and from 100 to 230\,nm
(for the absorbed power 4.7\,mW).
This exclusion holds at a 95\% confidence level.

Thus, in spite of the fact that the dc conductivity of a
dielectric-type semiconductor (Si in the absence of light) is
a real physical phenomenon, its inclusion in the model of
dielectric response makes the Lifshitz theory inconsistent
with the measurement data. This should be compared with
the violation
of the Nernst heat theorem in the Lifshitz theory which
occurs when the dc conductivity of a dielectric plate is taken
into account\cite{5,15,29}\cdash\cite{31} (see Sec.~1).
One can conclude that the inclusion of dc conductivity of
dielectric materials makes the Lifshitz theory both
theoretically and experimentally inconsistent in analogy
with the similar situation for metals discussed in Sec.~1.
Phenomenologically, the Lifshitz theory comes to agreement
with thermodynamics and is consistent with the experimental data
of all experiments performed up to date if one disregards
the relaxation processes of conduction electrons in the
case of metals\cite{5,15,69} and omits the contribution of
charge carriers in the case of dielectrics.\cite{5,15,70}
These prescriptions, however, remain unexplained from basic
principles of quantum statistical physics.\cite{21}

\begin{figure}[t]
\vspace*{-4.7cm}
\hspace*{-2.4cm}
\psfig{file=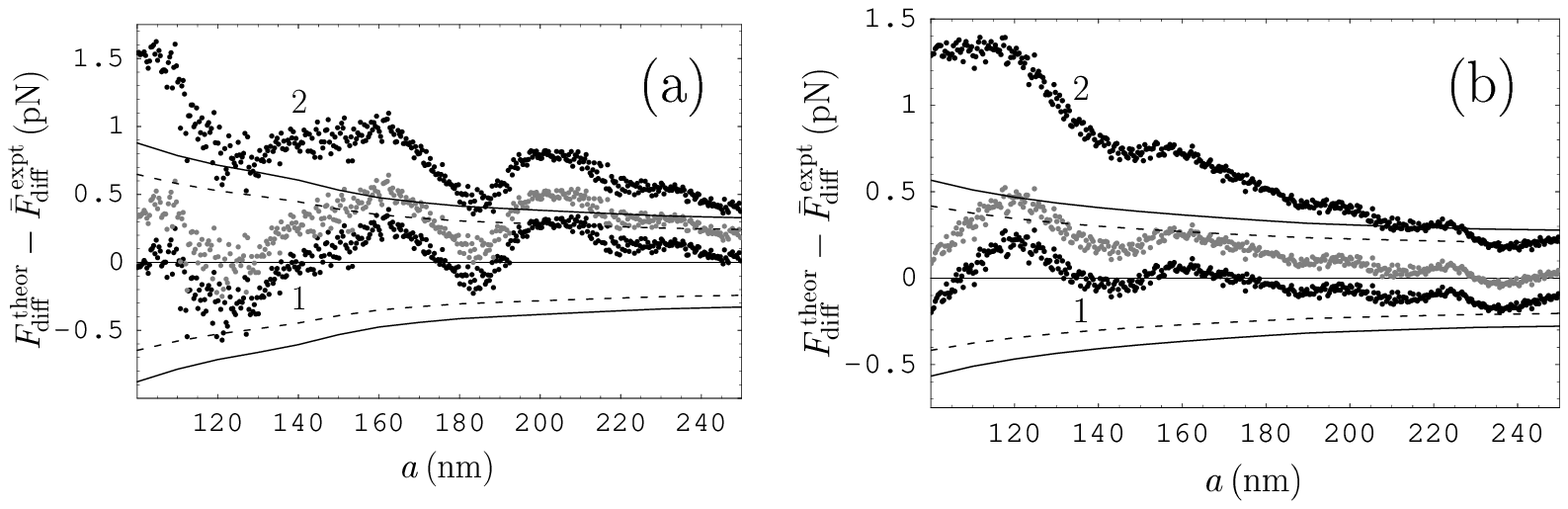,width=6.4in}
\vspace*{-14.2cm}
\caption{Theoretical minus mean experimental differences in the Casimir
forces are indicated by dots. For black dots labeled 1 and for
grey dots dc conductivity of Si in the dark phase is omitted,
and charge carriers in the bright phase are described by the
generalized plasma-like and Drude-like models, respectively.
For black dots labeled 2 dc conductivity of Si in the dark phase
is included and charge carriers in the bright phase are described by the
generalized  Drude-like model.
The pairs of the solid and dashed lines indicate the borders
of 95\% and 70\% confidence intervals, respectively.
The absorbed powers are (a) 9.3\,mW and (b) 4.7\,mW.}
\label{aba:fig16}
\end{figure}

The obtained conclusions were confirmed using another method of
comparison between experiment and theory (see
Figs.~\ref{aba:fig6} and \ref{aba:fig8}).
In Fig.~\ref{aba:fig16}(a,b) we plot as dots theoretical minus
mean experimental difference Casimir forces,
$F_{\!\rm diff}^{\,\rm theor}(a)-
\bar{F}_{\!\rm diff}^{\,\rm expt}(a)$,
for the absorbed powers 9.3\,mW and 4.7\,mW, respectively.
For the black dots labeled 1, the dc conductivity of a Si
plate in the dark phase is disregarded and charge carriers in
the bright phase are described by the generalized plasma-like
model (\ref{eqn29}). In so doing the Au coating on the sphere is
also described by means of the generalized plasma-like
model (\ref{eqn11}). For the grey dots, the dc conductivity
of Si in the dark phase is also omitted, but charge carriers
in the bright phase are described by the generalized Drude-like
model (\ref{eqn28}). To use a uniform approach, Au in this
case is described by the tabulated optical data extrapolated
by the Drude model. As to the black dots labeled 2, the dc
conductivity of a Si plate in the dark phase was taken into
account in accordance with Eq.~(\ref{eqn27}). Si in
the bright phase was described by the generalized Drude-like
model (\ref{eqn28}) and Au by the tabulated optical data extrapolated
by the Drude model.

The pairs of the solid and dashed lines in Fig.~\ref{aba:fig16}(a,b)
indicate the borders of the confidence intervals
$[-\Xi_{0.95}(a),\Xi_{0.95}(a)]$ and $[-\Xi_{0.7}(a),\Xi_{0.7}(a)]$
defined at a 95\% and 70\% confidence levels, respectively.
Here, the distribution law for the theoretical minus
experimental force differences remains unknown. Because of this,
we find the half-width of the confidence interval $\Xi_{0.7}(a)$
from the most conservative assumption of a uniform distribution,
i.e., $\Xi_{0.95}(a)/\Xi_{0.7}(a)=0.95/0.7\approx 1.357$
(compare with Sec.~2.1 where the normal distribution was used).

As can be seen in Fig.~\ref{aba:fig16}(a,b), all the black dots
labeled 1 and most of the grey dots belong to the confidence intervals
shown by the solid lines.
This means that the two versions of the theory with omitted
dc conductivity of Si in the dark phase are experimentally
consistent within a 95\%  confidence interval.
Most of the  black dots labeled 1 and the grey dots
within some separation intervals [especially in
Fig.~\ref{aba:fig16}(b)] also belong to the narrower 70\%
confidence intervals. This leads to a more definitive
conclusion that theoretical approaches with a dc
conductivity of Si in the dark phase omitted are
experimentally consistent within a 70\% confidence
interval. At the same time the experimental data of the
optical modulation experiment are not of sufficient
precision to convincingly discriminate between the
description of free charge carriers in Si in the bright
phase by means of the generalized plasma-like and
Drude-like models.
As to the black dots labeled 2, much more than 95\%
of them are outside the borders of the 95\% confidence
intervals. Because of this one can conclude that the
Lifshitz theory with the inclusion of the dc conductivity
of dielectric Si is experimentally excluded at a 95\%
confidence
level. This is  in agreement with similar results
obtained above using another method for the comparison
between experiment and theory.

Importantly, the same conclusion, that the Lifshitz
theory taking into account the dc conductivity of a
dielectric material contradicts with the experimental
data, was obtained from the measurement of the thermal
Casimir-Polder forces between ${}^{87}$Rb atoms and
a SiO${}_2$ plate.\cite{67}
The theoretical results for the Casimir-Polder force
computed\cite{67} with dc conductivity of
SiO${}_2$ omitted were found\cite{67} in a good agreement with
the measurement data. However, the theoretical results
computed\cite{71} taking the dc conductivity of
SiO${}_2$ into account were shown\cite{71}
to be excluded by the data at a 70\% confidence level.
Thus, at the present time there are three types
of experiments (with metallic,\cite{16}\cdash\cite{18a}
semiconductor,\cite{28,28a} and dielectric\cite{67,71}
materials) which demonstrate that serious problems
arise in the application of the Lifshitz theory in
connection with charge carriers.

{}From a technological point of view, the experiment
on optically modulated Casimir force is of special interest.
In Sec.~2 we discussed changes of the Casimir
force between an Au sphere and an Au plate when the plate
material is replaced with different semiconductors.
These changes are static and determined by the type of
semiconductor used. In the optical modulation experiment the
situation is quite different. Here, in the dark phase, the
magnitude of the Casimir force between a sphere and a plate
is approximately 66\% and 62\% of the Casimir force
between an Au sphere and Au plate at $a=100$ and 150\,nm,
respectively. The increase of the charge carrier density
by almost 5 orders of magnitude in the bright phase
results in a 3.8\% and 5.6\% increase in the magnitude of
the Casimir force at the same respective separations
($P^{\,\rm eff}=9.3\,$mW is assumed).
The Casimir force is changing periodically until the
laser pulses are on. The application
of this phenomenon to  microdevices
moving back and forth without use of mechanical spring
 under the influence of attractive and
repulsive Casimir forces are considered in Sec.~6.
The use of the optical modulation of the Casimir force in
an ambient environment would be most promising.
In this respect it was demonstrated\cite{72} recently
that an oxide film formed on a silicon surface in air
hardly affects the possibility of modulating the
Casimir force when distances between interacting
bodies are of the order of 100\,nm.

\subsection{Comparison with the modified Lifshitz theory}

In the previous section it was demonstrated that the Lifshitz theory
including the dc conductivity of dielectric materials is
not only thermodynamically inconsistent, but is in conflict with
the measurement data of two experiments. To solve this problem,
attempts were undertaken\cite{32}\cdash\cite{34} to modify
the Lifshitz theory by including into its formalism
screening effects and diffusion currents.  According to the
proposed modification,\cite{32} the electric field in the
dielectric material can be screened due to the presence of free
charge carriers with some small density $n$. As a result, the
potential around a point charge $e$, instead of a familiar
Coulombian form, takes the Yukawa-type form,
$e\,\exp(-\kappa r)/r$, where $\kappa$ is the inverse screening
length.  If $n$ is sufficiently low, the charge carriers can be
described by classical Maxwell-Boltzmann statistics and the
Debye-H\"{u}ckel approximation is valid where
\begin{equation}
\kappa=\kappa_{\rm DH}=\frac{1}{R_{\rm DH}}=
\sqrt{\frac{4\pi e^2n}{\varepsilon_0k_BT}}.
\label{eqn30}
\end{equation}
\noindent
Here, $\varepsilon_0=\varepsilon_{\rm ce}(0)$ is the dielectric
permittivity of the core electrons at zero frequency, and
$R_{\rm DH}$ is the screening length. It was assumed that the
effect of screening leads to noticeable changes of the
reflection coefficients only for the static field.
As a consequence, the TM reflection coefficient at zero
frequency is replaced with\cite{32}
\begin{equation}
r_{\rm TM}^{\rm mod}(0,k_{\bot})=
\frac{\varepsilon_0\sqrt{k_{\bot}^2+\kappa^2}-
k_{\bot}}{\varepsilon_0\sqrt{k_{\bot}^2+\kappa^2}+
k_{\bot}}.
\label{eqn31}
\end{equation}
\noindent
At the same time, all of the coefficients
$r_{\rm TM,TE}(i\xi_l,k_{\bot})$ with $l\geq 1$ and
$r_{\rm TE}(0,k_{\bot})$ defined
 in Eq.~(\ref{eqn4}) using the dielectric permittivity
$\varepsilon_{\rm ce}(i\xi_l)$ remain
unchanged.\cite{32}
The modified coefficient (\ref{eqn31}) should be compared with the
standard one, following from Eq.~(\ref{eqn4})
\begin{equation}
r_{\rm TM}(0,k_{\bot})=
\frac{\varepsilon_0-1}{\varepsilon_0+1}.
\label{eqn32}
\end{equation}
\noindent
It is evident that Eq.~(\ref{eqn32}) is obtained from  Eq.~(\ref{eqn31})
in the limiting case $n\to 0$.

The modified reflection coefficients for the TM and TE modes at any
frequency were obtained\cite{33} through the use of the Boltzmann
transport equation, which takes into account not only the drift
current {\boldmath$j$}, but also the diffusion current
$eD\nabla n$, where $D$ is the diffusion constant and $\nabla n$
is the gradient of the charge carrier density.
As expected, the TE reflection coefficient at zero frequency
remained the same as in  Eq.~(\ref{eqn4}) with the generalized
Drude-like dielectric permittivity (\ref{eqn12}).
All reflection coefficients
$r_{\rm TM}^{\rm mod}(i\xi_l,k_{\bot})$ with $l\geq 1$
were found approximately equal to
$r_{\rm TM}(i\xi_l,k_{\bot})$ defined in  Eq.~(\ref{eqn4})
to a high degree of accuracy. This approach was also
applied\cite{34} to metallic plates, i.e., to bodies with
high density of free charge carriers described by the
Fermi-Dirac statistics, by replacing the Debye-H\"{u}ckel
screening length with the Thomas-Fermi screening length
\begin{equation}
\kappa=\kappa_{\rm TF}=\frac{1}{R_{\rm TF}}=
\sqrt{\frac{6\pi e^2n}{\varepsilon_0E_{F}}},
\label{eqn33}
\end{equation}
\noindent
where $E_{F}=\hbar\omega_p$ is the Fermi energy.

The proposed modified theory of the van der Waals and Casimir
force includes the effect of free charge carriers of the
material plates in the reflection coefficients by means of
the microscopic quantity $n$ rather than by adding a
contribution like in Eq.~(\ref{eqn27}) in the dielectric
permittivity. Similar to the standard Lifshitz theory,
the modified theory was developed under a condition of
thermal equilibrium. However, in contradiction with this
condition, the modified theory takes into account the
screening effects and both drift and diffusion currents
described by means of the Boltzmann transport equation.
These physical phenomena are caused by a nonequilibrium
distribution of charge carriers in an external field,
i.e., by a situation out of thermal equilibrium. It is
pertinent to note that the Boltzmann transport equation
used to derive the reflection coefficients
$r_{\rm TM,TE}^{\,\rm mod}(i\xi)$ describes only
nonequilibrium processes which must be accompained by
an increase of entropy.\cite{73}

It was shown that the modified theory of the van der Waals
and Casimir force violates the Nernst heat theorem for
several classes of dielectric
materials\cite{5,15,35,37,38,74} and for metals with
perfect crystal lattices.\cite{5,15,36,38,75}
There were a few dissenters from this
result\cite{34,76}\cdash\cite{79} in application to
dielectric materials. It was noted,\cite{21} however,
that all suggested proofs\cite{34,76}\cdash\cite{79}
of the validity of the Nernst heat theorem in the
modified Lifshitz theory used an assumption that the
density of charge carriers in dielectric materials
vanishes when the temperature goes to zero.
By contrast, the proof of violation of the Nernst
theorem in the modified theory was formulated
for such dielectric materials as doped semiconductors
with doping concentration below critical,
dielecric-type semimetals and ionic conductors.
For all these materials the density of charge carriers
does not vanish when $T\to 0$ and conductivity
vanishes with temperature due to the vanishing mobility.
Thus, the modified Lifshitz theory is really in
contradiction with thermodynamics for wide classes
of materials.

\begin{figure}[t]
\vspace*{-4.7cm}
\hspace*{-2.4cm}
\psfig{file=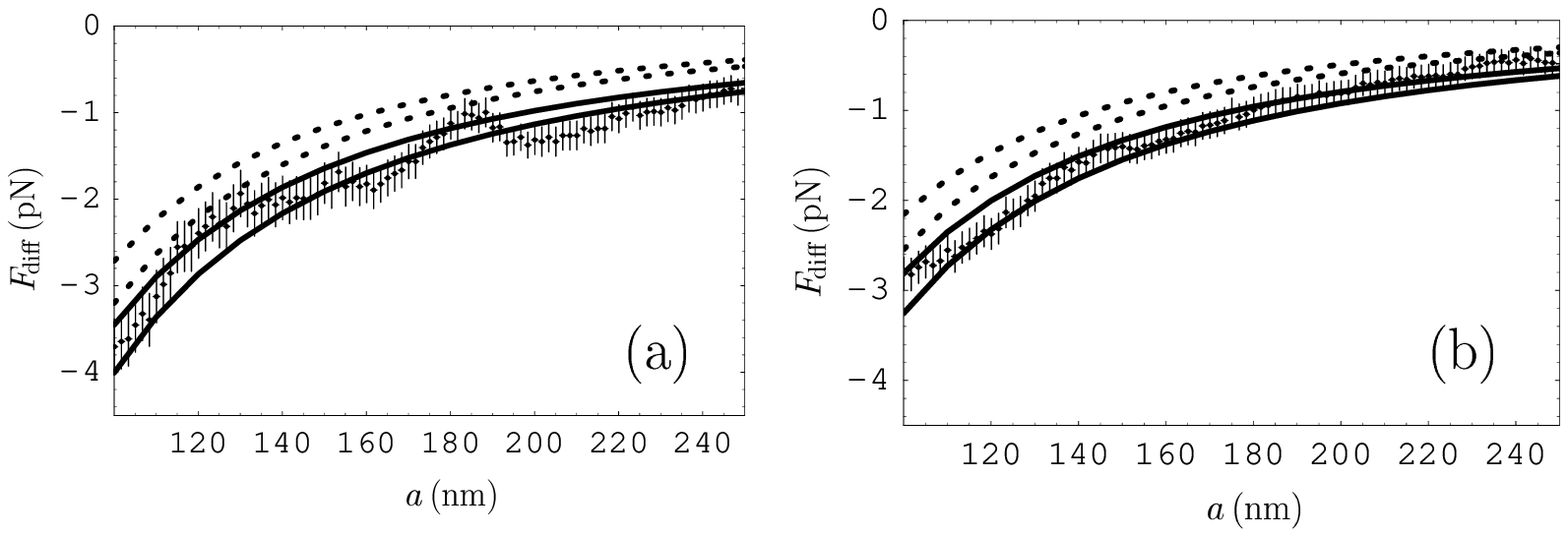,width=6.4in}
\vspace*{-14.2cm}
\caption{The experimental differences in the Casimir force
are presented as crosses plotted at a 70\% confidence level.
The theoretical bands confined between the pairs of solid
lines and dashed lines are found at a 70\% confidence level
using the standard Lifshitz theory
with the dc conductivity of a Si plate in the dark phase
neglected and the modified Lifshitz theory, respectively.
The absorbed powers are (a) 9.3\,mW and (b) 4.7\,mW.}
\label{aba:fig17}
\end{figure}

Here we compare theoretical predictions of the modified Lifshitz
theory with the experimental data of an experiment on the
optically modulated Casimir force. As was noticed in
Ref.~\refcite{35}, the comparison of this theory with the
experimental data at a 95\% confidence level is not informative
because it does not lead to a definite conclusion on the
incompatibility with the data. In Fig.~\ref{aba:fig17}(a,b)
the experimental data for the difference Casimir force for two
absorbed powers (9.3 and 4.7\,mW, respectively) are shown as
crosses. The arms of the crosses are twice the respective
errors determined at a 70\% confidence level. Keeping in mind
that the error $\Delta a$ in the measurement of absolute
separation is systematic, the value of this error at a 70\%
confidence level was found from the assumption of a uniform
distribution: $\Delta a=1\,\mbox{nm}/1.375\approx 0.74\,$nm
(see Sec.~3.3). The total errors in the force differences
are determined by the random errors which are characterized by the
normal (or Student with sufficiently large number of degrees
of freedom) distribution. Because of this, the absolute total
errors in the mean difference Casimir force at a 70\%
confidence level were obtained as one half of those shown
in Fig.~\ref{aba:fig13}(a) (see Sec.~3.2).
The theoretical bands confined between the pairs of two
solid lines were computed using the dielectric permittivity
of Si, $\varepsilon_{\rm ce}^{(2)}(i\xi)$, in the dark phase
and describing Si by the generalized plasma-like model in the
bright phase. Au was described by the generalized plasma-like
model.
The theoretical bands confined between the pairs of two
dashed lines were computed using the modified Lifshitz
theory.\cite{33} The widths of all theoretical bands were
found from the errors in the concentration of charge
carriers in the presence of light, $n_l$, determined at a
70\% confidence level. Taking into account that the errors in
charge carriers are of systematic nature, their values
at a 70\% confidence level were obtained using the most
conservative assumption of a uniform distribution.
The results are
$\Delta n_l=0.3\times 10^{19}\,\mbox{cm}^{-3}$ and
$\Delta n_l=0.22\times 10^{19}\,\mbox{cm}^{-3}$
for the absorbed powers 9.3 and 4.7\,mW, respectively
(compare with Sec.~3.1).

As can be seen in Fig.~\ref{aba:fig17}(a), the standard
Lifshitz theory with the dc conductivity neglected is
consistent with the data within a 70\% confidence interval
over the range of separations from 100 to 250\,nm.
The modified Lifshitz theory is excluded by the data at
a 70\% confidence level over the same separation region.
{}From Fig.~\ref{aba:fig17}(b) one can conclude that the
standard Lifshitz theory with neglected dc conductivity is also
consistent with the data over all separations, but the
modified theory is excluded over the separation region
from 100 to 210\,nm.

\begin{figure}[t]
\vspace*{-4.7cm}
\hspace*{-2.4cm}
\psfig{file=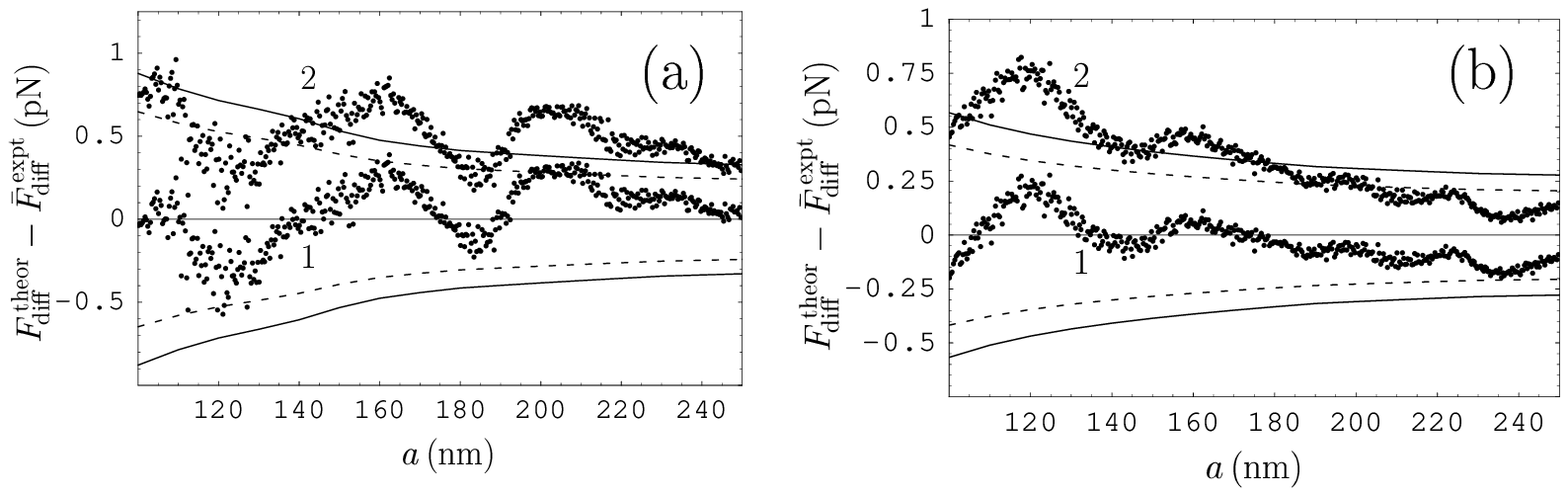,width=6.4in}
\vspace*{-14.2cm}
\caption{Theoretical minus mean experimental differences in the Casimir
forces are indicated by dots. For dots labeled 1
the standard Lifshitz theory is used with
dc conductivity of Si in the dark phase omitted,
and charge carriers in the bright phase described by the
generalized plasma-like model.
For dots labeled 2 theoretical difference Casimir forces are
computed using the modified Lifshitz theory.
The pairs of the solid and dashed lines indicate the borders
of 95\% and 70\% confidence intervals, respectively.
The absorbed powers are (a) 9.3\,mW and (b) 4.7\,mW.}
\label{aba:fig18}
\end{figure}

The same results were confirmed using another method of
comparison between experiment and theory.
In Fig.~\ref{aba:fig18}(a,b) the theoretical minus mean
experimental difference Casimir forces are indicated as
dots.  For dots labeled 1, the theoretical differences
were computed using the standard Lifshitz theory with neglected
dc conductivity of Si in the dark phase.
For Si in the bright phase and for Au the
generalized plasma-like model has been used.
For dots labeled 2, Si in both phases and Au were described by the
modified Lifshitz theory.\cite{33}
The pairs of the solid and dashed lines indicate the borders
of 95\% and 70\% confidence intervals, respectively.
As is seen in Fig.~\ref{aba:fig18}(a,b), the dots labeled 1
belong both to 95\% confidence intervals and [most of dots
in Fig.~\ref{aba:fig18}(a) and all dots in Fig.~\ref{aba:fig18}(b)]
to 70\% confidence intervals. This means that the standard
Lifshitz theory with neglected dc conductivity of a Si plate
in the dark phase is experimentally consistent not only
within a 95\% confidence interval, but also within a narrower
70\% confidence interval. At the same time, from Fig.~\ref{aba:fig18}(a)
it is seen that most of dots labeled 2 are outside a 70\%
confidence interval (and many of them even outside a 95\%
confidence interval) over the entire range of separations from
100 to 250\,nm. The same conclusion follows from Fig.~\ref{aba:fig18}(b)
at separations below 205\,nm. This leads to the conservative
conclusion that the modified Lifshitz theory is excluded by the
measurement data of the experiment on optically modulated
Casimir forces at a 70\% confidence level.

It was claimed\cite{34} that the measurement data of the
experiment on the optical modulation of the Casimir force are
equally consistent with the modified Lifshitz theory and with the
standard Lifshitz theory with the dc conductivity neglected in the
dark phase. To prove this, the experimental data of
Fig.~\ref{aba:fig17}(a) at a 70\% confidence level were used,
but the theoretical band between the two dashed lines was plotted
at a 95\% confidence level using the corresponding uncertainty
in the charge carrier density
$\Delta n_l=0.4\times 10^{19}\,\mbox{cm}^{-3}$.
Such an mismatched comparison was recognized by the author as an
error. In the Erratum,\cite{80} however, both the experimental
data and the theoretical bands were plotted at the 95\% confidence
level, and the same claim concerning an equal consistency of both
theoretical approaches with the data was repeated.
This claim disregards the fact that at a 70\% confidence level the
theoretical prediction of the modified Lifshitz theory is excluded
by the data of experiment on the optical modulation of the Casimir force
[see Figs.~\ref{aba:fig17}(a,b) and \ref{aba:fig18}(a,b)].
As to the comparison with the modified Lifshitz theory at a 95\%
confidence level, the data were earlier recognized\cite{35} to
lack precision for this comparison to be informative.
Note also that in application to metallic test bodies the modified
Lifshitz theory was excluded\cite{36}\cdash\cite{38} at a 99.9\%
confidence level by dynamic measurements of the Casimir pressure
using a micromachined oscillator.\cite{18,18a}

\section{Changes of the Casimir force in phase transitions}

An interesting possibility to control the magnitude of the Casimir
force due to a change in the charge carrier density is offered by
semiconductor materials that undergo a phase transition with
change in temperature. This might be used in microdevices as one
further possibility to achieve periodic variations of the
Casimir force. {}From a fundamental point of view, the
modulation of the Casimir force by phase transitions of various
types offers additional test of the role of charge carriers in
the Lifshitz theory.
It has been known\cite{3} that the variation of the Casimir
free energy of a metal associated with the phase transition
to the superconducting state is very small. Nevertherless,
the magnitude of this variation can be comparable to the
condensation energy of a superconducting film and causes a
measurable increase in the value of the critical magnetic
field.\cite{80b,80c} It was proposed\cite{80d} also to measure
the change of the Casimir force in a superconducting cavity due
to a small change in temperature.
In semiconductors much greater influence of phase transitions
on the Casimir force is expected.
 Below we consider measurements\cite{39}
of Casimir forces between an Au sphere and AgInSbTe plates in
amorphous and crystalline phases. The phase transition occurs
with an increase of temperature (annealing). We also discuss
the proposal to measure the changes of the Casimir force
between an Au sphere and a vanadium dioxide film which undergoes
a dielectric to metal phase transition when the temperature
increases.

\subsection{Amorphous and crystalline  plates}

This experiment\cite{39} is of the same type, as discussed in
Sec.~2.2, but it was performed in the dynamic regime.
The gradient of the Casimir force was measured in succession
between an Au-coated sphere of $20.2\,\mu$m diameter attached
to the end of a cantilever of an AFM (see Fig.~\ref{aba:fig1})
and two plates consisting of AgInSbTe films deposited onto Al
coated Si wafers. In the first plate a $1\,\mu$m thick
 AgInSbTe film was amorphous, and in the second it was
annealed to the crystalline state. The cantilever of an AFM
was vibrated at its resonant frequency equal to 83.6\,kHz.
When sphere approached to the plate, the frequency shift induced
by the Casimir force was measured which is proportional to the
gradient of the Casimir force (see Sec.~2.3 for the description
of the dynamic regime).

The calibration of the setup was performed by measuring the
gradient of the Casimir force with different applied voltages.
The residual potential difference $V_0$ was found to vary
between 0--20\,mV for separations 40--150\,nm for both
amorphous and crystalline samples. The correction from the
residual electrostatic force was subtracted from the data
to obtain the measured gradient of the Casimir force for both
plates. Typical roughness of the samples was measured
using an AFM. It was described by a several
nanometers rms with a few
isolated peaks. The relative experimental error in the force
measurements was estimated as about 7\% for both plates at
all separations considered. This is in contrast with other dynamic
measurements of the Casimir force (see Sec.~2.3 and
Refs.~\refcite{17,18}). The two experimental force curves were
obtained (one for the first plate and another for the second)
each of which was an average of 13 measurements taken in
different areas of both plates. It was concluded that the
gradient of the Casimir force for the crystalline plate is
greater in magnitude by approximately 20\% than for an
amorphous plate within the separation region from 55 to 150\,nm.

The experimental results were compared with the
theoretically computed
using the Lifshitz theory. For this purpose the imaginary parts
of the dielectric functions for the crystalline and amorphous
films of the AgInSbTe were measured with ellipsometry as a
function of frequency in the frequency range from 0.04 to
8.9\,eV. The ${\rm Im}\,\varepsilon(\omega)$ for a crystalline
film exhibiting metallic conductivity was extrapolated to
frequencies below 0.04\,eV by means of the Drude model.
For an amorphous film the contribution from the dc conductivity
was neglected. The dielectric permittivities of both films
along the imaginary frequency axis were found by means of the
Kramers-Kronig relations. The obtained $\varepsilon(i\xi)$
for both films in the frequency region below 20\,eV have
drastic differences. This was attributed\cite{39} to the
structural transformation from the amorphous to the
crystalline state alone. It should be noted, however, that
the chosen description of dielectric properties disregards
free charge carriers that are present in the amorphous state
at any nonzero temperature and, thus, interprets this phase
transition as occuring from an ideal dielectric to
metallic state.

Computations of the gradient of the Casimir forces between
an Au-coated sphere and each of the plates were performed
using the Lifshitz theory. It is not reported whether the
Lifshitz theory at zero temperature or at
$T=300\,$K (as it was at the laboratory)
 has been used. This question
is in fact of importance for the comparison between the
results of computations and the measurement data.\cite{81}
In computations using the Lifshitz theory the surface
roughness was not taken into account. As a justification
for the neglect of roughness in computations it was noted
that the small roughness of the used samples ``is negligible
for the Casimir-force computations at separations above
70\,nm [\refcite{82}]''. As a result it was found that
the Lifshitz theory based on the measured optical
properties predicts a force smaller that the measured one
by 8--18\%. After discussing several reasons which might
be responsible for this deviation, which is larger than
the experimental and theoretical errors, it was stated
somewhat contradictorily that ``the observed deviation
between theory and experiment can be attributed to
surface roughness as discussed recently in
Ref.~[\refcite{83}].''

To conclude, the phase transition from amorphous to
crystalline states is a promising subject for further
investigations. In principle a semiconductor material
can be periodically switched between an amorphous and
a crystalline phases by using an appropriate power and
duration of the laser pulse.\cite{39} This opens
opportunities for measuring the gradient of the
difference Casimir force in the configuration of an
Au-coated sphere above a semiconductor plate undergoing
periodic structural changes similar to the experiment
on the optically modulated Casimir forces considered
in Sec.~3. The combination of a dynamic measurement
scheme with the advantages of a difference force
measurement can bring much more exact experimental
data which should be compared with different theoretical
approaches proposed in the literature taking into
account nonzero temperature, surface roughness and all
other relevant factors.

\subsection{Dielectric to metal transition in {\rm VO}${}_2$}

One more phase transition which leads to a change of the carrier
density of order $10^{4}$ is the transition from dielectric to
metallic state which occurs with the increase of temperature.
Although in the literature it is common to speak about dielectric
(insulator) to metal transition, this is in fact a transition
between two semiconductor plates with lower ($n$) and higher
($\tilde{n}$) charge carrier densities. As was shown above,
an increase of $n$ by a factor of $10^{4}$ leads to measurable
changes in the Casimir force. Thus, the phase transition from
dielectric to metal phase can be used to control the Casimir
force in a sphere-plate geometry by mere temperature change.

It was proposed\cite{40} to measure the change of the Casimir
force acting between an Au-coated sphere and a vanadium
dioxide (VO${}_2$) film deposited on a sapphire substrate.
VO${}_2$ films are monoclinic at  room temperature and can be
considered as semiconductors of the dielectric type as their
conductivity vanishes with vanishing temperature.
At 68${}^{\circ}$C the resistivity of VO${}_2$ films abruptly
decreases by a factor of $10^{4}$ from 10 to
$10^{-3}\,\Omega\,$cm. This is a consequence of the phase
transition to a metallic tetragonal phase.\cite{84}
In the process, the optical transmission of VO${}_2$ for
wavelength from $1\,\mu$m to greater than $10\,\mu$m
decreases by a factor 10--100. Thus, in this case we deal with
the structural transformations between the two different
crystalline phases, whereas in Sec.~4.1 the transformation
from an amorphous to a crystalline phase was considered.

The schematic of the experimental setup suitable for the
observation of the difference Casimir force in the dielectric
to metal phase transition is similar to that shown in
Fig.~\ref{aba:fig10}. Main differences are that the
specially fabricated Si plate should be replaced with
VO${}_2$ film on a sapphire substrate and, instead of a 514\,nm
Ar laser, a chopped 980\,nm laser should be used to heat the
VO${}_2$ film.\cite{85,86} Estimations show that approximately
10--100\,mW of power of the 980\,nm laser are required to
bring about all the optical switching of a VO${}_2$ film.
Here, as opposed to Sec.~3, the wavelength of the laser was
chosen in such a way that light only leads to heating of the
VO${}_2$ film, but does not change the number of free
charge carriers.

The proposed experiment on the measurement
of the difference Casimir force in the dielectric
to metal phase transition is of much interest for actuation
of nanodevices through a modulation of the Casimir force,
and for further tests of the Lifshitz theory. The latter
demands precision measurements of the Casimir force in the
configuration of an Au-coated sphere and VO${}_2$ film
and careful comparison between the experimental data
and theory.

\begin{figure}[t]
\vspace*{-5.7cm}
\hspace*{-.45cm}
\psfig{file=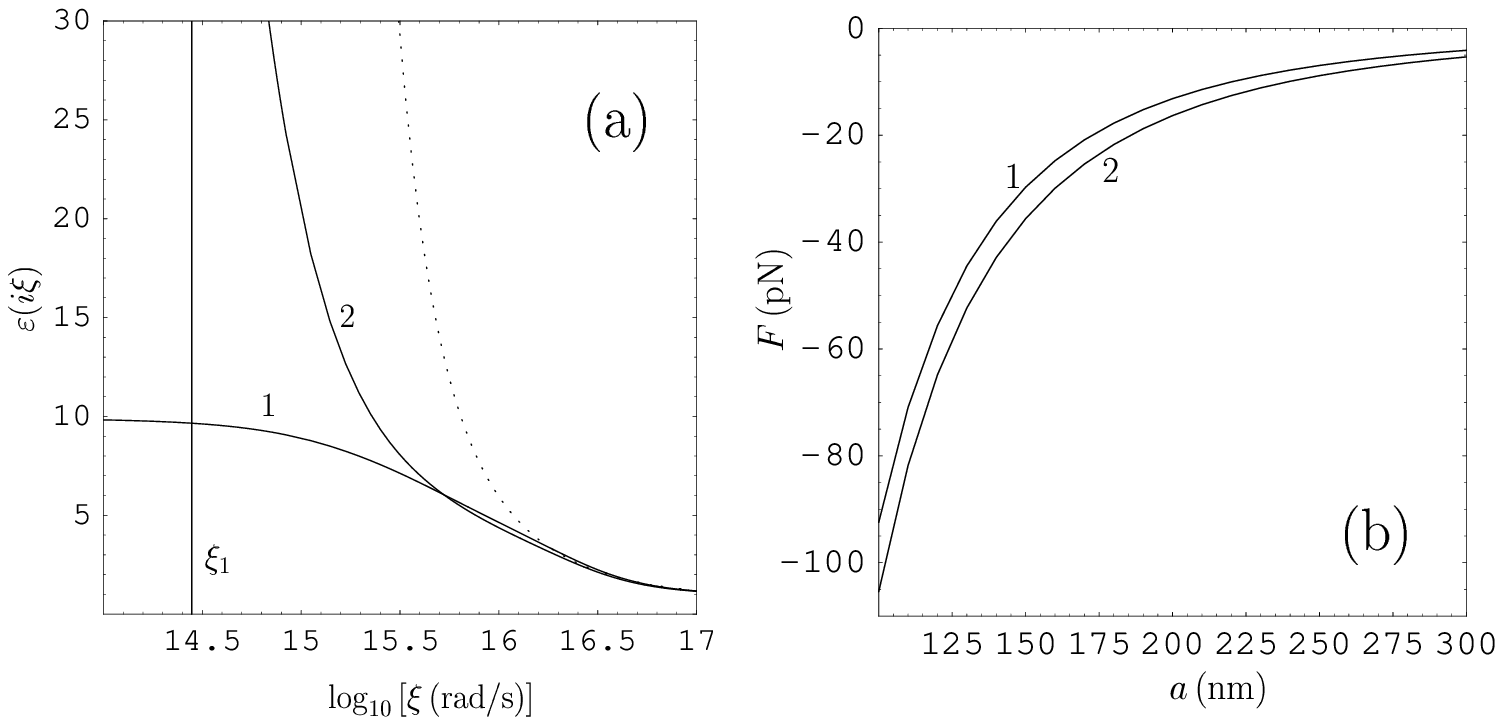,width=6.5in}
\vspace*{-12.cm}
\caption{(a) The effective dielectric permittivity of VO${}_2$ film
on sapphire substrate along the
imaginary frequency axis before and after the phase
transition are shown by the solid lines 1 and 2,
respectively. The permittivity of Au is indicated by the
dotted line.
(b) The Casimir force between an Au coated sphere and VO${}_2$
film on sapphire substrate
versus separation before and after the phase transition
are shown by the solid lines 1 and 2, respectively.}
\label{aba:fig19}
\end{figure}

The Lifshitz theory expresses the Casimir force both before
(the density of charge carriers $n$) and after
(the density of charge carriers $\tilde{n}$) the phase
transition by Eqs.~(\ref{eqn3}) and (\ref{eqn7}).
To compute the Casimir force both before and after the phase
transition one needs the optical data of VO${}_2$ on a
sapphire plate in a wide frequency region. Such data were
measured\cite{87} and used in the fit to the oscillator
model of the dielectric permittivity in the frequency
region from 0.25 to 5\,eV. The additional terms were also
added\cite{40} to the dielectric permittivity taking into
account the contributions from the optical data at higher
frequencies up to about 10\,eV.
As a result, the effective dielectric permittivity of the
VO${}_2$ film on a sapphire substrate at the imaginary Matsubara
frequencies before the phase transition (at $T=300\,$K)
takes the form
\begin{equation}
\varepsilon_n(i\xi_l)=1+\sum_{i=1}^{7}
\frac{s_{n,i}\omega_{n,i}^2}{\omega_{n,i}^2+\xi_l^2+
\Gamma_{n,i}\xi_l}+
\frac{(\varepsilon_{\infty}^{(n)}-1)\omega_{\infty}^2}{\omega_{\infty}^2
+\xi_l^2}.
\label{eqn34}
\end{equation}
\noindent
Here, $\varepsilon_{\infty}^{(n)}=4.26$, $\omega_{\infty}=15\,$eV
and the values of all oscillator parameters can be found in
Refs.~\refcite{40} and \refcite{87}.
In Fig.~\ref{aba:fig19}(a) the dielectric permittivity (\ref{eqn34})
as a function of frequency is shown by the solid line labeled 1.
In the same figure, the dielectric permittivity of Au computed
using the generalized plasma-like model is shown as dots.
The vertical line indicates the position of the first Matsubara
frequency at $T=340\,$K, i.e., in the region of the phase
transition.

After the phase transition the VO${}_2$ film is in metallic state with
increased charge carrier density $\tilde{n}$. In this case
the effective dielectric permittivity of the
VO${}_2$ film on a sapphire substrate is given by
\begin{equation}
\varepsilon_{\tilde n}(i\xi_l)=1+
\frac{\omega_{p,\tilde{n}}^2}{\xi_l(\xi_l+\gamma_{\tilde n})}+
\sum_{i=1}^{7}
\frac{s_{\tilde n,i}\omega_{n,i}^2}{\omega_{\tilde n,i}^2+\xi_l^2+
\Gamma_{\tilde n,i}\xi_l}+
\frac{(\varepsilon_{\infty}^{(\tilde n)}-1)\omega_{\infty}^2}{\omega_{\infty}^2
+\xi_l^2}.
\label{eqn35}
\end{equation}
\noindent
The values of the oscillator parameters used here can be found in
Refs.~\refcite{40} and \refcite{87}, whereas the other parameters
are\cite{87}
$\varepsilon_{\infty}^{(\tilde n)}=3.95$, $\omega_{p,\tilde n}=3.33\,$eV,
and $\gamma_{\tilde n}=0.66\,$eV.
In Fig.~\ref{aba:fig19}(a) the dielectric permittivity (\ref{eqn35})
 is shown by the solid line labeled 2.

The computational results for the Casimir force
 between an Au-coated sphere and VO${}_2$
film deposited on a sapphire plate versus separation were
obtained\cite{40} using the dielectric permittivity (\ref{eqn34})
before the phase transition and dielectric permittivity
(\ref{eqn35})
after the phase transition.
They are shown in Fig.~\ref{aba:fig19}(b) by the solid lines 1
and 2, respectively. As can be seen in Fig.~\ref{aba:fig19}(b),
after the phase transition the magnitude of the Casimir force
increases due to an increase in the charge carrier density and
structural changes of the material. Specifically, the difference
Casimir force from a phase transition changes from 13\,pN at
$a=100\,$nm to 1.2\,pN at $a=300\,$nm. Note that the Casimir
forces between an Au-coated sphere and VO${}_2$ film deposited
on a sapphire were also computed\cite{88} not using the
effective dielectric permittivities, as above, but considering
instead a two layer system of VO${}_2$ and sapphire with their
individual permittivities. However, computations of Casimir
forces in Ref.~\refcite{88} were performed in the framework
of the Lifshitz theory at zero temperature.

The difference Casimir force in the dielectric-to-metal phase
transition can be used as one more test of the Lifshitz theory.
Similar to Sec.~3, we arrive at different results for the
difference Casimir force depending on whether or not the dc
conductivity of the dielectric VO${}_2$ before the phase
transition was taken into account in computations.
The quantity equal to the difference Casimir force with
neglected dc conductivity of dielectric VO${}_2$ minus
the difference Casimir force with dc conductivity included
varies from 1.6\,pN at $a=100\,$nm to 0.2\,pN at $a=300\,$nm.
This can be used to experimentally discriminate between the
two approaches and deeply probe the role of material
properties in the Lifshitz theory at nonzero temperature.

\section{Semiconductor plate with rectangular corrugations}

In this section we discuss two experiments\cite{41,42}
on the measurement of the gradient of the Casimir force
between an Au-coated sphere of a radius $R=50\,\mu$m
and a Si surface that has been structured with nanoscale
rectangular corrugations (trenches). Measurements were
performed in the dynamic regime using a micromechanical
torsional oscillator. This means that the directly measured
quantity was the shift of the resonant frequency of the oscillator,
which is proportional to the gradient of the Casimir force.
Using Eq.~(\ref{eqn14}) this gradient can be recalculated
into the Casimir pressure between the two plates one of which
is flat and another one rectangular corrugated. In both
experiments the trenches were fabricated in $p$-type silicon
with the density of charge carriers
$2\times 10^{18}\,\mbox{cm}^{-3}$ and respective resistivity
equal to $0.028\,\Omega\,$cm. Note that this density of charge
carriers is a bit smaller than the critical value above which
Si becomes a metallic-type semiconductor.

\begin{figure}[t]
\vspace*{-11.5cm}
\hspace*{-2.4cm}
\psfig{file=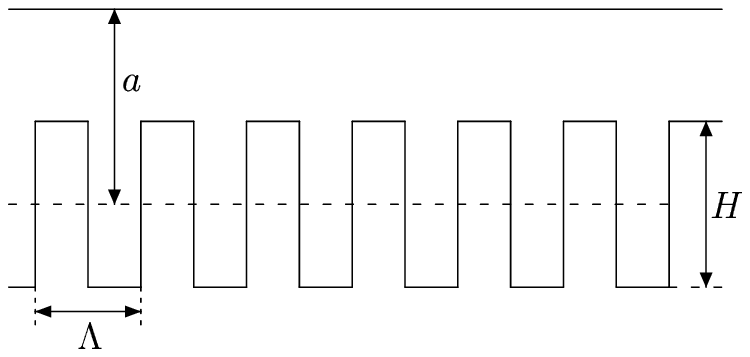,width=7.5in}
\vspace*{-12.cm}
\caption{The periodic uniaxial rectangular corrugations
on one of the plates.}
\label{aba:fig20}
\end{figure}
The configuration of two plates with periodic rectangular
corrugations on one of them used in the first experiment\cite{41}
is shown in Fig.~\ref{aba:fig20}. The depth of the trenches
was $H\approx 1\,\mu$m. Three types of lower plate of size
$0.7\times0.7\,\mbox{mm}^2$ were made: sample A with a period
$\Lambda_{\rm A}=1\,\mu$m, sample B with a period
$\Lambda_{\rm B}=400\,$nm and one additional sample with a flat
surface. The sidewalls of the trenches A and B were described as
having angles of 90.3${}^{\circ}$ and 91.0${}^{\circ}$,
respectively, with the top surfaces. The bottom corners were
reported to have some degree of rounding in comparison with
the top corners. Perfect corners of 90${}^{\circ}$ were assumed
in all of the analyses. Fractional areas of the top surfaces
of the unetched regions were found to be
$p_{\rm A}=0.478\pm 0.002$ and $p_{\rm B}=0.510\pm 0.001$
for the samples A and B using a scanning electron microscope.

The electrostatic calibrations and the measurements of the
gradient of the Casimir force were done in a vacuum of
$10^{-6}\,$Torr. A residual potential difference
$V_0\sim -0.43\,$V between the sphere and the flat Si plate
was noted to vary by 3\,mV over the range of separations
from 600\,nm to $2.5\,\mu$m. Since no analytic expression
for the electrostatic force is available for a trench
geometry, a $2D$ numerical solution of the Poisson equation
was used to calculate the electrostatic energy between a
flat plate and a trench surface. This energy was then
converted to a force between a sphere and the trench surface
using the PFA.

The Casimir force gradients between the sphere and the flat
plate and samples A and B were measured at the laboratory
temperature $T=300\,$K  after the application of compensating
voltages to the plates. The main uncertainty in these
measurements was from thermodynamical noise with a value
of about $0.64\,\mbox{pN\,$\mu$m}^{-1}$ at $a=800\,$nm.
The gradient of the Casimir force between the flat plate and
an Au sphere $F_{\rm flat}^{\prime}$ was measured first.
Good agreement was claimed between these measurement results
and the Lifshitz theory at zero temperature using the
tabulated optical data\cite{46} for Au and Si extrapolated
to low frequencies by means of the Drude model.
There are, however, some deviations between the data and
the computed force gradient (note that
large effect from the inclusion of dc conductivity is more
pronounced when the Lifshitz formula at nonzero temperature
is used in computations). The roughness correction was taken
into account employing an rms roughness of 4\,nm on the
sphere and 0.6\,nm on the silicon surface measured by means
of an AFM. Next, the force gradients $F_{\rm A,expt}^{\prime}$
and $F_{\rm B,expt}^{\prime}$ were measured on the
corrugated surfaces.

The measurement results were compared with theory
at zero temperature. For the
configuration with deep rectangular corrugations one can
neglect the contribution of the remote bottom parts of the
trenches. Then the PFA leads to the following results for
the samples A and B
\begin{equation}
F_{\rm A,PFA}^{\prime}(a)=-2\pi Rp_{\rm A}
P_r\left(a-\frac{H}{2}\right), \quad
F_{\rm B,PFA}^{\prime}(a)=-2\pi Rp_{\rm B}
P_r\left(a-\frac{H}{2}\right),
\label{eqn36}
\end{equation}
\noindent
where $P_r$ is the Casimir pressure between two parallel
noncorrugated plates calculated by the
Lifshitz formula  (\ref{eqn6c})
taking into account stochastic roughness.
To compare the experimental data with theory the
following ratios were considered\cite{41}
\begin{equation}
\rho_{\rm A}(a)=
\frac{F_{\rm A,expt}^{\prime}(a)}{F_{\rm A,PFA}^{\prime}(a)}, \quad
\rho_{\rm B}(a)=
\frac{F_{\rm B,expt}^{\prime}(a)}{F_{\rm B,PFA}^{\prime}(a)}.
\label{eqn37}
\end{equation}
\noindent
It was found that for sample A there were deviations of
$\rho_{\rm A}$ from unity of up to 10\% over the measurement
range from $a=650$ to 750\,nm, exceeding the experimental errors.
For sample B, there were deviations of $\rho_{\rm B}$ from
unity up to 20\% over the same measurement range. These
deviations are natural as $a/\Lambda_{\rm A}=0.7$ and
$a/\Lambda_{\rm B}=1.75$ at the typical separation considered,
$a=700\,$nm. As discussed in the literature,\cite{5,15,89}
the application region of the PFA to periodic structures is
restricted by an additional condition $a/\Lambda\ll 1$,
where $\Lambda$ is the period. Thus, for sample B the applicability
condition of the PFA is violated to a larger extent than
for sample A.

The experimental data were also compared with exact theoretical
values\cite{90} computed for ideal metal corrugations at zero
temperature. In so doing the PFA was not applied to corrugations,
but used only to convert the plate-plate to sphere-plate case.
However, the above measured deviations from the PFA were found
to be 50\% less than calculated for ideal metal boundaries.
This discrepancy was reported as being quite natural, owing to
the fact that exact calculation did not take into account
nonzero skin depth of both Au and Si surfaces. Furthermore,
exact calculations for a corrugated plate were
performed\cite{91} at zero temperature where Au was described by
the simple plasma model and the Si grating by the generalized
Drude-like model.\cite{88} This allowed to decrease discrepancies
between experiment and theory.

The second experiment\cite{42} on the measurement of the Casimir
force gradient between an Au sphere and a plate covered with
rectangular corrugations was performed with smaller corrugation
depth. In this case, unlike the first experiment, both top and
bottom surfaces of the corrugations contributed to the force
gradient. For the second esperiment, the same $p$-type Si was
used as in the first with the parameters indicated above.
The period of corrugations was $\Lambda=400\,$nm. The sidewalls of
corrugations were at 94.6${}^{\circ}$ to the top surface, i.e.,
not exactly vertical. As a result, the lengths of the top surface
and the bottom surface in one period were measured to be
$l_1=185.3\,$nm and $l_2=199.1\,$nm, respectively. The average
depth of corrugations was $H=97.8\pm 0.7\,$nm (i.e., an order
of magnitude less than in the first experiment).
Another sample consisted of a flat surface with no corrugations
was also used with the same optical properties.

The same micromechanical torsional oscillator was used in the
dynamic regime to measure the gradient of the Casimir force.
The sphere radius was $R=50\,\mu$m. Voltages were applied to
the flat plate for calibration purposes while the sphere
remained grounded. The residual voltage was found to be
$V_0=-0.499\,$V from the fit to the exact force-distance
relation. It was found to change only by less than 3\,mV
for $a$ changing from 150 to 650\,nm. Similar calibration
procedure was repeated for a corrugated Si plate, but
instead of an analytic force-distance relation the fit to
a numerical solution of the Poisson equation was performed.
Next, the gradients of the Casimir force,
$F_{f,\rm expt}^{\prime}(a)$ and $F_{c,\rm expt}^{\prime}(a)$,
on the flat and corrugated surfaces, respectively, were
measured by applying the compensating voltages to the plate
in order to compensate the residual potential difference
$V_0$.

The obtained experimental data were compared with the results
of theoretical computations.  For a flat plate
$F_{f,\rm theor}^{\prime}(a)$ was computed using Eq.~(\ref{eqn14})
and the Lifshitz formula for the Casimir pressure at zero
temperature. In so doing Si was described by the generalized
Drude-like model (\ref{eqn12}) with the Drude parameters
$\omega_p^{(2)}=1.36\times 10^{14}\,$rad/s and
$\gamma^{(2)}=4.75\times 10^{13}\,$rad/s.
Au was described by the tabulated optical data extrapolated
to low frequencies by means of the Drude model with the
parameters indicated in Sec.~2.1. The surface roughness
($\sim 4\,$nm rms for an Au surface and
$\sim 0.6\,$nm rms for a Si surface) was taken into account
by the method of geometrical averaging.\cite{5,15,17,24}
The comparison between the data, $F_{f,\rm expt}^{\prime}(a)$,
and the computational results, $F_{f,\rm theor}^{\prime}(a)$,
corrected to a presence of surface roughness, shows that there
is some disagreement (the measured force gradients are less
than the computed ones). This can be explained by the use of
the Lifshitz formula at zero temperature in computations,
whereas the experiment was performed at $T=300\,$K.
The related differences are of about --0.5\% or 1\%
depending on the model of conductivity properties of Si
used at nonzero temperature.

Now we discuss the comparison between experiment and theory for
a corrugated plate. Here, the approximate expression for the
gradient of the Casimir force can be simply obtained using
the PFA from Eq.~(1) of Ref.~\refcite{42}. The result is:
\begin{eqnarray}
&&
F_{c,r;\rm PFA}^{\prime}(a)=-2\pi R\left\{
\vphantom{\frac{1-p_1-p_2}{H}}
p_{1}P_r\left(a-H_1\right)+p_{2}
P_r\left(a+H_2\right)\right.
\nonumber \\
&&~~~~~~~~~~~~~~~~~~~~~~
\left.
+\frac{1-p_1-p_2}{H}
\left[E_r\left(a-H_1\right)-
E_r\left(a+H_2\right)\right]\right\}.
\label{eqn38}
\end{eqnarray}
\noindent
Here, similar to Eq.~(\ref{eqn36}), $P_r$ is the Casimir
pressure between two parallel noncorrugated plates
at $T=0$ calculated
by the Lifshitz formula (\ref{eqn6c})
 with account of surface roughness. The Casimir energy
$E_r$ is given by the Lifshitz formula (\ref{eqn6b})
 and is also corrected for the
presence of roughness using the method of geometrical
averaging. The quantities $p_1$ and $p_2$ are defined as
$p_1=l_1/\Lambda$, $p_2=l_2/\Lambda$, and
$H_i=(\Lambda-l_i)/(2\Lambda-l_1-l_2)$.
The first and second
contributions on the right-hand side of Eq.~(\ref{eqn38})
describe the role of the top and bottom surfaces of the Si
grating, respectively. They contribute about 97\%
of the Casimir force computed using the PFA.

The comparison between the data, $F_{c,\rm expt}^{\prime}(a)$,
and the computational results, $F_{c,\rm theor}^{\prime}(a)$,
corrected to a presence of surface roughness, shows that the
measured force gradients are a bit larger than the computed
ones. This again might be partially connected with the use
of the Lifshitz formula at zero temperature. Note that the
PFA result for the gradient of the Casimir force at any
$T\neq 0$ is obtained from Eq.~(\ref{eqn38}) by using the
pressures $P_r$ calculated using the Lifshitz formula at
nonzero temperature and by replacing the energy $E_r$
with the free energy ${\cal F}_r$.
With the increase of
separation from $a=0.2$ to $0.55\,\mu$m the experimental
data were found to be in better agreement with the
PFA results.
Specifically, the ratio
\begin{equation}
\rho(a)=
\frac{F_{c,\rm expt}^{\prime}(a)}{F_{c,r;\rm PFA}^{\prime}(a)}
\label{eqn39}
\end{equation}
\noindent
was equal\cite{42} to approximately 1.15 at $a=200\,$nm and
1 at $a=0.53\,\mu$m (our definition of separation $a$
counted from the mean level of corrugations is connected
with that used in Ref.~\refcite{42} as
$a\approx z+50\,$nm).

The gradient of the Casimir force between an Au sphere and
corrugated Si plate, $F_{c,\rm theor}^{\prime}(a)$, was also
computed using the exact scattering approach at $T=0$.
It was shown that the ratio
\begin{equation}
\rho^{\rm theor}(a)=
\frac{F_{c,\rm theor}^{\prime}(a)}{F_{c,r;\rm PFA}^{\prime}(a)}
\label{eqn40}
\end{equation}
\noindent
varies from 1.1 at $a=200\,$nm to 1.05 at $a=550\,$nm.
This was characterized as a good agreement with variations of
the quantity $\rho(a)$ indicated above.
It should be noted, however, that both quantities,
$F_{c,r;\rm PFA}^{\prime}(a)$ in the denominator of
$\rho(a)$ and $\rho^{\rm theor}(a)$, were computed at $T=0$
while the measurements were performed at $T=300\,$K.
For real materials the quantity
 $\rho^{\rm theor}(a)$ decreases towards unity
 with the increase of $a$ just as for the case of ideal
 metals.\cite{90} Note that at separations $a<270\,$nm
 there is some disagreement between the exact theory and
 the experimental data [because the differences
$\rho(a)-\rho^{\rm theor}(a)$ exceed the experimental
errors]. This
might be explained by the slow convergence of computations
in the exact theory and insufficient number of iterations used.

All in all, the second experiment using an Au sphere
interacting with a Si corrugated plate provided further
evidence concerning an interplay between geometry effects and
material properties in the case of rectangular corrugated
surfaces. The Casimir force measured at $T=300\,$K was found
to deviate from the PFA computations at $T=0\,$K by about 10\%.
It is pertinent to note that the quantitative comparison at
a 95\% confidence level between the measurement data, the
exact computational results using the scattering approach, and
the predictions of the PFA was performed\cite{92,93} in an
experiment measuring the lateral Casimir force in the
configuration of a sinusoidally corrugated sphere and
a sinusoidally corrugated plate. In this experiment both
test bodies were metallic (coated with Au). The experimental
data were found to be in excellent agreement with the
results of exact computations using the scattering approach,
but deviate significantly from the results of the PFA
computations. In so doing the exact computations were
performed at the laboratory temperature $T=300\,$K.
This provided the quantitative confirmation of the
observation of diffraction-type effects in the Casimir
force that are disregarded within the PFA approach.
Note that for a lateral Casimir force the quantity analogous
to $\rho^{\rm theor}$ in Eq.~(\ref{eqn40}),
\begin{equation}
\rho_{\rm lat}^{\rm theor}(a)=
\frac{F_{c,\rm theor}^{\rm lat}(a)}{F_{c,r;\rm PFA}^{\rm lat}(a)},
\label{eqn40a}
\end{equation}
\noindent
is less than unity. According to the computational
results,\cite{92,93} $\rho_{\rm lat}^{\rm theor}(a)$ decreases
with the increase of $a$, i.e., for the lateral Casimir force
the accuracy of the PFA computations
decreases with the increase of separation.
This is because the lateral force is determined
by the dependence of the Casimir energy on the
phase shift between the corrugations.

\section{Semiconductors in three-layer systems:
Pulsating Casimir force}

In all experiments using semiconductor test bodies discussed
above, the Casimir force was always attractive.
The possibility to realize a repulsive Casimir force attracts
much attention due to potential applications in nanotechnology
to avoid stiction of closely positioned microelements and
to actuate the periodic movement in electro- and
optomechanical micromachines. Up to the present, however,
most of the proposals on this subject were far from
experimental realization. The repulsive Casimir forces have
long been discussed for an ideal metal sphere\cite{5} and
ideal metal rectangular boxes with appropriate ratio of
side lengths.\cite{5,94}  These results, however,
were not generalized for spheres and boxes made of real
metals. Numerous speculations on the possibility of Casimir
repulsion between artificial metamaterials have recently
been proven to be not realizable.\cite{95,96}

The single realistic possibility for the observation of the
Casimir repulsion in three-layer systems was proposed long
ago. It was noticed\cite{7} that repulsion happens when the
two semispaces (plates) with the dielectric permittivities
along the imaginary frequency axis
$\varepsilon^{(2)}(i\xi)<\varepsilon^{(1)}(i\xi)$
are separated with a material layer having the dielectric
permittivity $\varepsilon^{(0)}(i\xi)$ such that
\begin{equation}
\varepsilon^{(2)}(i\xi)<\varepsilon^{(0)}(i\xi)
<\varepsilon^{(1)}(i\xi).
\label{eqn41}
\end{equation}
\noindent
To obtain repulsion, inequality (\ref{eqn41}) should hold over a
sufficiently wide frequency region. This can be achieved by
using a liquid intermediate layer with the dielectric
permittivity $\varepsilon^{(0)}(i\xi)$.
Experimentally the effect of Casimir repulsion was
demonstrated\cite{97}\cdash\cite{100} in sphere-plate
geometry using $\alpha$-alumina and Au spheres,
amorphous SiO${}_2$ plates and different liquids (cyclohexane,
bromobenzene etc.) at separations up to about 10\,nm.
Note that at such separations the relativistic retardation effects
are already important. The Casimir repulsion between
an Au sphere and a SiO${}_2$ plate with an intermediate layer
of bromobenzene was observed\cite{101} at separations of
about 30\,nm. Some problems which add complexity to
measurements of the Casimir force in liquids, specifically,
the formation of a double layer owing to salt impurities,
are discussed in Ref.~\refcite{102}.
Here, we consider how semiconductor can be employed in three-layer
systems for the realization of a pulsating Casimir
force,\cite{43} i.e., interchangeable replacement of attraction
with repulsion and vice versa without use of mechanical springs.

Let us consider three pairs of plates immersed in ethanol. The first
pair includes one Au and one Si plate and the Si plate is
illuminated by laser pulses as in the experiment on the
optically modulated Casimir force discussed in Sec.~3.
The second pair consists of two similar Si plates, and only one
of them is illuminated with laser pulses.
According to Sec.~3, the illumination of Si with light of
appropriate power and wavelength increases the charge carrier
density by four orders of magnitude and changes the dielectric
permittivity along the imaginary frequency axis over a wide
frequency range. This can be used to satisfy Eq.~(\ref{eqn41})
or, on the contrary, to violate it when it was satisfied in the
absence of light. In the third pair of plates immersed in
ethanol, we consider one plate made of $\alpha$-Al${}_2$O${}_3$
and the other of Si. The latter plate is also illuminated with
laser pulses. Computations of the Casimir force show that
for the first pair of plates there is a repulsive Casimir force
within a wide range of separations when the light is off and an
attractive force when the light is on. For the second and third
pair of plates, the force is repulsive when the light is on
and attractive when the light is off.

To perform these computations the Lifshitz formula for the Casimir
pressure in three-layer systems was applied.\cite{5,43}
It has exactly the same form as Eq.~(\ref{eqn6}) but the
reflection coefficients (\ref{eqn4}) should be replaced
with
\begin{eqnarray}
&&
r_{\rm TM}^{(n)}=\frac{\varepsilon^{(n)}(i\xi_l)k_l^{(0)}-
\varepsilon^{(0)}(i\xi_l)k_l^{(n)}}{\varepsilon^{(n)}(i\xi_l)k_l^{(0)}
+\varepsilon^{(0)}(i\xi_l)k_l^{(n)}},
\nonumber \\
&&
r_{\rm TE}^{(n)}=\frac{k_l^{(0)}-k_l^{(n)}}{k_l^{(0)}+k_l^{(n)}},
\qquad
k_l^{(0)}\equiv\left[k_{\bot}^2+\varepsilon^{(0)}(i\xi_l)
\frac{\xi_l^2}{c^2}\right]^{1/2}\!\!.
\label{eqn42}
\end{eqnarray}

To calculate the Casimir pressure for the above three pairs of
plates, in addition to the dielectric permittivities of Au and
Si discussed in Secs.~2 and 3, one needs the dielectric
permittivities of $\alpha$-Al${}_2$O${}_3$ and ethanol along
the imaginary frequency axis. They can be presented in the
Ninham-Parsegian form\cite{104,105}
\begin{equation}
\varepsilon^{(n)}(i\xi_l)=1+
\frac{C_n^{\rm IR}}{1+\xi_l^2/\omega_{{\rm IR},n}^2}+
\frac{C_n^{\rm UV}}{1+\xi_l^2/\omega_{{\rm UV},n}^2},
\label{eqn43}
\end{equation}
\noindent
where the values of all parameters for $\alpha$-Al${}_2$O${}_3$
($n=1$) and for ethanol ($n=0$) can be found in
Refs.~\refcite{43,104,105}.
\begin{figure}[t]
\vspace*{-10.5cm}
\hspace*{0.5cm}
\psfig{file=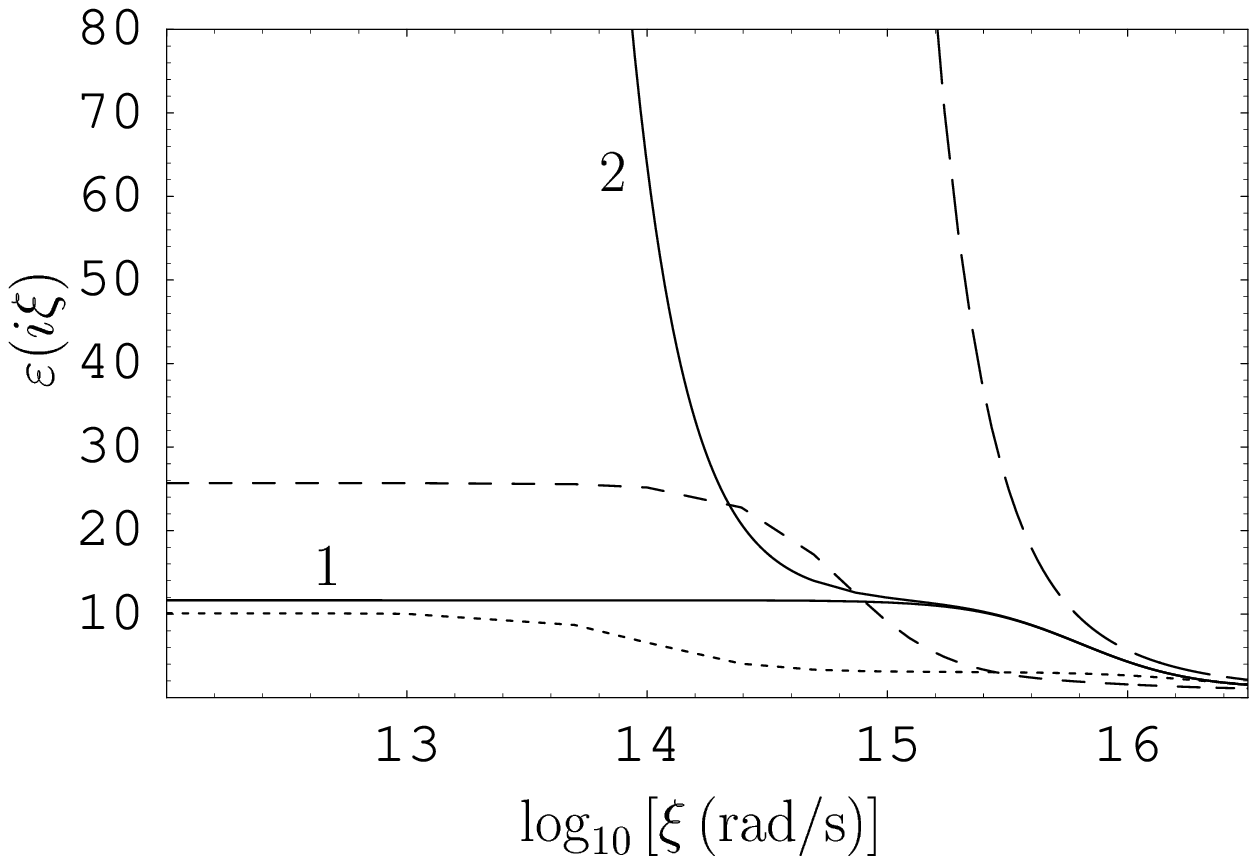,width=6.in}
\vspace*{-5.cm}
\caption{The dielectric permittivities of different materials
along the imaginary frequency axis are shown with solid
lines 1 and 2 for Si in the absence and in the presence of laser
light, respectively, with a long-dashed line for Au, with a
short-dashed line for ethanol, and with a dotted line for
$\alpha$-Al${}_2$O${}_3$.}
\label{aba:fig21}
\end{figure}
In Fig.~\ref{aba:fig21}
we show all related dielectric permittivities along the
imaginary frequency axis. The dielectric permittivities of Si
 in the absence and in the presence of light are indicated by
 the solid lines 1 and 2, respectively.
The line 2 is plotted for the absorbed power
$P^{\rm eff}=9.3\,$mW (see Sec.~3).
 The dielectric
 permittivities of Au, ethanol, and $\alpha$-Al${}_2$O${}_3$
 are presented by the long-dashed, short-dashed, and dotted
 respective lines.

\begin{figure}[b]
\vspace*{-5.5cm}
\hspace*{-2.75cm}
\psfig{file=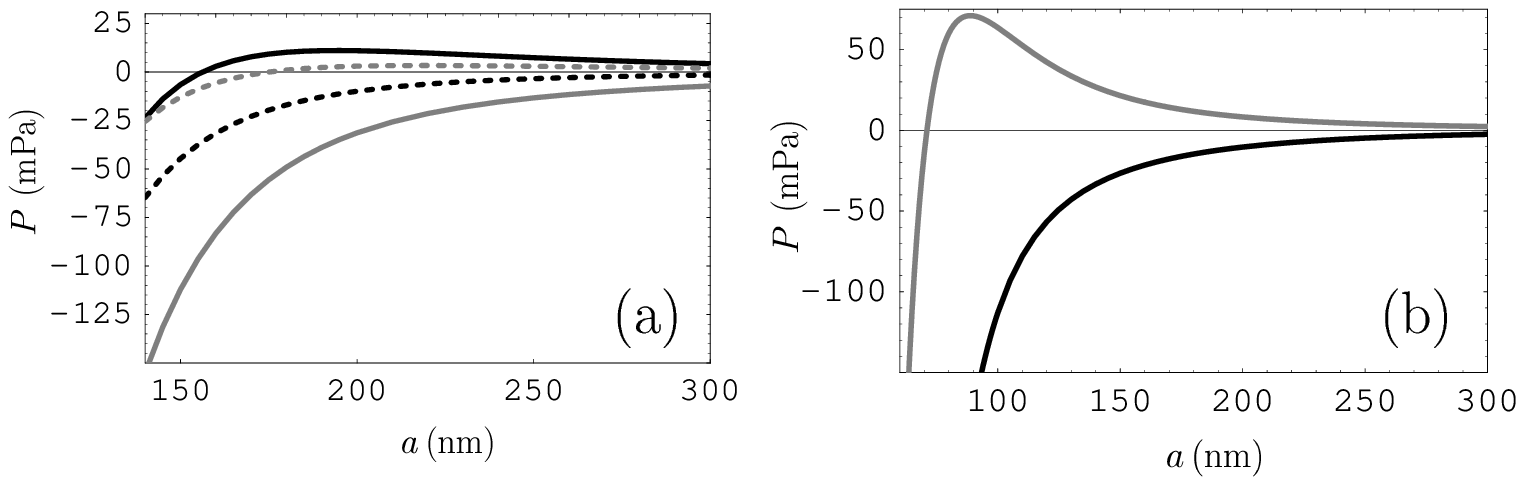,width=6.8in}
\vspace*{-14.8cm}
\caption{The Casimir pressure versus separation in
three-layer systems (a) Au--ethanol--Si (the black
and grey solid lines are for the absence and presence
of light on the Si plate, respectively) and
Si--ethanol--Si (the black
and grey dashed lines are for the absence of light on
both Si plates and for the presence
of the light on one of them);
(b) $\alpha$-Al${}_2$O${}_3$--ethanol--Si (the black
and grey lines are for the absence and presence
of light on the Si plate, respectively).}
\label{aba:fig22}
\end{figure}

In Fig.~\ref{aba:fig22}(a) we present the computational results
for the Casimir pressure versus separation for the first (the
solid lines) and second (the dashed lines) pairs of plates.
For the first pair of plates, i.e., for Au and Si plates
separated by ethanol, the pressure-distance dependence in the
absence of light on a Si plate is shown as the black solid
line. As can be seen in Fig.~\ref{aba:fig22}(a), for
separations larger than 156\,nm the Casimir pressure shown by
the black solid line is repulsive. This reflects the fact that
in Fig.~\ref{aba:fig21} the dielectric permittivity of ethanol
is sandwiched between the dielectric permittivities of Si
and Au in a wide frequency range, i.e., inequalities
(\ref{eqn41}) are satisfied. The pressure-distance dependence
in the presence of laser light on a Si plate is shown by the
grey solid line in Fig.~\ref{aba:fig22}(a). This line corresponds to
attraction at all separation distances. Physical explanation
of this fact is seen in Fig.~\ref{aba:fig21}, where the
dielectric permittivity of Si in the presence of light (the
solid line 2) is larger than the dielectric permittivity of
ethanol (the short-dashed line) within a wide frequency region.
Thus, the illumination of a Si plate with laser light violates
inequalities (\ref{eqn41}) and changes repulsion for attraction.

For the second pair of plates, i.e., for the two Si plates
separated by ethanol, the Casimir force in the absence of
laser light is attractive at all separations [see the
dashed black line in Fig.~\ref{aba:fig22}(a)]. This
result is expected because the permittivities of both
plates are equal. When, however, one Si plate is illuminated
with a laser light of appropriate wavelength and power,
its dielectric permittivity is described by the solid
line labeled 2 in in Fig.~\ref{aba:fig21}, and the inequalities
(\ref{eqn41}) are satisfied. In this case, the computational
results for the Casimir pressure versus separation are shown
by the dashed grey line in Fig.~\ref{aba:fig22}(a).
As can be seen in this figure, at $a<175\,$nm the Casimir
force is attractive, but at $a>175\,$nm it is repulsive.
Thus, the illumination of one of the two Si plates
separated by ethanole with light changes the Casimir attraction
to Casimir repulsion.

For the first and second pairs of plates, the magnitudes of the
repulsive forces were several times less than the magnitudes of
the attractive forces. This is not the case for the third pair
of plates, i.e., for the three-layer system
$\alpha$-Al${}_2$O${}_3$--ethanol--Si. In the absence of laser
light the inequalities (\ref{eqn41}) are not satisfied
(see Fig.~\ref{aba:fig21}). The computational results for the
Casimir pressure in this case are shown in Fig.~\ref{aba:fig22}(b)
by the black line. It is seen that the Casimir force is
attractive as expected. Now let the Si plate be illuminated with
the laser light.
Then the dielectric permittivity of Si is described by the solid
line labeled 2 in Fig.~\ref{aba:fig21}, and the inequalities
(\ref{eqn41}) are satisfied.
The respective computational results for the
Casimir pressure versus separation are shown
by the solid grey line in Fig.~\ref{aba:fig22}(b).
As can be seen in this figure, at separations $a>71.5\,$nm
the corresponding Casimir force is repulsive. In this case,
the Casimir repulsion and attraction are of the same order
of magnitude within a wide  range of separations. Thus, the
third pair of plates provides an example where the
illumination of the Si plate changes the Casimir attraction to
the Casimir repulsion of the same order of magnitude.

The possibility to observe the pulsating Casimir force was
discussed\cite{43} for the two plates completely immersed in
the liquid far away from any air-liquid interface in order to
prevent the occurrence of capillary forces. It is expected
that special procedures for surface preparation of the plates
will be necessary to bring about intimate contact between the
plates and liquid. Then the only liquid-based force is the drag
force due to the movement of the plates in response to the
change of the Casimir force. It can be seen, however, that the
drag force is negligibly small. Thus, for the Casimir pressure
values of around 10\,mPa and typical spring constants of
$0.02\,\mbox{N\,m}^{-1}$, the corresponding drag pressure from
plate movement would be six orders of magnitude less in value.
The proposed effect of the pulsating Casimir force can be
used to actuate the periodic movement of electrodes and mirrors
in electro- and optomechanical devices. This can be achieved
by using the standard frequency generators and modulators
to select the appropriate duration and time between the laser
pulses.

In fact the effect of a pulsating Casimir force is the
combination of the familiar properties of three-layer
systems\cite{5} and the recently discovered\cite{28,28a}
modulation of the Casimir force with light. The possibility
of pulsation owing to vacuum fluctuations alone without
use of mechanical springs
to a large measure depends on the use of
the repulsive Casimir force. According to the introductory
part of this section, three-layer systems are presently the
single configuration where the Casimir repulsion is
realized experimentally. In this respect it is pertinent
to mention that the repulsive Casimir force was recently
predicted\cite{106,107} in the configuration of two
parallel plates separated by an empty gap, one of which
is made of a ferromagnetic dielectric and another of
a metal described by the plasma model. By now the influence
of magnetic properties of plate materials onto the Casimir
force in not confirmed experimentally. Keeping in mind that
magnetic films are widely used in different
microelectromechanical devices, such a confirmation would
have an important impact on nanotechnology.

\section{Conclusions and discussions}

In the foregoing we have considered all experiments performed
up to date and a few proposals where the magnitude of the
Casimir force was controlled using semiconductor test bodies.
It is common knowledge that Casimir\cite{1} discovered his
famous effect for ideal metal planes and the most of modern
experiments on measuring the Casimir force were performed with
metallic test bodies.\cite{15} However, applications of the
Casimir effect in nanotechnology, which is heavily based on
the use of semiconductor materials, demand extensive study
of the Casimir force in configurations with semiconductor
surfaces. This opens considerable opportunities to change the
magnitude of the Casimir force in a desirable way and to use
this force for operation of microdevices. The first steps in
this direction were made by the many performed experiments
discussed in this review.

In Sec.~2 we  considered the results of two pioneer
experiments\cite{23}\cdash\cite{25} on measuring the Casimir
force between an Au-coated sphere and semiconductor plates
with different densities of charge carriers. Already the
first experiment of this kind\cite{23,24} using a plate
made of $p$-type Si demonstrated that the Casimir force was
markedly different from the calculated forces both in the
case of two metallic test bodies and an Au-coated sphere
interacting with high-resistivity dielectric Si.
In the second experiment,\cite{25} the Casimir forces
between an Au-coated sphere and two plates made of $n$-type
Si with different charge carrier densities were directly measured
in succession. In this way it was convincingly shown that the
Casimir force depends on the density of charge carriers and this
can be used to control its magnitude. In both experiments it
was confirmed that the replacement of an Au plate with a Si
plate leads to decrease in the magnitude of the Casimir force
from about 25\% to 35\% depending on the value of separation
in comparison with the case of two Au test bodies. Later it
was shown\cite{26} that the use of other semiconductor materials
can result in even larger decrease in the magnitude of the
Casimir pressure. Thus, the magnitude of the Casimir pressure
between an Au plate and an ITO plate was found\cite{26} to vary
by from 40\% to 50\% smaller than for two Au plates.
We have also discussed the experiment\cite{27} on measuring the Casimir
force between a Ge plate and a Ge spherical lens with about
15\,cm curvature radius. With respect to this experiment several
disadvantages were remarked which make the obtained results for
the Casimir force indefinite. Specifically, the subtracted
asymptotic expression for the electric force might be not
applicable at short separations, and the comparison with
theory was based on the version of PFA which does not
take into account surface imperfections which are always present
on lenses of centimeter-size radii.

The most important experiment with semiconductor surfaces
performed up to date is an experiment on the optical modulation
of the Casimir force considered in Sec.~3. This was the
measurement of the difference Casimir force between an
Au-coated sphere and a $p$-type Si plate illuminated with
laser pulses.\cite{28,28a}
According to the measurement scheme employed not the individual
forces, but only the difference of the forces in the presence
and in the absence of light on a Si plate has been measured. Here,
similar to the experiments considered in Sec.~2, the Casimir
force in the absence of light on a Si plate was of about
35\%--40\% smaller in magnitude than for two Au test bodies.
However, in the presence of light on a Si plate the force
magnitude increased by about 4\%--6\% depending on separation.
The increase and then the decrease of the force were repeated
periodically in line with the laser  pulses.
This opens new prospective opportunities for the use of the
Casimir force in the operation of micromachines.

The comparison of the measurement data of the experiment on
optically modulated Casimir force with the Lifshitz theory
resulted in an unexpected conclusion\cite{28,28a} that if the
dc conductivity of a Si plate in the absence of laser light
is taken into account in the model of dielectric response
the theory is excluded by the data at  a 95\% confidence
level. If the dc conductivity of Si in the dark phase
is omitted, the Lifshitz theory is found to be consistent
with the data within both 95\% and more narrow 70\%
confidence intervals. Keeping in mind that the dc conductivity
is a real physical effect and that the Lifshitz theory is a
fundamental theory based on first principles of quantum
statistical physics and quantum electrodynamics, the
obtained conclusion should be considered as challenging.
The hypothesis\cite{21} that some concepts of statistical
physics concerning interaction of real and fluctuating
electromagnetic fields might need a reconsideration
has raised a heated debate.\cite{79}
However, the experimental data of one more experiment\cite{67}
led to the same conclusion that the Lifshitz theory is in
disagreement with the data when the dc conductivity of a
dielectric plate is taken into account.\cite{71}
The attempts\cite{32}\cdash\cite{34} to modify the Lifshitz
theory were also found to be in disagreement with the data
of the experiment on the optically modulated Casimir force.
Thus, this experiment takes on great significance as a test
for fundamental physical theories.

In Sec.~4 we have discussed changes in the Casimir force which
occur when the material of the plate undergoes a phase
transition. Both already performed\cite{39} and
proposed\cite{40} experiments were considered. Specifically,
it was observed\cite{39} that the gradient of the Casimir
force between an Au-coated sphere and a crystalline plate
is greater by approximately 20\% than for an amorphous plate
made of the same semiconductor material (AgInSbTe).
In spite of rather large deviations between the experimental
data and theory in this experiment, the conclusion was made
that phase transitions in semiconductor materials are very
promising for the control of the Casimir force.

Another possibility to control the Casimir force discussed
in Sec.~5 is connected with the use of corrugated semiconductor
surfaces. We have discussed the two experiments\cite{41,42}
on the measurement of the Casimir force gradient between an
Au-coated sphere and Si plates covered with rectangular
corrugations of different profiles. These experiments
demonstrated that there is a nontrivial interplay between
material properties and geometrical profiles of semiconductor
surfaces which cannot be taken into account with sufficient
precision by using the PFA. Although the comparison of
experiment and theory in Refs.~\refcite{41,42} was shown to
be not complete (specifically, theory at zero temperature was
used for the interpretation of the room-temperature
measurement data) the obtained results show that geometrical
shape of interacting semiconductor surfaces is a promising
tool for the control of the Casimir force in addition to
material properties.

A more sophisticated possibility for control of the Casimir
force than all those listed above was considered in Sec.~6.
While in all the above experiments and proposals only the attractive
Casimir force was exploited, here the attraction is replaced
with repulsion and vice versa. Such a pulsating Casimir
force\cite{43} using no mechanical springs can be realized
by a combination of the optical modulation experiment with the
advantages of three-layer systems. As at least one of the layers,
a semiconductor material which changes its charge
carrier density by about five orders of magnitude under the
influence of a laser pulse can be used.

The investigation of possibilities to control the Casimir
force using semiconductor test bodies is presently
in the very beginning and
much should be done before some of them will find application in
industrially produced devices. In future the influence of many
different semiconductor materials with different kind of doping
on the Casimir force should be explored. It will be of much
interest to repeat the experiment on the optical modulation of
the Casimir force with plates made of different semiconductors
and in an ambient environment in order to determine the possibility
of its application in microelectromechanical
 devices. Till the moment, the
use of phase transitions in semiconductors in Casimir physics
is in a very initial state. A great variety of phase transitions
and their influence on the Casimir force will be investigated in
the near future. Last but not the least is the prospect to achieve
a deeper understanding of the role of charge carriers in the
theory of thermal Casimir force. Such an outstanding problem,
when it remains unsettled for a long time, may considerably retard
further experimental developments.

\section*{Acknowledgements}
This work was supported by the NSF grant PHY0970161 and DOE grant
DEF010204ER46131.


\end{document}